\pdfoutput=1
 \RequirePackage{fixltx2e}
\documentclass[twocolumn,showpacs,aps,epsfig,nofootinbib,floatfix]{revtex4}

\usepackage{graphicx}
\usepackage{epstopdf}
\usepackage{latexsym}
\usepackage{amssymb}
\usepackage[center]{subfigure}

\usepackage[center]{subfigure}

\begin{document}

 \newcommand{\bq}{\begin{equation}}
 \newcommand{\eq}{\end{equation}}
 \newcommand{\bqn}{\begin{eqnarray}}
 \newcommand{\eqn}{\end{eqnarray}}
 \newcommand{\nb}{\nonumber}
 \newcommand{\lb}{\label}
\newcommand{\PRL}{Phys. Rev. Lett.}
\newcommand{\PL}{Phys. Lett.}
\newcommand{\PR}{Phys. Rev.}
\newcommand{\CQG}{Class. Quantum Grav.}

\title{Static and rotating universal horizons and black holes in gravitational theories with broken Lorentz invariance}

\author{Kai Lin$^{a, b}$}
\email{lk314159@hotmail.com}

 \author{V. H. Satheeshkumar$^{c}$}
\email{vhsatheeshkumar@gmail.com}

\author{Anzhong Wang$^{a, c, d}$\footnote{Corresponding author}}
\email{anzhong_wang@baylor.edu}
\affiliation{$^{a}$Institute  for Advanced Physics $\&$ Mathematics,
Zhejiang University of
Technology, Hangzhou 310032,  China\\
$^{b}$Universidade Federal de Itajub\'a, Instituto de F\'isica e Qu\'imica, CEP 37500-903, Itajub\'a, Brazil\\
$^{c}$Departamento de F\'isica Te\'orica, Universidade do Estado do Rio de Janeiro, Rua S\~ao Francisco Xavier 524, Maracan\~a,
CEP 20550Ð013, Rio de Janeiro, RJ, Brazil\\
$^{d}$GCAP-CASPER, Physics Department, Baylor
University, Waco, TX 76798-7316, USA }

\date{\today}

\begin{abstract}

In this paper,  we show the existence of static and rotating universal horizons and black holes in gravitational theories with broken Lorentz invariance.  We pay particular attention to  the ultraviolet regime, and show that  universal horizons and black holes exist not only in the low energy limit but also 
at  the ultraviolet energy scales.  This is realized by presenting various static and stationary exact solutions of the full theory of the projectable  Ho\v{r}ava gravity with an extra  U(1) symmetry in (2+1)-dimensions, which, by construction,  is power-counting renormalizable.

\end{abstract}


\maketitle

\section{Introduction}
\renewcommand{\theequation}{1.\arabic{equation}} \setcounter{equation}{0}

Lorentz invariance (LI) has been the cornerstone of modern physics and is strongly supported by observations \cite{Kostelecky:2008ts}. In fact, all the experiments carried out so far are consistent with it, and there is no evidence to show that such a symmetry needs to be broken at a certain energy scale, although it is arguable that the constraints in the matter sector are much stronger than those in the gravitational sector \cite{M-L}. 

Nevertheless, there are various reasons to construct gravitational theories with broken LI. In particular, when spacetime is quantized, as what we currently understand from the point of view of quantum gravity \cite{Carlip,Kiefer}, space and time emerge from some discrete substratum.  Then,  LI, as a continuous spacetime symmetry, cannot apply to such discrete space and time any more. Therefore,  it cannot be a fundamental symmetry, but instead  an emergent one at the low energy physics.
 Following this line of thinking, various gravitational theories that violate LI have been proposed, such as ghost condensation \cite{ArkaniHamed:2003uy}, Einstein-aether theory \cite{TJ}, and more recently, Ho\v{r}ava  theory of gravity \cite{Horava}. While the ghost condensation and Einstein-aether theory are considered as the low energy effective theories of gravity, the Ho\v{r}ava gravity is supposed to be ultraviolet (UV) complete \cite{Horava}. In particular, in this theory the LI is broken in the UV, so the theory can include higher-dimensional spatial derivative operators. As a result, the UV behavior of the theory is dramatically improved and can be made power-counting renormalizable. On the other hand, the exclusion of higher-dimensional time derivative operators prevents the ghost instability, whereby the unitarity problem of the theory, known since 1977 \cite{Stelle:1976gc},
  is resolved. In the infrared (IR), the lower dimensional operators take over, whereby a healthy low-energy limit is presumably resulted \cite{Horava:2011gd}. Recently, it was shown   that the Ho\v{r}ava theory is not only power-counting renormalizable but also perturbatively renormalizable \cite{Barvinsky:2015kil}. In addition, it is also very encouraging that the theory is canonically quantizable in (1+1)-dimensional spacetimes with  \cite{Lia} or without  \cite{Lib} the projectability condition.

However, once LI is broken different species of particles can travel with different velocities, and in certain theories, including the Ho\v{r}ava theory  mentioned above, they can be even arbitrarily large. This suggests that black holes may exist only at low energies \cite{Wang:2012nv}. At high energies, any signal initially trapped inside the horizon may be able to escape out of it and propagate to infinity, as long as the signal has sufficiently large velocity (or energy). This seems in a sharp conflict with current observations that support the existence of rotating black holes in our universe \cite{Narayan}. 

\begin{figure}[tbp]
\centering
\includegraphics[width=1\columnwidth]{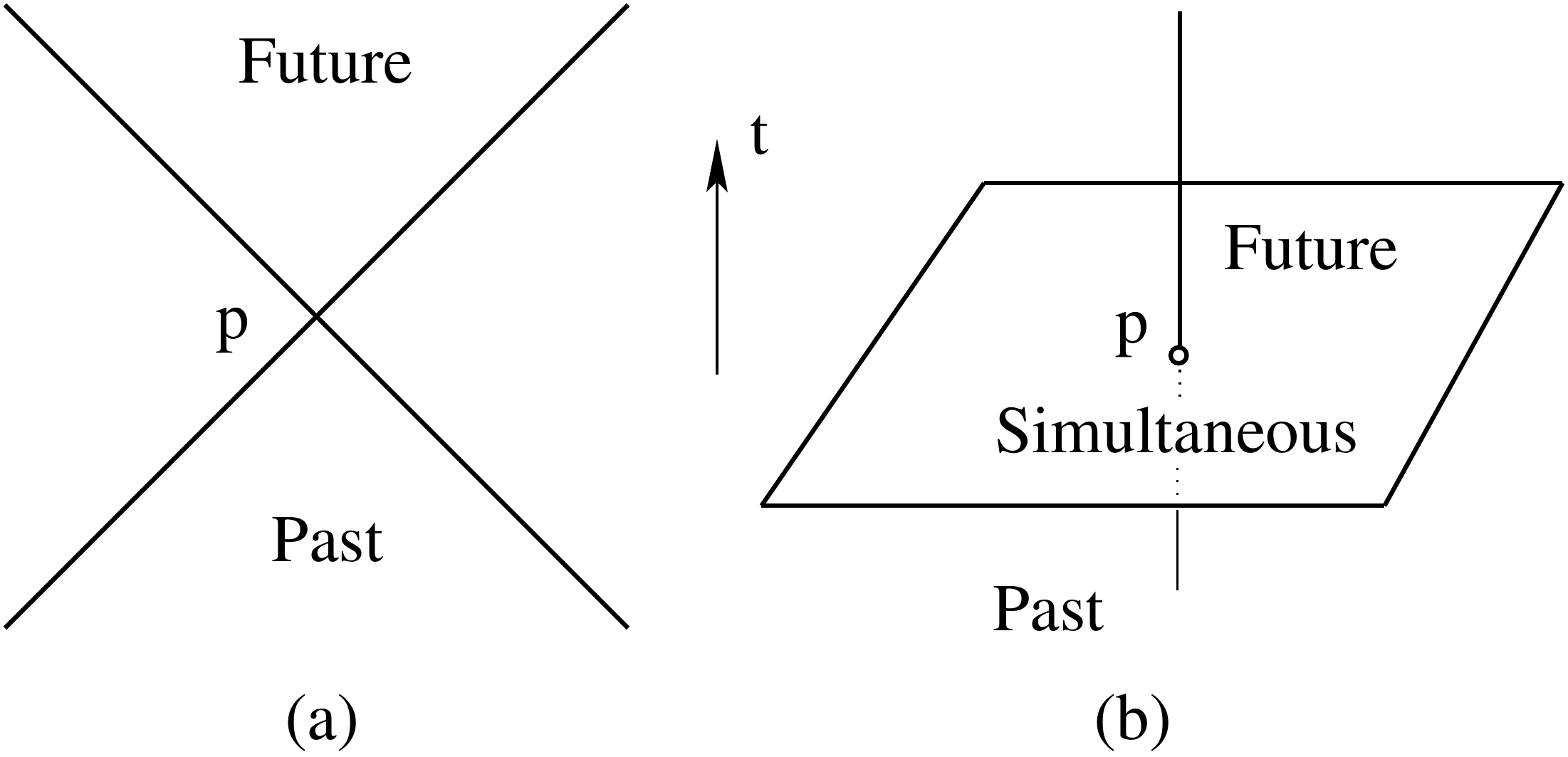}
\caption{(a) The light cone of the event $p$ in special relativity. (b) The causal structure of the point $p$ in Ho\v{r}ava theory.} 
\label{Fig1}
\end{figure}

\begin{figure}[tbp]
\centering
\includegraphics[width=1\columnwidth]{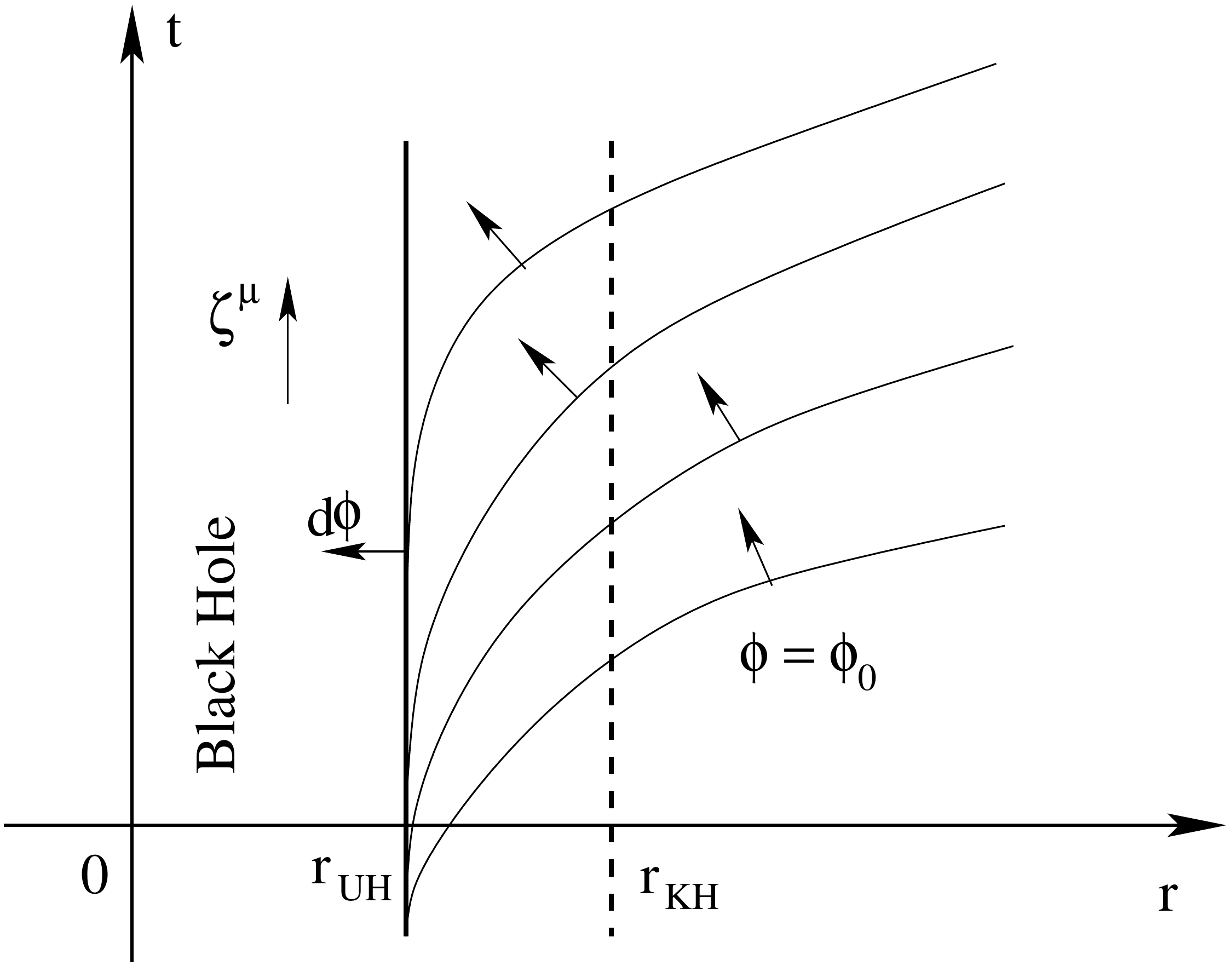}
\caption{Illustration of the bending of the $\phi$ = constant surfaces, and the existence of the universal horizon  in the Schwarzschild spacetime \cite{Lin:2014eaa}, where $\phi$ denotes the khronon field, and $t$ and $r$ are the Painlev\'{e}-Gullstrand coordinates. Particles move always along the increasing direction of $\phi$. The Killing vector $\zeta^{\mu} = \delta^{\mu}_{t}$ always points upward at each point of the plane. The vertical dashed  line is   the location of the Killing horizon, $r=r_{KH}$. The universal horizon, denoted by the vertical solid  line, is  located at $r = r_{UH}$, which is always inside the Killing horizon.} 
\label{Fig2}
\end{figure}

The above situation was dramatically changed in 2011 \cite{Blas:2011ni,Barausse:2011pu}, in which it was found that there still exist absolute causal boundaries, the so-called universal horizons, and particles even with infinitely large velocities would just move around on these boundaries and cannot escape to infinity. The main idea is as follows. In a given spacetime, a globally timelike scalar field, the so-called khronon \cite{Blas:2011ni}, might exist. Then, similar to the Newtonian theory, this khronon field defines a global absolute time, and all particles are assumed to move along the increasing direction of the khronon, so the causality is well defined [Cf. Fig.~\ref{Fig1}]. In such a spacetime, there may exist a surface as shown in  Fig.~\ref{Fig2}, denoted by the vertical solid line. Given that all particles move along the increasing direction of the khronon, from Fig.~\ref{Fig2} it is clear that a particle must cross this surface and move inward, once it arrives at it. This is an one-way membrane, and particles even with infinitely large speed cannot escape from it, once they are trapped inside it. So, it acts as an absolute horizon to all particles (with any speed),  which is often called {\em the universal horizon} \cite{Blas:2011ni,Barausse:2011pu}.  Since then, this subject has already attracted lots of attention \cite{Bhattacharyya:2015uxt}.

However, in most studies of universal horizons carried out so far the khronon plays a part of the gravitational theory involved \cite{Bhattacharyya:2015uxt}. To generalize the conception of the universal horizons to any gravitational theory with broken LI, recently we considered  the khronon as a test field, and assumed it to play the same  role as a Killing vector, so its existence does not affect the spacetime considered, but defines the properties of it  \cite{Lin:2014ija}. By this way, such a field is no longer part of the gravitational theory and it may or may not exist in a given spacetime, depending on the   spacetime considered. Then, we showed that the universal horizons indeed exist, by constructing concrete static charged solutions of the Ho\v{r}ava gravity.  Taking the khronon field as a test field,  we further showed that the  universal horizons  exist  and are always inside the Killing horizons \cite{Lin:2014eaa} in the three well-known black hole solutions: the Schwarzschild, Schwarzschild anti-de Sitter, and  Reissner-Nordstr\"om. It should be  noted that these solutions  are often also solutions of gravitational theories with the broken LI, such as the Ho\v{r}ava theory \cite{GLLSW}, and the Einstein-aether theory \cite{TJ}.

At the universal horizon, a slightly modified first law of black hole mechanics exists for the neutral Einstein-aether black holes  \cite{Berglund:2012bu}, but for the charged Einstein-aether black holes, such a first law is still absent  \cite{Ding:2015kba}. Using the tunneling method, the Hawking radiation at the universal horizon for a scalar field that violates the local LI was studied, and found that the universal horizon radiates as a blackbody at a fixed temperature  \cite{Berglund:2012fk}. A different approach was taken in  \cite{Cropp:2013sea}, in which ray trajectories in such black hole backgrounds were studied, and evidence was found, which shows that Hawking radiation is associated with the universal horizon, while the ``lingering" of low-energy ray trajectories near the Killing horizon hints a reprocessing there. However, the study of a collapsing null shell showed that the mode passing across the shell is adiabatic at late time  \cite{Michel:2015rsa}. This implies that large black holes emit a thermal flux with a temperature fixed by the surface gravity of the Killing horizon. This, in turn, suggests that the universal horizon should play no role in the thermodynamic properties of these black holes, although it should be noted that in such a setting, the khronon field is not continuous across the collapsing null shell. As mentioned above,  a  globally-defined khronon plays an essential role in the existence of a universal horizon, so it is not clear how the results presented in \cite{Michel:2015rsa} will be affected once the continuity of the khronon field is imposed.  On the other hand, using the Hamilton-Jacobi  method, recently we studied quantum tunneling of both relativistic and non-relativistic particles at Killing as well as  universal horizons of Einstein-Maxwell-aether black holes, after higher-order curvature corrections are taken into account  \cite{Ding:2015fyx}.  Our results showed that only relativistic particles are created at the Killing horizon, and the corresponding radiation is thermal with a temperature exactly the same as that found in general relativity. In contrary, only non-relativistic particles are created at the universal horizon and are radiated out to infinity with a thermal spectrum. However, different species of particles, in general, experience different temperatures. 
 
 In this paper, our main purpose is twofold. First, we shall show that universal horizons exist not only  in the low energy limit, but also in the UV regime. To show this, we consider solutions of the full theory of Ho\v{r}ava gravity, that is, with all the higher-order derivative terms. In general, these calculations are very cumbersome. 
  To make the problem tractable, we restrict ourselves to the (2+1)-dimensional case in the framework of the projectable Ho\v{r}ava theory with an extra U(1) symmetry \cite{HMT,WW,Silva,HW}. 
It should be noted that in \cite{MM15},  the effects of higher-order derivative terms on the existence of universal horizons were studied, and found that, if a three-Ricci curvature squared term is joined in the ultraviolet modification,  the universal horizon appearing in the low energy limit was turned into  a spacelike singularity. While this is possible, as the universal horizons might not be stable against nonlinear perturbations \cite{Blas:2011ni}, the results presented in this paper show that they do exist even in the UV regime.  Second, we shall show that universal horizons exist not only in static spacetime, but also in the ones with rotation. This is important for both theory and observation, as we expect that the majority of astrophysical 
black holes should be the ones with rotation \cite{Narayan}.  In \cite{Sotiriou:2014gna}, it was shown that rotating universal horizons exist in the IR limit, in the framework of the non-projectable Ho\v{r}ava gravity without the extra U(1) symmetry \cite{BPS}.  In this paper, we shall show that  this is true not only in the IR limit but also in  the UV.

The rest of the paper is organized as follows. In Section II, we give a brief consideration  of the stability of the (2+1)-dimensional Ho\v{r}ava theory with both projectability condition and the extra U(1) symmetry, while a more complete review of the theory in 
(d+1)-dimensions is presented in Appendix A. In section III, we present various static and stationary solutions by working in the Painlev\'{e}-Gullstrand (PG) coordinates \cite{PG}. The main reason to work with these coordinates is that the solutions are free of coordinate singularities across the Killing horizons. However, a price to pay is that the field equations become mathematically more complicated. 
 Fortunately, they still allow us to find analytical solutions in closed forms. In Section IV, we study the locations of Killing and universal horizons, and find that such horizons indeed exist, even when the higher-order curvature terms are included. We end this paper with Section V, in which our main conclusions presented. There are also two more appendices,  Appendix B and Appendix C, in which some mathematical expressions are presented.

Before proceeding further, we would like to note  that the study of black holes in gravitational theories with the broken LI is also crucial in the understanding of quantization of gravity \cite{Carlip,Kiefer} and the non-relativistic AdS/CFT correspondence \cite{Kachru:2008yh,LSWW,Janiszewski:2014ewa,Basu:2016vyz}. But, such studies are all  in its infancy, and more detailed investigations are highly demanded.  

\section{Projectable Ho\v{r}ava theory with U(1) symmetry  in (2+1) dimensions }
\renewcommand{\theequation}{2.\arabic{equation}} \setcounter{equation}{0}

In the 2+1 dimensional spacetimes,  the Riemann and Ricci tensors $R_{ijkl}$ and $R_{ij}$   of the two-dimensional (2d) leaves of $t = $ constant have only one independent component, and are given by \cite{Carlip},
\bqn \lb{3.0}
R_{ijkl} &=& \frac{1}{2}\left(g_{ik}g_{jl} - g_{il}g_{jk}\right)R,\nb\\
R_{ij} &=& \frac{1}{2}g_{ij}R.
\eqn
Then,   the potential part of the action of the Ho\v{r}ava theory up to the fourth-order  is given by
 \bqn \lb{7}
 {\cal{L}}_{{V}}=  2\Lambda +  g_1 R  + \frac{1}{\zeta^{2}}\left(g_2 R^{2} + g_3 \Delta R\right),
 \eqn
where $\Lambda$ is the cosmological constant,  and $g_n$'s   are dimensionless coupling  constants, and $\zeta$ has the dimension of $({\mbox{mass}})^{-1}$. However, in 2d spaces the Ricci scalar $R$ always takes a complete derivative form. Then, when $N = N(t)$, the action can be integrated once and this term can be expressed as a boundary term. The same is true for the $g_3 \Delta R$ term. Therefore, in the case with the projectability condition, without loss of generality, we can always drop the $g_1$ and $g_3$ terms.

In Appendix A, we provide a brief introduction to the (d+1)-dimensional Ho\v{r}ava theory with  the projectability condition (\ref{1.6}) and the  Diff($M, \; {\cal{F}}$) symmetry (\ref{symmetry}). Setting $d=2$ and taking the above potential (with $g_1 = g_3 = 0$) into account, one can obtain the field equations.  In particular, the relativistic case is recovered by setting
\bq
\lb{3.0a}
(\lambda, g_2)^{GR} = (1, 0).
\eq
In addition, one can show that  the Minkowski spacetime
\bq
\lb{Minkowski}
(N, N^i, g_{ij}, A, \varphi) = (1, 0, \delta_{ij}, 0, 0),
\eq
is a solution of the field equations with $\Lambda = \Lambda_g = 0$. Then, its linear perturbations can be cast in the form\footnote{In (2+1)-dimensions, there are no vector and tensor perturbations \cite{Carlip}. This is true also in the Ho\v{r}ava theory.},
 \bqn
N=1+\phi,\;\;N_i=\partial_i B,\nb\\
g_{ij}=(1-2 \psi) \delta_{ij}+2 \partial_i\partial_j E,\nb\\
A=\delta A,\;\;\;\;\varphi= \delta \varphi,
 \eqn
 where $\phi, B,\psi, E, \delta{A}$ and $\delta\varphi$ represent the scalar perturbations, and the projectability condition requires $\phi = \phi(t)$. Using the gauge freedom, without loss of generality, we can always  set
\cite{ZWWS}
 \bqn
  \phi = E=\delta\varphi = 0,
 \eqn
which uniquely fixes the gauge. Then,   the quadratic action  without matter takes the form,
\bqn
\lb{S2}
S^{(2)}&=&\zeta^2 \int dtd^3 x \Bigg\{2(1-2 \lambda) (\dot{\psi}^2+\dot{\psi} \partial^2 B)\nb\\
&&+(1-\lambda) (\partial^2 B)^2 -
2\Big(A+2\frac{g_2}{\zeta^2} \psi
\partial^2\Big)\partial^2 \psi\Bigg\},\nb\\
\eqn
where $\partial^2 = \delta^{ij}\partial_i \partial_j$. Now,
variations of $S^{(2)}$ with respect to $A$, $B$, and $\psi$
yield, respectively,
\bqn
\lb{eq.a1}
\partial^2\psi &=& 0,\\
\lb{eq.a2}
(1-2 \lambda) \dot{\psi}+(1-\lambda)  \partial^2 B &=& 0,\\
\lb{eq.a4} 
\ddot{\psi}+\frac{1}{2} \partial^2 \dot{B} +\frac{1}{2(1-2 \lambda)} \left(4\frac{g_2}{\zeta^2}\partial^2
\psi+\partial^2A\right) &=& 0.\nb\\
\eqn
From Eq.(\ref{eq.a1}) it can be seen that  the scalar $\psi$ satisfies the Laplace equation. Thus, it does
not represent a propagative mode, and with proper boundary conditions, one can always set it to zero. Similarly, this is also true for other scalars. Hence,   the spin-0 gravitons are not present in (2+1)-dimensions, similar to the (3+1)-dimensional case  \cite{HW}.

\section{Static and stationary vacuum solutions}
\renewcommand{\theequation}{3.\arabic{equation}} \setcounter{equation}{0}

In this section we are going to study   vacuum solutions of the projectable Ho\v{r}ava theory with the extra U(1) symmetry introduced in the last section in (2+1)-dimensional  static and stationary spacetimes. Since our main purpose is to study the existence of universal horizons, which are always located inside the Killing horizons\footnote{In this paper we define a Killing horizon as the location at which the time-translation Killing vector $\zeta^{\mu}$ becomes null. In the spacetimes with rotations, this coincides with the ergosurface (ergosphere), while in the static spacetimes it coincides with the event horizon \cite{Visser07,Lin:2014eaa}}, we shall choose the gauge such that the solutions do not have coordinate singularities outside of universal horizons.  In the spherically symmetric spacetimes with a time-like foliation, this is quite similar to the PG coordinates \cite{PG}. Therefore, in this paper we shall refer such a coordinate system as the PG coordinates. To proceed further, let us first consider static spacetimes.

\subsection{Static Spacetimes}

The general static  solutions with the projectability condition $N = N(t)$ can be cast in the form,
 \bqn
\lb{8}
&& N = 1, \;\;\; N^{i} = \delta^i_{r} h(r),\;\;\;
g_{ij} = \left(\frac{1}{f(r)}, r^2\right),\nb\\
&& \varphi = \varphi(r),\;\;\;
A = A(r),
 \eqn
in the  spatial coordinates $x^i = (r, \theta)$. Using the U(1) symmetry, without loss of the generality, we choose the gauge,
\bq
\lb{gaugeA}
\varphi = 0,
\eq
so that $F^{ij}_{(\varphi, n)}  = 0 = f_{\varphi}^{ij}$. Then, we find that the quantities $K_{ij},\; R_{ij}, \pi_{ij},\; F_{ij}^A, \; F_{ij}$
and ${\cal{L}}_{i}$  are given by Eq.(\ref{2.1a}) in Appendix B.  Then,   Eqs.(\ref{eq1}), (\ref{eq2}), (\ref{eq4a}), (\ref{eq4b}), (\ref{eq3}), (\ref{eq5a}), and (\ref{eq5b}) reduce, respectively, to
Eqs.(\ref{B-2}) - (\ref{2.2e3}) given in Appendix B.

When the spacetime is vacuum, Eqs.(\ref{2.2b}) - (\ref{2.2e2}) reduce, respectively, to,
\bqn
\lb{2.2bA}
\left(\lambda - 1\right)\Big[h'' - a(r)h'\Big] + b(r) h = 0,\\
 \lb{2.2cA}
(\lambda-1)\left(h''' + \frac{2}{r}h''\right) + c(r) h' + d(r) h = 0, \\
\lb{2.2dA}
{f'}+2\Lambda_g r = 0,\\
\lb{2.2e1A}
A' + P(r) A + Q(r) = 0,\\
\lb{2.2e2A}
A'' + U(r)A' + V(r) A +  W(r) = 0,
\eqn
where $a, b, c, d, P, Q, U, V$ and $W$ are given by Eq.(\ref{coefficients}) in Appendix B.

 It should be noted that not all of the above equations are independent. In fact, Eq.(\ref{2.2cA}) can be obtained from Eqs.(\ref{2.2bA}) and (\ref{2.2dA}), while
 Eq.(\ref{2.2e2A}) can be obtained from Eqs.(\ref{2.2e1A}) and (\ref{2.2dA}). Therefore, in the present case there are three independent equations,
 (\ref{2.2bA}),  (\ref{2.2dA}) and  (\ref{2.2e1A}), for the three unknowns, $f, \; h$ and $A$. In particular, one can first find $f$ from Eq.(\ref{2.2dA}),
 \bq
\lb{9}
f(r)= {C_{1}-{\Lambda}_{g} r^2},
 \eq
 where $C_1$ is an integration constant. Substituting it into Eq.(\ref{2.2bA}), one can find $h(r)$. Once $f$ and $h$ are known, from Eq.(\ref{2.2e1A}), we find that
 \bq
 \lb{gaugeB}
 A(r) = \sqrt{C_1 - \Lambda_{g}r^2}\left(A_0 - \int^{r}{\frac{Q(r')dr'}{\sqrt{C_1 - \Lambda_{g}{r'}^2}}}\right),
 \eq
 where $A_{0}$ is another integration constant. Therefore, our main task now becomes to solve Eq.(\ref{2.2bA}) for $h$ with $f$ given by Eq.(\ref{9}). Once $h$ is known,
 the gauge field $A$ can be obtained by quadrature from Eq.(\ref{gaugeB}). To solve Eq.(\ref{2.2bA}), we consider the two cases $\Lambda_{g} = 0$ and $\Lambda_{g} \not= 0$, separately.

\subsubsection{$\Lambda_{g} = 0$ }

 When $\Lambda_{g} = 0$,  we have
 \bq
\lb{15}
f(r)=C_{1} > 0,
\eq
and Eq.(\ref{2.2bA}) simply reduces to
\bq
\lb{15.a}
\left(\lambda - 1\right)\left(h'' + \frac{1}{r}h' - \frac{1}{r^2}h\right) = 0.
\eq
Therefore, we need to consider the two cases $\lambda = 1$ and $\lambda \not=1$, separately.

{\bf Case with {$\lambda = 1$}:} Then,  Eq.(\ref{15.a}) is satisfied identically, and $h$ is undetermined. This is similar to the (3+1)-dimensional case \cite{GSW}. Inserting Eq.(\ref{15}) into the expression for
$Q(r)$ defined in Eq.(\ref{coefficients}) with $\lambda = 1$, we find that
\bq
\lb{15.aa}
Q(r) = \frac{1}{C_1}\left(hh' - \Lambda r\right),
\eq
for which Eq.(\ref{gaugeB}) yields,
\bq
\lb{15.ab}
A(r) = A_0\sqrt{C_1} - \frac{1}{2C_1}\left(h^2 - \Lambda r^2\right).
\eq

{\bf Case with {$\lambda \not= 1$}:} In this case,  Eq.(\ref{15.a}) has the general solutions,
\bq
\lb{16}
h(r)=C_{2}r + \frac{C_{3}}{r},
\eq
where $C_{2}$ and $C_3$ are other integration constants.
 Inserting Eqs.(\ref{15}) and (\ref{15.a}) into Eq.(\ref{gaugeB}), we obtain
 \bq
\lb{17}
A(r)= A_0-\frac{C_3^2}{2C_1r^2}-\frac{(2\lambda-1)C_2^2-\Lambda}{2C_1}r^2.
 \eq

\subsubsection{$\Lambda_{g} \not = 0$}

 When $\Lambda_{g} \not= 0$, it is also found convenient to   study the two cases, $\lambda=1$ and $\lambda \not=1$, separately.

{\bf Case with {$\lambda = 1$}:}  In this case, Eq.(\ref{2.2bA}) yields
 \bq
 \lb{10}
h(r)=0.
 \eq
Then,  from Eqs.(\ref{coefficients}) and (\ref{9}), we find that
 \bq
 \lb{11.a}
Q(r) = \frac{2g_{2}\Lambda_g^2- \zeta^2\Lambda}{\zeta^2\left(C_1-\Lambda_g r^2\right)}r.
 \eq
Inserting it into Eq.(\ref{gaugeB}), we obtain
 \bqn
 \lb{11}
A(r)&=& A_0\sqrt{C_1-\Lambda_g r^2} - A_1,
 \eqn
where   $A_1 \equiv 2g_{2}\Lambda_g/\zeta^{2}- {\Lambda}/{\Lambda_g}$. It can be shown that this is the static BTZ solution with mass $M=-C_1$,  provided that $A_1=-1$.
When $A_1\not=-1$, we provide  the main properties of the corresponding 
solution in Appendix C.

{\bf Case with {$\lambda \not= 1$}:} In this case,    Eq.(\ref{2.2bA}) takes the form,
\bqn
\lb{12.aA}
&& r^2\left(C_1 - \Lambda_g r^2\right)^2 h'' + C_1 r \left(C_1 - \Lambda_g r^2\right)h' \nb\\
&& - \left[C_1\left(C_1 - 3\Lambda_g r^2\right) + \frac{\Lambda_g r^2\left(C_1 - \Lambda_g r^2\right)}{\lambda -1}\right]h = 0, \nb\\
\eqn
which has the general solution,
\bqn
\lb{12}
h(r)&=& r\left(C_1-\Lambda_g r^2\right)\Bigg(C_4 \; F\left(a,b;2;z\right)\nb\\
&& +C_5\int{\frac{dr}{r^3(C_1- \Lambda_g r^2)^{\frac{3}{2}}F\left(a,b;2;z\right)}}\Bigg), ~~~
\eqn
where $z \equiv \Lambda_g r^2/C_1$,  $C_{4}$ and $C_{5}$ are constants,  $F(a, b; c; z)$  is the hypergeometric function, and now
 \bq
 \lb{12.a}
 a\equiv\frac{1}{4}\left(5+\sqrt{\frac{\lambda-5}{\lambda-1}}\right),\;\;\;
 b\equiv\frac{1}{4}\left(5- \sqrt{\frac{\lambda-5}{\lambda-1}}\right).
 \eq
Inserting it, together with $f$ given by Eq.(\ref{9}), into Eq.(\ref{gaugeB}), we can obtain $A$. However, because of the
complexity of $h$, it is found that no explicit expression for $A$ can be obtained, except for the case where $h = 0$ (or $C_4  = C_5 = 0$),
for which we find that $Q(r)$ and $A(r)$ are given exactly by Eqs.(\ref{11.a}) and (\ref{11}).

In addition, Eq.(\ref{12}) holds only for $C_1 \not= 0$. When $C_1 = 0$, Eq.(\ref{2.2bA}) has the general solution,
\bqn
\lb{13}
h(r)&=&C_+ r^{\delta_{+}}+C_- r^{\delta_{-}},
\eqn
where $C_{\pm}$ are the integration constants, and
 \bq
 \lb{13.a}
 \delta_{\pm} =  \frac{1}{2}\left(1 \pm
\sqrt{\frac{\lambda-5}{\lambda-1}}\right).
 \eq
Note that to have $\delta_{\pm}$ real, the parameter $\lambda$ must be either (i) $\lambda < 1$ or (ii) $\lambda \ge 5$.
 Inserting it into Eq.(\ref{gaugeB}), we find that
 \bqn
 \lb{13.b}
 A(r)&=& A_0 r+1-\Lambda_2+\frac{\hat{C}_{+}}{r^{ 2\delta_{-}}}+\frac{\hat{C}_{-}}{r^{2\delta_{+} }},
  \eqn
  with
  \bqn
  \lb{13.c}
  \hat{C}_{\pm}&\equiv&\frac{8\mp(\lambda-1)\left[3\sqrt{\lambda^2-6\lambda+5}\mp\left(13-3\lambda\right)\right]}{4\Lambda_g(3\lambda+1)C_{\pm}^{-2}},\nb\\
  \Lambda_2&=&1-\frac{\zeta^2\Lambda-2g_{2}\Lambda_g
  ^2}{\zeta^2\Lambda_g}.
 \eqn

\subsection{Stationary  Spacetimes}

To find rotating black holes, let us consider the stationary spacetimes described by,
  \bqn
 \lb{Eq5A1}
 N &=& 1,\;\;\; N^{i} = h_r(r)\delta_r^i+h_\theta(r)\delta_\theta^i,\nb\\
  g_{ij}&=&\frac{1}{f(r)}\delta_i^r\delta_j^r+r^2\delta_i^\theta\delta_j^\theta.
 \eqn
 To obtain the general analytic solutions in this case, it is found very difficult, and instead let us first consider  the case where
$R_{ij}=0 = \Lambda_g$. Then, depending on the values of $\lambda$, we find three classes of solutions.
The first class is for $\lambda = 1$,  given by
 \bqn
 \lb{Eq5A5}
 f(r) &=& C_1,\nb\\
 h_{\theta}(r)&=&\frac{h_a}{r^2}+h_b,\nb\\
 A(r)&=&1-A_0-\frac{h_a^2}{2r^2}+\frac{\Lambda
 r^2}{2C_1}-\frac{h_r(r)^2}{2C_1},
 \eqn
where $C_1, h_a, h_b$ and $A_0$ are all integration constants, and similar to the static case the function $h_r(r)$ is arbitrary.

The second class is for $\lambda \not = 1$,  given by
 \bqn
 \lb{Eq5A8}
 f(r) &=& C_1,\nb\\
 h_{\theta}(r)&=&\frac{h_a}{r^2}+h_b,\nb\\
 h_r(r)&=&\frac{H_A}{r}+H_Br,\nb\\
 A(r)&=&1-A_0-\frac{1}{2C_1r^2}\left(H_A^2+C_1h_a^2-\Lambda_C r^4\right),  \nb\\
 \eqn
where $\Lambda_C\equiv \Lambda+(1-2\lambda)H_B$, and $C_1, H_A, H_B, h_a, h_b$ and $A_0$ are all constants.

 The third class of rotating solutions can be obtained by considering  the ansatz
  \bqn
 \lb{E1}
  N&=&1,\;\;\;  N^{\mu} = h_\theta(r)\delta_\theta^\mu,\nb\\
 g_{ij}&=&\frac{1}{f(r)}\delta_i^r\delta_j^r+r^2\delta_i^\theta\delta_j^\theta,
 \eqn
for which we find the following rotating solution
  \bqn
 \lb{E2}
 f(r)&=&f_0-\Lambda_gr^2,\nb\\
 h_\theta(r)&=&h_B-\frac{h_A}{2f_0^{3/2}r^2}\left[\sqrt{f_0\left(f_0-\Lambda_gr^2\right)}\right.\nb\\
 &&\left.+\Lambda_gr^2\left(\frac{f_0+\sqrt{f_0(f_0-\Lambda_gr^2)}}{r}\right)\right],\nb\\
 A(r)&=&A_0\sqrt{f_0-\Lambda_gr^2}+\frac{\Lambda}{\Lambda_g}-2g_2\Lambda_g\nb\\
 &&-\frac{3h_A^2\Lambda_g\sqrt{f_0-\Lambda_gr^2}\arctan\left(\sqrt{\frac{f_0}{f_0-\Lambda_gr^2}}\right)}{8f_0^{5/2}}\nb\\
 &&-\frac{h_A^2(f_0-3\Lambda_gr^2)}{8f_0^2r^2},
 \eqn
where $h_0$, $A_0$, $h_A$ and $h_B$ are constants. 

\section{Universal Horizons and Black Holes  Without or with Rotations}
\renewcommand{\theequation}{4.\arabic{equation}} \setcounter{equation}{0}

As mentioned above, the fundamental variables of the gravitational field in the Ho\v{r}ava theory with Diff($M, \; {\cal{F}}$) and $U(1)$ symmetry (\ref{symmetry}) are
$$
(N, N^i, g_{ij}, A, \varphi).
$$
In the framework of the universal coupling \cite{LMWZ}, they are related to the spacetime line element $ds^2$ via the relations,
\bqn
\lb{equ3.1}
ds^2 = \gamma_{\mu\nu} dx^{\mu}dx^{\nu}, \; (\mu, \nu = 0, 1, 2, 3)
\eqn
where  $\gamma_{\mu\nu}$ is given by Eq.(\ref{Pmetric}), that is,
\bqn
\lb{equ3.2}
\left(\gamma_{\mu\nu}\right) &\equiv& \left(\matrix{-{\cal{N}}^2 +{\cal{N}}^{i}{\cal{N}}_{i} &{\cal{N}}_{i}\cr
{\cal{N}}_{i} & {\gamma}_{ij}\cr}\right),
\eqn
where ${\gamma}^{ij}{\gamma}_{ik} = \delta^{j}_{k},\;{\cal{N}}_{i} \equiv {\gamma}_{ij}{\cal{N}}^{j}$, and
\bqn
\lb{equ3.3}
& &{\cal{N}} \equiv \left(1 - a_1\sigma\right)N,\;\;\;
{\cal{N}}^i \equiv N^i + Ng^{ij} \nabla_j\varphi, \nb\\
&&  {\gamma}_{ij} \equiv \left(1 - a_2\sigma\right)^2g_{ij},\;\;\;
\sigma  \equiv \frac{A - {\cal{A}}}{N}, \nb\\
&&  {\cal{A}} \equiv - \dot{\varphi}  + N^i\nabla_i\varphi
+\frac{1}{2}N\left(\nabla^i\varphi\right)\left(\nabla_i\varphi\right).
\eqn
Here $a_1$ and $a_2$ are two arbitrary coupling constants. The solar system tests in (3+1)-dimensions require that they must satisfy the conditions (\ref{eq8-2}). In particular, for $(a_1, a_2) = (1, 0)$, the PPN parameters will be the same as those given in general relativity \cite{Will}. Although in (2+1)-dimensions, no such constraints exist, in order to compare with those obtained in (3+1)-dimensions, we shall impose these
conditions also in the (2+1)-dimensional spacetimes considered in this paper. In particular, we shall only consider the case with
\bq
a_1 = 1,\;\;\;
a_2 = 0.
\eq
Therefore, with the gauge choice $\varphi = 0$, Eq.(\ref{equ3.3}) reduces to
\bqn
\lb{equ3.4}
&& {\cal{N}} = N-A,\;\;\; {\cal{N}}^i = N^i,
\;\;\; {\gamma}_{ij} = g_{ij}.
\eqn
This is also consistent with the one adopted  in \cite{HMT} in (3+1)-dimensions, when the solar system tests were considered.

On the other hand, the critical point for a universal horizon to be present is the existence of a globally defined khronon field $\phi$, which is always
timelike \cite{BS11}. Then, the causality is assured by assuming that all the particles move along the increasing direction of $\phi$. In this sense,
$\phi$ serves as an absolute time introduced in the Newtonian theory. Setting
\bq
\lb{equ3.5}
u_\mu \equiv \frac{\partial_\mu\phi}{\sqrt{X}},
\eq
one can see that $u_{\mu}$ is always timelike, $\gamma^{\mu\nu} u_\mu u_\nu = -1$, where $X\equiv-g^{\alpha\beta}\partial_\alpha\phi\partial_\beta\phi$.
In addition, such defined $u_\mu$ is invariant under the gauge transformation,
\bq
\lb{equ3.6}
\phi = {\cal F}(\tilde\phi),
\eq
provided that ${\cal F}(\tilde\phi)$ is a monotonically increasing (or decreasing) and otherwise arbitrary function of
$\tilde\phi$. Such defined  $u_\mu$ also satisfies the hypersurface-orthogonal condition,
  \bqn
  \lb{4.1}
  u_{[\nu}D_\alpha u_{\beta]}=0,
  \eqn
where $D_{\alpha}$ denotes the covariant derivative with respect to $\gamma_{\mu\nu}$.

In (2+1)-dimensional spacetimes, the most general form of action of
khronon is described as \cite{Jacobson}
  \bqn
  \lb{4.2}
  S_\phi&=&\int d^{2+1}x\sqrt{|\gamma|}{\cal L}_\phi\nb\\
  &=&\int d^{2+1}x\sqrt{|\gamma|}\left[c_1\left(D_\mu u_\nu\right)^2+c_2\left(D_\mu
  u^\mu\right)^2\right.\nb\\
  &&\left.+c_3\left(D^\nu u^\mu\right)\left(D_\mu u_\nu\right)-c_4a^\mu a_\mu\right],
  \eqn
where $a_\mu\equiv u^\alpha D_\alpha u_\mu$, and $c_i$'s denote the coupling constants of the khronon field. However,
due to the identity (\ref{4.1}), not all the four terms are independent. In fact,  from Eq.(\ref{4.1}) we find that
$$
  \Delta{\cal L}_\phi\equiv a^\mu a_\mu+\left(D^\mu u^\nu\right)\left(D_\mu
  u_\nu\right)-\left(D^\nu u^\mu\right)\left(D_\mu
  u_\nu\right)=0.
  $$
  Then,  we can always add this term into $S_\phi$ with
arbitrary coupling constant $c_0$, so that the coupling constants
$c_i$ in ${\cal L}_\phi$ can be redefined as
  \bqn
  \lb{4.4}
  c'_1&=&c_1+c_0,~~~c'_2=c_2,\nb\\
  c'_3&=&c_3-c_0,~~~c'_4=c_4-c_0.
  \eqn
Thus,  one can always set one  of the terms $c'_1$, $c'_3$ and $c'_4$ to zero by properly choosing $c_0$. In the following, we shall leave this possibility open.

Then,   the variation of $S_\phi$ with respect to $\phi$ yields the khronon equation \cite{Wang:2012at},
  \bqn
  \lb{4.5}
D_\mu {\cal A}^\mu=0,
  \eqn
  where
  \bqn
  \lb{4.6}
  {\cal A}^\mu &\equiv& \frac{\delta^\mu_\nu+u^\mu
u_\nu}{\sqrt{X}}{\AE}^\nu,\nb\\
{\AE}^\mu&\equiv&D_\nu J^{\nu\mu}+c_4 a_\nu D^\mu u^\nu,\nb\\
{J^\alpha}_\mu&\equiv&\left(c_1g^{\alpha\beta}g_{\mu\nu}+c_2\delta^\alpha_\mu\delta^\beta_\nu\right.\nb\\
&&\left.+c_3\delta^\alpha_\nu\delta^\beta_\mu-c_4u^\alpha u^\beta
g_{\mu\nu}\right)D_\beta u^\nu.
  \eqn
 From the above expressions, we find
 \bq
 \lb{4.6a}
 u_\mu {\cal A}^\mu=0,
 \eq
 that is, ${\cal A}^\mu$ is always orthogonal to $u_\mu$.

 Eq.(\ref{4.5}) is a second-order differential equation for $u_{\mu}$, and to uniquely determine it, two boundary conditions are needed.
These two conditions in stationary and asymptotically flat  spacetimes can be chosen as follows \cite{BS11,Lin:2014eaa}:
(i)  $u^{\mu}$ is
aligned asymptotically with the time translation Killing vector $\zeta^{\mu}$,
\bq
\lb{1.13}
u^{\mu} \propto \zeta^{\mu}.
\eq
(ii) The khronon has a regular future sound horizon, which
  is a null surface of the effective metric \cite{EJ},
\bq
\lb{1.14}
g^{(\phi)}_{\mu\nu} = g_{\mu\nu} - \left(c_{\phi}^2 -1\right)u_{\mu}u_{\nu},
\eq
where $c_{\phi}$ denotes the speed of the khronon  given by,
\bq
\lb{1.15}
c_{\phi}^2 = \frac{c_{123}}{c_{14}},
\eq
where $c_{123}\equiv c_1+c_2+c_3,\; c_{14}\equiv c_1+c_4$.
It is interesting to note that such a speed does not depend on the redefinition of the new parameters $c_i'$ given by Eq.(\ref{4.4}), as it is expected.

The universal horizon is the location at which $u^{\mu}$ and $\zeta^{\mu}$ are orthogonal \cite{BS11,Lin:2014eaa},
\bq
\lb{1.15a}
\gamma_{\mu\nu} u^{\mu}\zeta^{\nu} = 0.
\eq
Since $u^{\mu}$ is always timelike, and $\zeta^{\mu}$ is also timelike outside the Killing horizon, Eq.(\ref{1.15a}) is possible  only inside the Killing horizon, in which $\zeta^{\mu}$ becomes spacelike.

With all the above in mind, now we are ready to consider the locations of universal horizons in the solutions found in the last section for static or stationary spacetimes.

\subsection{Universal Horizons and Black Holes in Static Spacetimes}

 In the static spacetimes of the solutions found in the last section,   the general form of metric is
\bq
 \lb{4.7ab{4.8}}
 ds^2=-{\cal{N}}^2dt^2+\frac{1}{f(r)}[dr+h(r)dt]^2+ r^2d\theta^2,
\eq
as can be seen from Eqs.(\ref{8}) and (\ref{equ3.4}). Following \cite{BS11}, it can be shown that
Eq.(\ref{4.5}) is equal to
\bq
\lb{4.7aa}
{\cal A}^\mu=0,
\eq
 in asymptotically flat spacetimes, in which we have
  \bqn
  \lb{4.8aa{eq8-2}}
  &&V\rightarrow0,~~{\cal{N}} \rightarrow1,~~ F\rightarrow1,~~u_t\rightarrow1,\nb\\
  &&X\rightarrow1,~~h\rightarrow0,
  \eqn
as $ r\rightarrow\infty$. In the following we shall assume that this is also true  to other spacetimes.
To simplify Eqs.(\ref{4.7aa}), in the following we only consider the case
\bq
\lb{CDa}
c_{14} = 0,
\eq
for which the speed $c_{\phi}$ of the khronon field becomes infinitely large, as one can see form Eq.(\ref{1.15}). In this case, the sound horizon of the khronon coincides with the universal horizon,
and the requirement that the khronon has a regular future sound horizon reduces to that of the universal horizon.

It can be shown that   Eq.(\ref{4.7aa}) has only one independent component, and with the assumption (\ref{CDa}), it reduces to
  \bqn
  \lb{4.11a}
  &&\frac{V''}{V} +\left(\frac{{\cal N}'}{{\cal N}}-\frac{f'}{2f}+\frac{1}{r}\right)\frac{V'}{V} +\frac{c_p}{r}\left(\frac{{\cal N}'}{{\cal N}}-\frac{f'}{2f}\right) -\frac{1}{r^2}\nb\\
  && +\frac{{\cal N}''}{{\cal N}}-\frac{f'}{2f}-\left(\frac{{\cal N}'}{{\cal N}}\right)^2+\frac{1}{2}\left(\frac{f''}{f}\right)^2= 0,
   \eqn
where $V \equiv u^r,\; c_p\equiv {c_{13}}/{c_{123}}$.

The timelike Killing vector $\zeta^{\mu}$ now is given by $\zeta^{\mu} = \delta^{\mu}_{t}$, so the location of the universal horizon is at
\bq
\lb{1.15aA}
 u_{\mu}\zeta^{\mu} = u_t \equiv \sqrt{G},
\eq
where  
 \bq
 \lb{1.15aB}
G\equiv f^{-1}\left[{\cal N}^2\left(f+V^2\right)-h^2\right].
 \eq
Note that $G$ is not necessary to be always non-negative.  However, to have the khronon field well-defined, we must assume that $G(r) \ge 0$ for any $r \in (0, \infty)$. 
Then, one can see that the location of the universal horizon must be the minimum of
the function $G(r)$, so that at $r = r_{UH}$, we must have \cite{Lin:2014eaa},
  \bqn
  \lb{4.17}
  G(r)|_{r=r_{UH}}=0=G'(r)|_{r=r_{UH}}.
  \eqn

On the other hand, at the Killing horizon $r = r_{KH}$  we have $\zeta^\mu\zeta_\mu=0$,  or equivalently  $ {\cal{K}} \left(r_{KH}\right) = 0$, where
  \bqn
  \lb{4.19}
  {\cal{K}} (r) \equiv {\cal{N}}(r)^2-\frac{h(r)^2}{f(r)}.
  \eqn
 Then, $ \gamma_{\mu\nu}$ and $ \gamma^{\mu\nu}$ are given by,
  \bqn
  \lb{4.20}
  \gamma_{\mu\nu} \left(r_{KH}\right)&=&\left.\left(
  \begin{array} {ccc}
  0 & h(r)/f(r) & 0 \nb\\ h(r)/f(r) & 1/f(r) & 0 \nb\\ 0&
0 & r^2
\end{array}
\right)\right|_{r= r_{KH}},\nb\\
  \gamma^{\mu\nu}\left(r_{KH}\right)&=&\left. \left(
  \begin{array} {ccc}
  -f(r)/h(r)^{2} & f(r)/h(r) & 0 \nb\\ f(r)/h(r) & 0 & 0 \nb\\ 0&
0 & r^{-2}
\end{array}
\right)\right|_{r= r_{KH}}.\nb\\
  \eqn
Therefore, in order for the metric to be free from coordinate singularities across the Killing horizon, we must require that  {\em both $h \left(r_{KH}\right)$ and $f \left(r_{KH}\right)$  are finite and non-zero.} In the (3+1)-dimensional case, we know that the Schwarzschild and Schwarzschild-de Sitter solutions satisfy these conditions, but not for the Schwarzschild-anti-de Sitter and Reissner-Nordstr\"om solutions \cite{Lin:2014eaa}. For the latter, one needs first to make extensions across those horizons, and then study the existence of universal horizons inside of those Killing horizons. In the following, we shall show that even with such strong conditions solutions that harbor universal horizons still exist.

\subsubsection{$\Lambda_g=0,\; \lambda =1$}

In this case the solutions are given by Eqs.(\ref{15}) and (\ref{15.ab}) with $h(r)$   being an arbitrary function.  In order to have the metric regular across the Killing horizon, we assume that $h(r) \not= 0$, for which  the metric takes the form,
  \bqn
  \lb{4.21a}
  ds^2&=&-\left(1-A_0\sqrt{C_1}+\frac{h^2-\Lambda
  r^2}{2C_1}\right)^2dt^2\nb\\
  &&+\frac{1}{C_1}\left(dr+h(r)dt\right)^2+r^2d\theta^2,
  \eqn
where   $C_1\not=0$. Rescaling the coordinates, without loss of generality, we can always set $C_1 = 1$, so the metric take the final form,
  \bqn
  \lb{4.21c}
  ds^2&=&-\left(\bar{A}_0+\frac{h^2-\Lambda
  r^2}{2}\right)^2dt^2\nb\\
  &&+\left(dr+h(r)dt\right)^2+r^2d\theta^2,
  \eqn
where $\bar{A}_0=1-A_0$. To study the existence of universal horizons and black holes, let us consider the case where the function $h(r)$ is given by,
  \bqn
  \lb{4.21cA}
  h=\frac{H}{r^\beta},
  \eqn
where $\beta$ and $H$ are two constants. Then, Eq.(\ref{4.11a}) becomes
  \bqn
  \lb{4.21d}
  &&\frac{V''}{V}+\frac{1}{r}\left[1-2\frac{H^2\beta+\Lambda r^{2+2\beta}}{H^2+r^{2\beta}(2\bar{A}_0-r^2\Lambda)}\right]\frac{V'}{V}\nb\\
  &&-\frac{2c_p}{r^2}\frac{H^2\beta+\Lambda
  r^{2+2\beta}}{H^2+r^{2\beta}(2\bar{A}_0-r^2\Lambda)}+\frac{1}{r^{2}}\nb\\
  &&\times\left[H^2+r^{2\beta}(2\bar{A}_0-r^2\Lambda)\right]^{-2}\left[(2\beta-1)H^4\right.\nb\\
  &&+2H^2r^{2\beta}\left(2\bar{A}_0(2\beta^2+\beta-1)-\beta(2\beta+5)\Lambda
  r^2\right)\nb\\
  &&\left.-r^{4\beta}(4\bar{A}_0^2+3\Lambda^2r^4)\right]=0.
  \eqn

On the other hand, the scalar and extrinsic curvatures $R$ and   $K$ are given by,
 \bqn
 \lb{4.21cB}
 R&=&\frac{8}{r^{2}D^3} \left\{-H^6\beta^2 +r^{6\beta}\Lambda\left(r^3\Lambda-2\bar{A}_0r\right)^2\right.\nb\\
 &&+H^4r^{2\beta}\left[\beta+r^2\Lambda +\beta^2(2\Lambda
 r^2-4\bar{A}_0)\right]\nb\\
 &&-H^2r^{4\beta}\left[4\bar{A}_0^2\beta^2\right.\nb\\
 &&-2\bar{A}_0\left(2\Lambda r^2-\beta+2\beta^2(1+\Lambda r^2)\right)\nb\\
 &&\left.\left.+r^2\Lambda\left(\beta-2+2\Lambda r^2+\beta^2(2+\Lambda r^2)\right)\right]\right\},\nb\\
 K&=&\frac{2H(1-\beta)r^{\beta-1}}{D(r)},\nb\\
 D(r) &\equiv& H^2-(\Lambda r^2-2\bar{A}_0)r^{2\beta}.
 \eqn
Clearly, to avoid spacetime singularities occurring at finite and non-zero $r$, we must assume that
$D(r) \not= 0$ for $r \in (0, \infty)$.

When $\Lambda=0$, we have $ D(r) =  H^2 + 2\bar{A}_0 r^{2\beta}$. Therefore, for
 $A_0>0$, we always have $D(r) > 0$. In this case, if we further set $H = 2$ and $\beta = 1/2$, we find that  Eq.(\ref{4.21d}) has the asymptotic solution,
 \bq
 \lb{4.21cD}
 V|_{r\rightarrow\infty}\rightarrow \frac{u_0}{r},
 \eq
which satisfies the boundary condition $u^\mu\propto \xi^\mu$ at infinity. But, an analytical solution  of Eq.(\ref{4.21d}) for any $r$ is still absent
even in this simple case.   The corresponding metric takes the form,
  \bqn
  \lb{4.21e}
ds^2 = -  \left(\bar{A}_0+\frac{2}{r}\right)^2dt^2 + \left(dr + \frac{2}{\sqrt{r}} dt\right)^2 + r^2 d\theta^2. \nb\\
  \eqn
Since even in this simple case, analytically solving Eq.(\ref{4.21d}) is not trivial, instead in the following we shall
use the shooting method first to solve it, and then localize  the positions of the Killing and universal
horizons, which satisfies, respectively, the equation   ${\cal{K}}(r_{KH})=0$, and (\ref{4.17}), where ${\cal{K}}(r)$ is given by Eq.(\ref{4.19}).

\begin{figure*}[tbp]
\includegraphics[width=5cm]{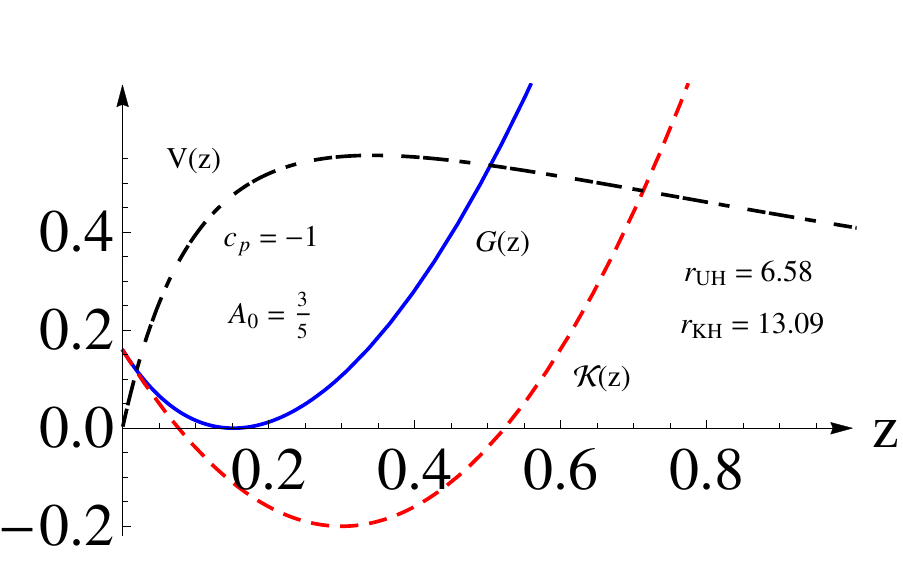}\includegraphics[width=5cm]{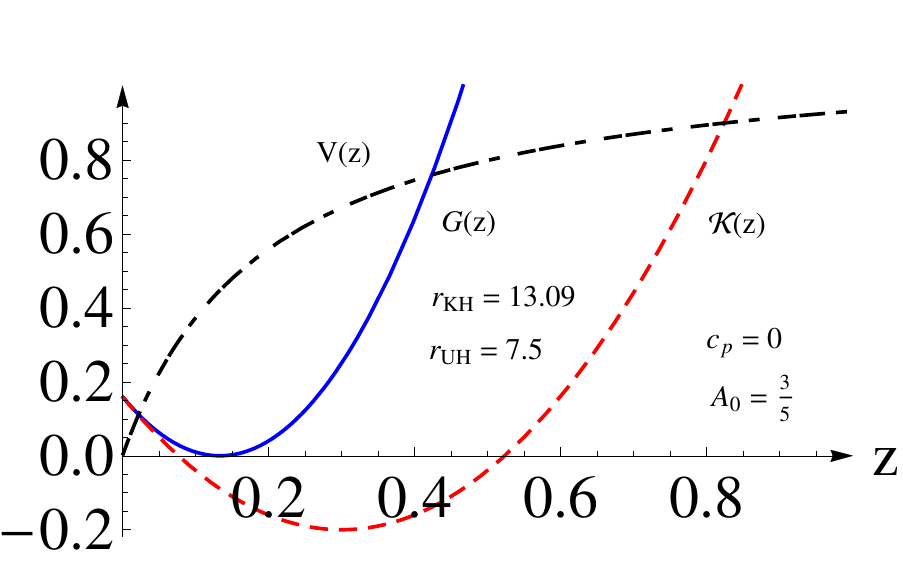}\includegraphics[width=5cm]{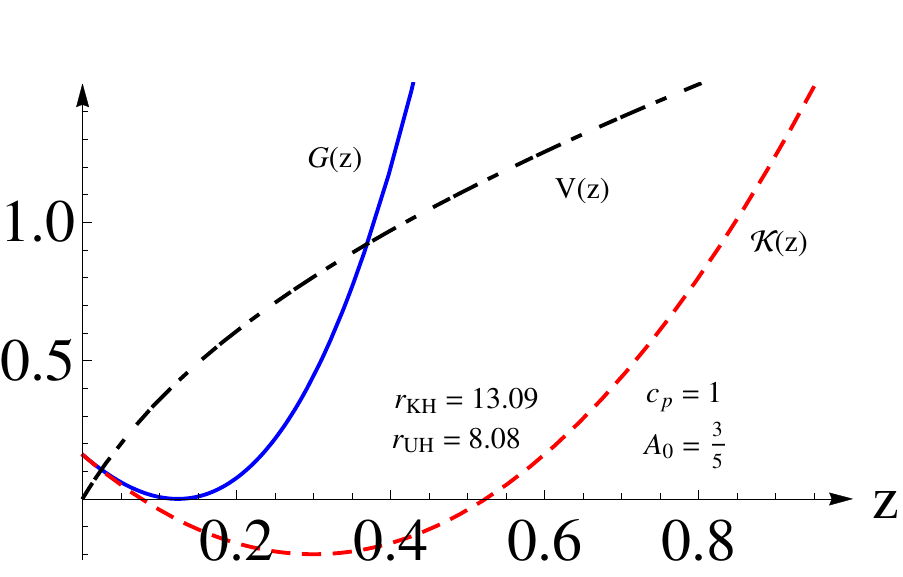}
\caption{The   functions $G, V \left(\equiv u^r\right)$ and ${\cal{K}}$ vs $z \equiv 1/r$ and the locations of the Killing ($r = r_{KH}$)
and universal   ($r = r_{UH}$) horizons, for the spacetime given by Eq.(\ref{4.21e}) with $ A_0 = 3/5$ and various choices of $c_p$. } 
\label{Fig3}
\end{figure*}

In Fig.~\ref{Fig3}, we show the curves of $G(z), V(z)$ and ${\cal{K}}(z)$ for various choices of the parameter $c_p \equiv c_{13}/c_{123}$, and find the locations
of the Killing and universal horizons, denoted, respectively, by $r_{KH}$ and $r_{UH}$, where $z \equiv 1/r$.  From this figure one can see that the locations of the universal horizons
depend on $c_p$ as it is expected.

When $\Lambda\not=0$, the mathematics becomes more involved. In the following we shall consider some representative choices of the parameter $\beta$.

\begin{figure*}[h]
\includegraphics[width=5cm]{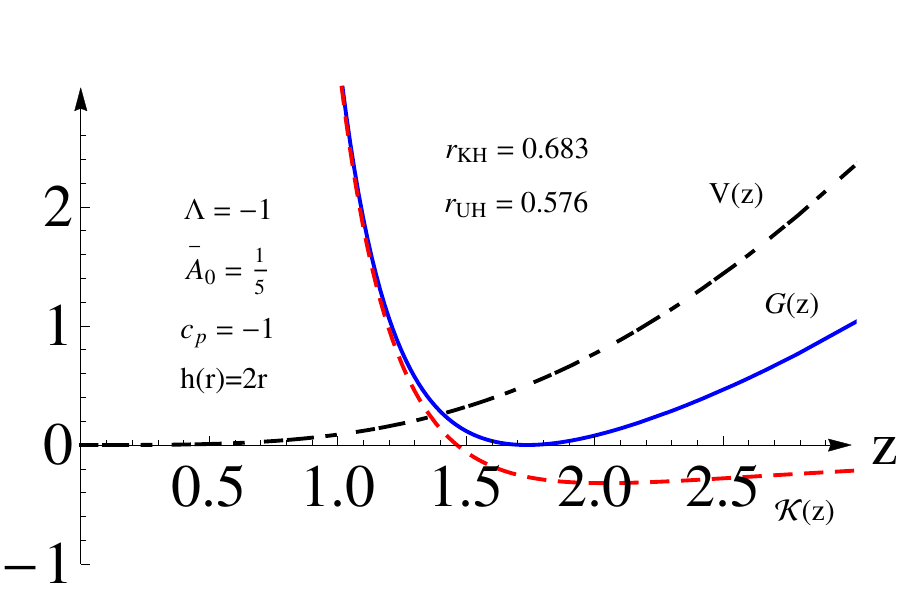}\includegraphics[width=5cm]{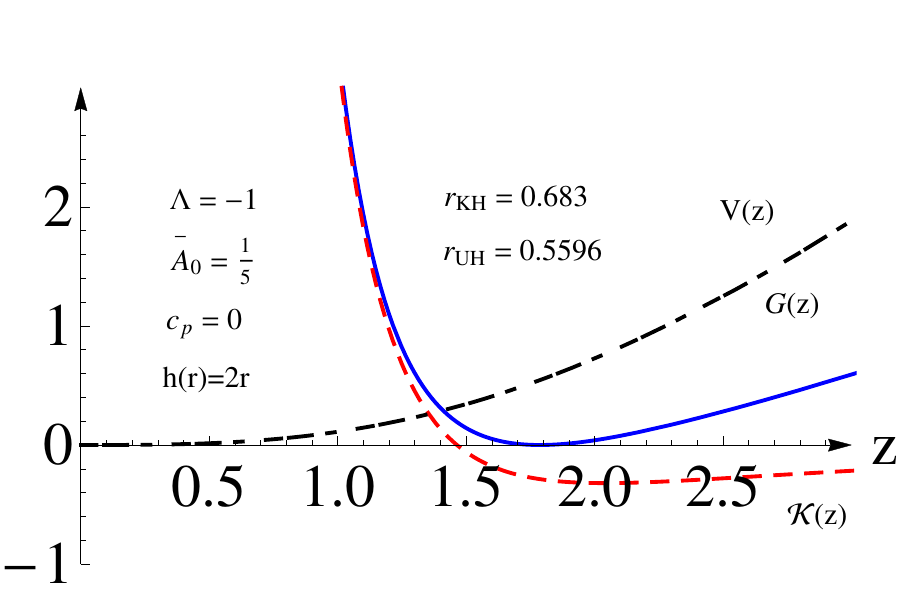}\includegraphics[width=5cm]{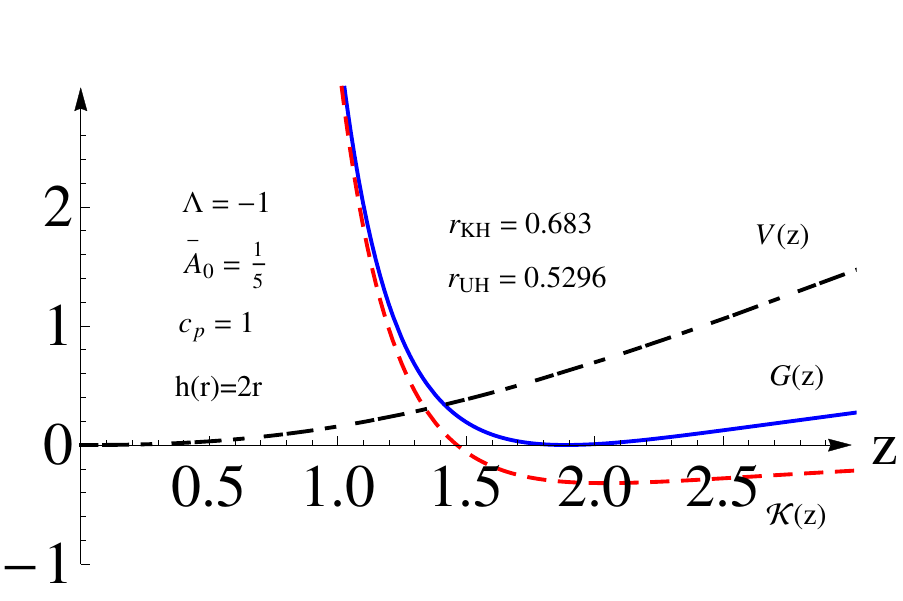}
\caption{The   functions $G, V \left(\equiv u^r\right)$ and ${\cal{K}}$ vs $z$, for the spacetime given by Eq.(\ref{4.21c}) with $ \Lambda = -1, A_0 = 3/5,
H = 2, \beta = -1$ and various choices of $c_p$.} 
\label{Fig4}
\end{figure*}

{\bf Case 1.a $\beta=-1$:}  In the case, to have $D(r) \not= 0$ for $r \in (0, \infty)$, we assume that   $\frac{\bar{A}_0}{\Lambda-H^2}\le0$. In
Fig.~\ref{Fig4}, we show the functions $G, V$ and ${\cal{K}}$  and the locations of the Killing
and universal horizons for various choices of $c_p$.

\begin{figure*}[h]
\includegraphics[width=5cm]{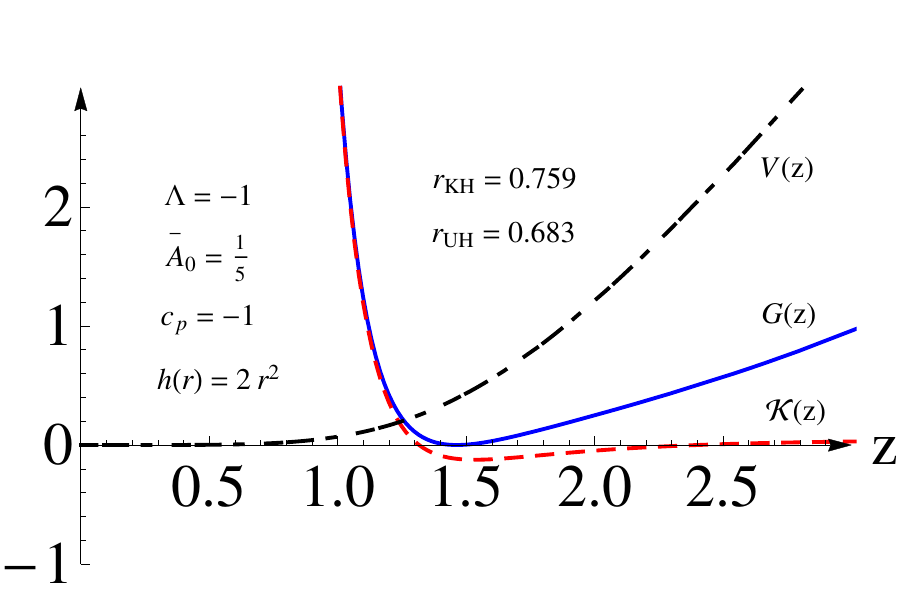}\includegraphics[width=5cm]{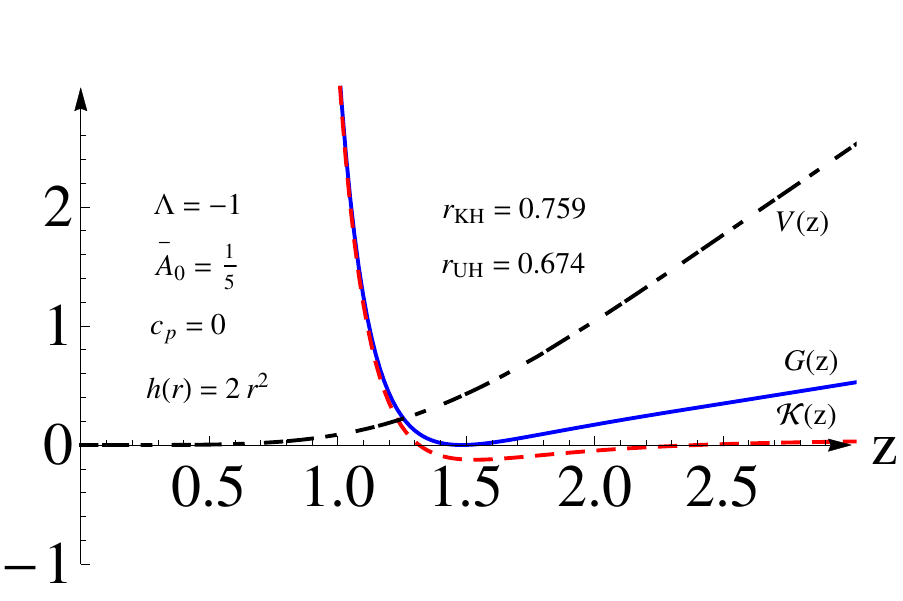}\includegraphics[width=5cm]{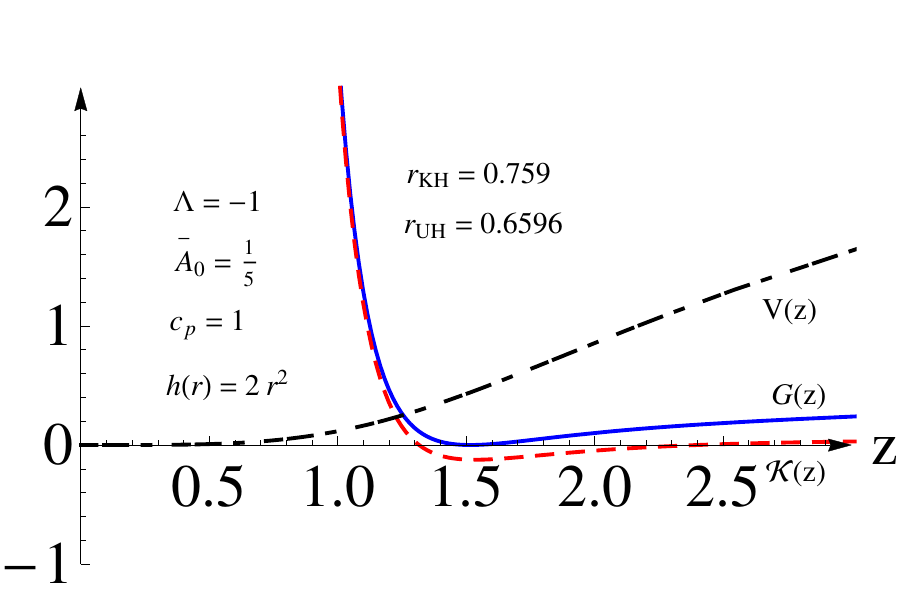}
\caption{The   functions $G, V \left(\equiv u^r\right)$ and ${\cal{K}}$ vs $z$, for the spacetime given by Eq.(\ref{4.21c}) with $ \Lambda = -1, A_0 = 3/5,
H = 2, \beta = -2$ and various choices of $c_p$.} 
\label{Fig5}
\end{figure*}

{\bf Case 1.b  $\beta<-1$:}  In the case, to have $D(r) \not=0$ for $r \in (0, \infty)$, we must assume that  either  $\bar{A}_0\ge0$ and $\Lambda<0$, or $\bar{A}_0\ge0$, $\Lambda>0$.
Then, in Fig.~\ref{Fig5} we show the functions $G, V$ and ${\cal{K}}$  and the locations of the Killing
and universal horizons for various choices of $c_p$ with $\beta = -2$.

\begin{figure*}[h]
\includegraphics[width=5cm]{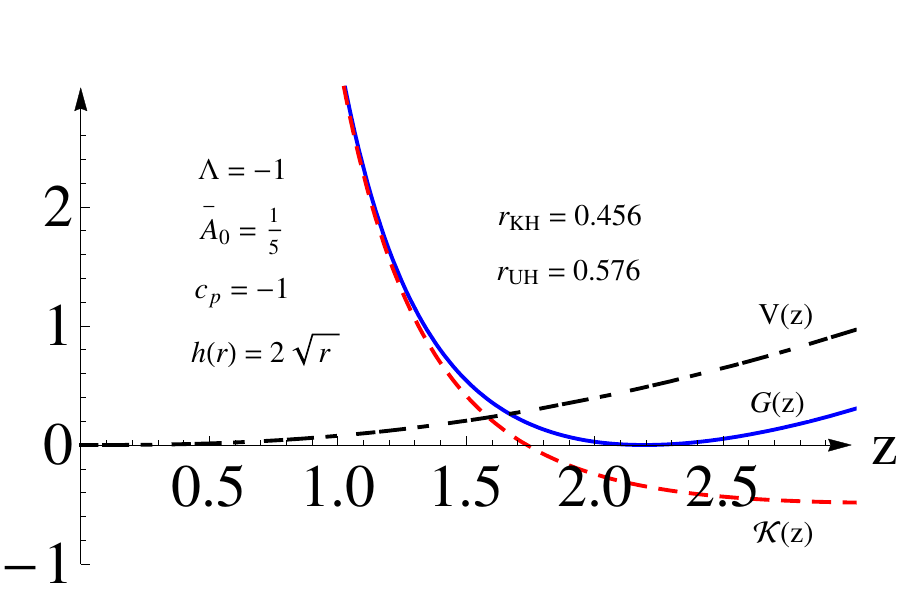}\includegraphics[width=5cm]{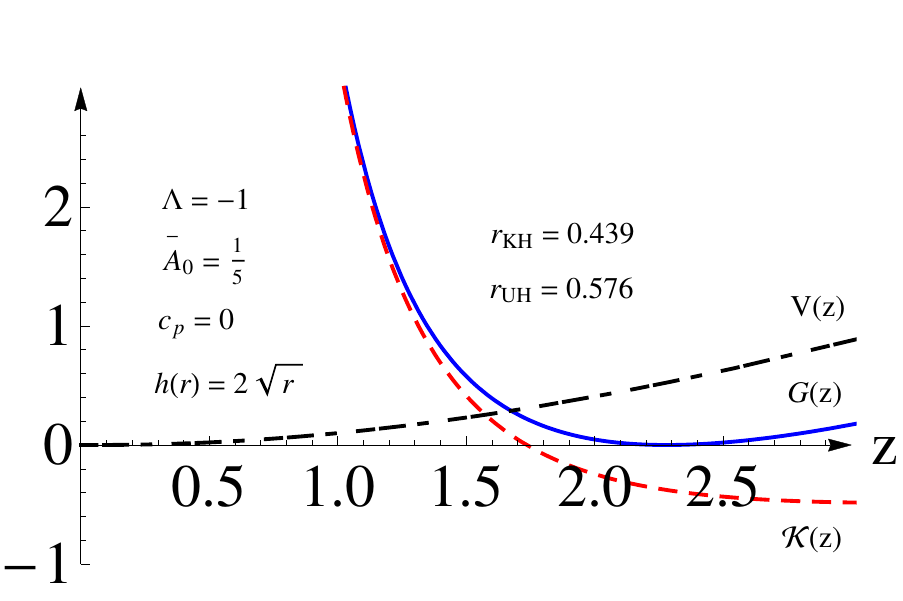}\includegraphics[width=5cm]{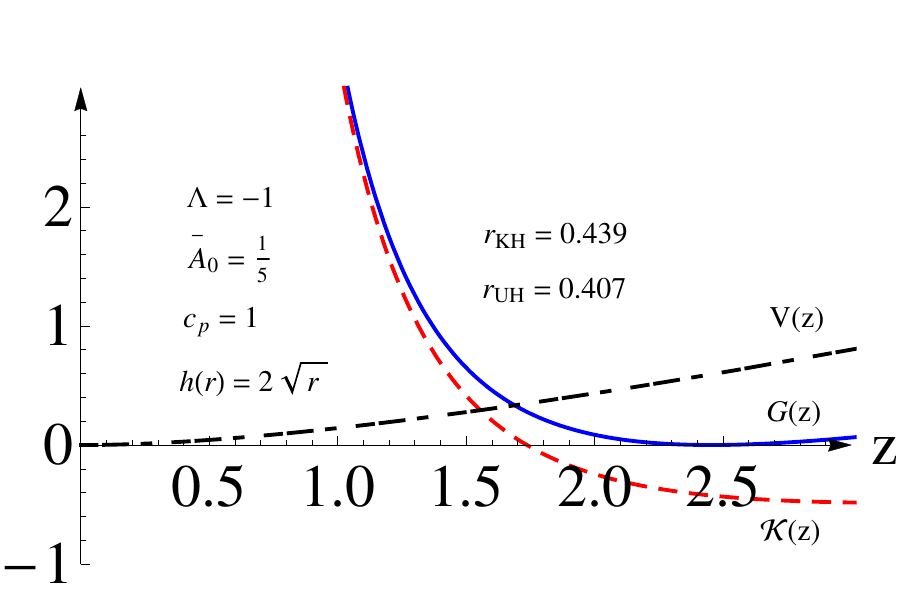}
\caption{The   functions $G, V \left(\equiv u^r\right)$ and ${\cal{K}}$ vs $z$, for the spacetime given by Eq.(\ref{4.21c}) with $ \Lambda = -1, A_0 = 3/5,
H = 2, \beta = -1/2$ and various choices of $c_p$.} 
\label{Fig6}
\end{figure*}

{\bf Case 1.c  $0>\beta>-1$:} In this case, we find that we must assume that  either  $\bar{A}_0>0,\; \Lambda<0$, or $\bar{A}_0>0,\; \Lambda>0$, in order not to have spacetime singularities at
a finite and non-zero $r$. In   Fig.~\ref{Fig6},  we show the functions $G, V$ and ${\cal{K}}$  and the locations of the Killing
and universal horizons for various choices of $c_p$ with $\beta = -1/2$.

\begin{figure*}[h]
\includegraphics[width=5cm]{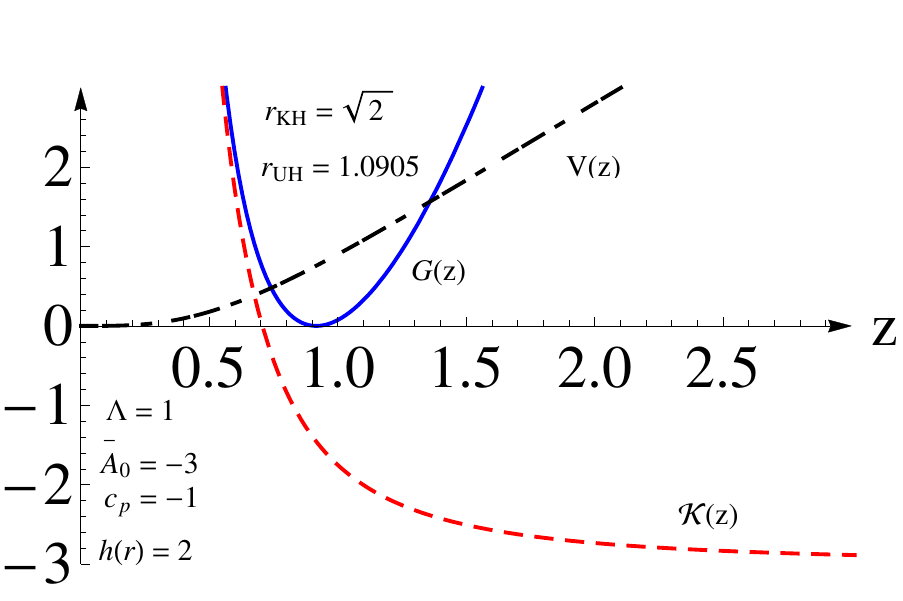}\includegraphics[width=5cm]{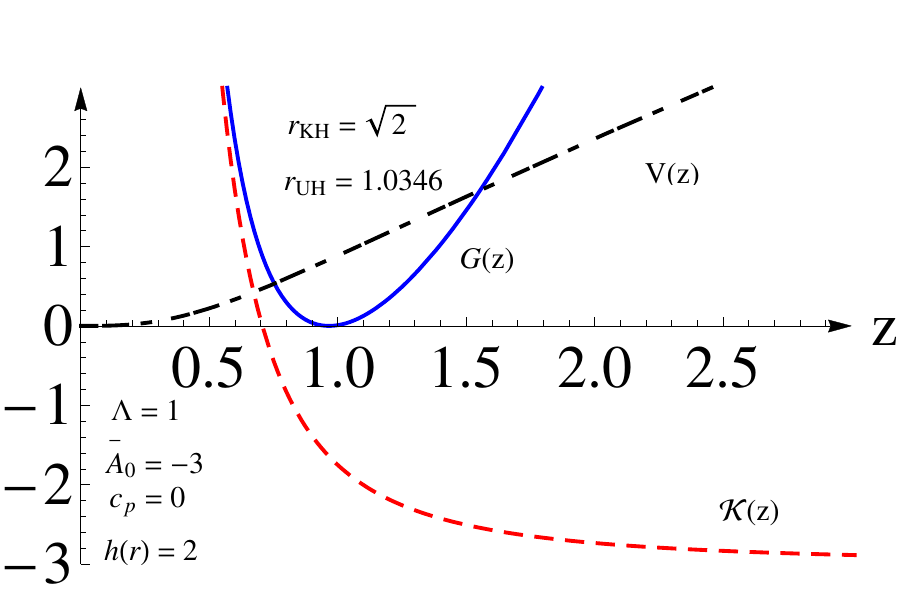}\includegraphics[width=5cm]{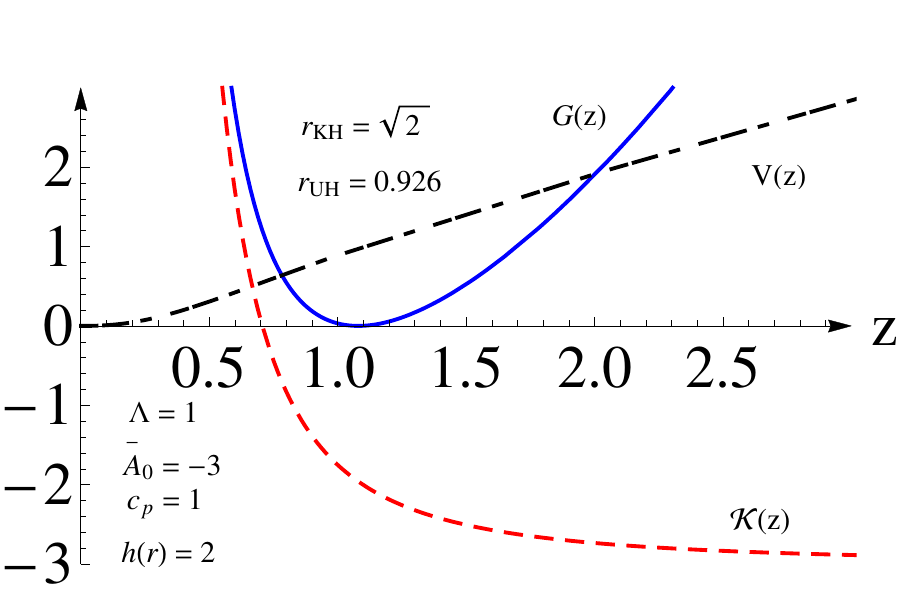}
\caption{The   functions $G, V \left(\equiv u^r\right)$ and ${\cal{K}}$ vs $z$, for the spacetime given by Eq.(\ref{4.21c}) with $ \Lambda = -1, A_0 = 3/5,
H = 2, \beta = 0$ and various choices of $c_p$. } 
\label{Fig7}
\end{figure*}

{\bf Case 1.d $\beta=0$:} In the case, we must assume that  $\frac{2\bar{A}_0+H^2}{\Lambda}\le0$, and   in
Fig.~\ref{Fig7} we show the functions $G, V$ and ${\cal{K}}$  and the locations of the Killing
and universal horizons.

\begin{figure*}[h]
\includegraphics[width=5cm]{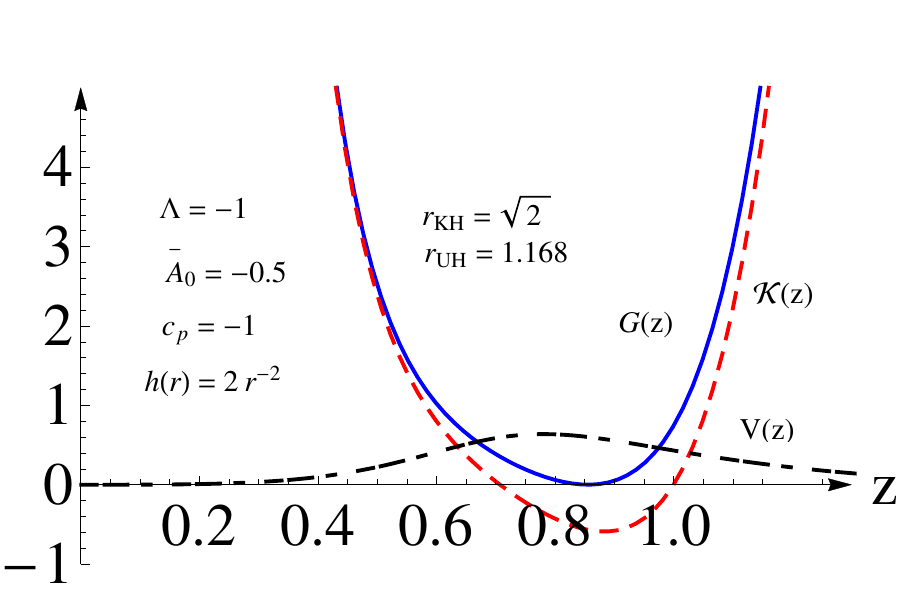}\includegraphics[width=5cm]{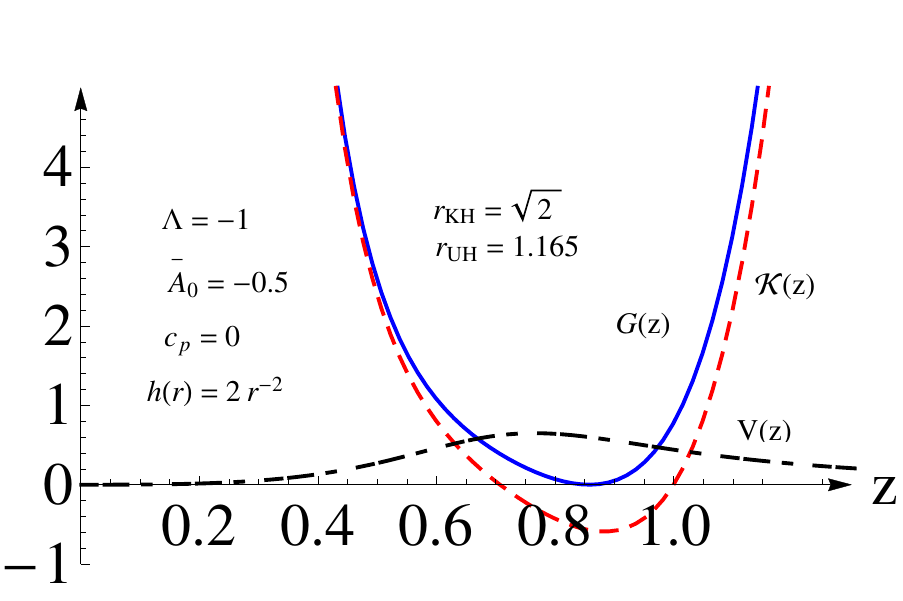}\includegraphics[width=5cm]{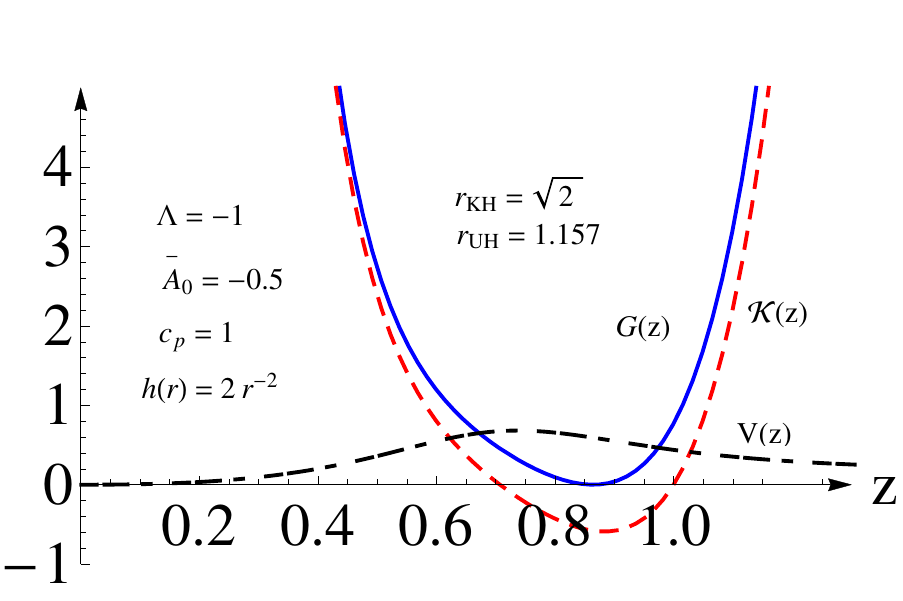}
\caption{The   functions $G, V \left(\equiv u^r\right)$ and ${\cal{K}}$ vs $z$, for the spacetime given by Eq.(\ref{4.21c}) with $ \Lambda = -1, A_0 = 3/5,
H = 2, \beta = 2$ and various choices of $c_p$. } 
\label{Fig8}
\end{figure*}

{\bf Case 1.e $\beta>0$:}  In this case, we require that   $\Lambda<0$. Then,    in
Fig.~\ref{Fig8} we show the functions $G, V$ and ${\cal{K}}$  and the locations of the Killing
and universal horizons for $\beta = 2$.

\subsubsection{$\Lambda_g=0,\; \lambda \not=1$}

In this case, the solutions are given by Eqs.(\ref{15}), (\ref{16}) and (\ref{17}). Similar to the last case, without loss of the generality, we can always set $C_1 = 1$, and the metric takes the form,
  \bqn
  \lb{4.22b}
  ds^2&=&-\left(A_0+\frac{C_3^2}{2r^2}-\Lambda r^2\right)^2dt^2\nb\\
  &&+\left[dr+\left(C_2r+\frac{C_3}{r}\right)dt\right]^2+r^2d\theta^2.
  \eqn
 To study these solutions further, let us consider the cases $C_3 = 0$ and $C_3 \not= 0$, separately.

\begin{figure*}[h]
\includegraphics[width=5cm]{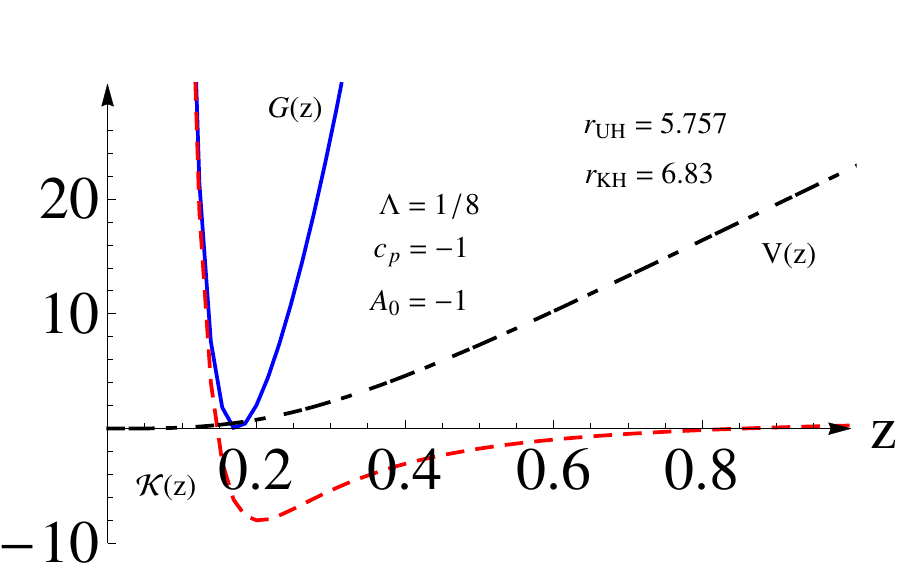}\includegraphics[width=5cm]{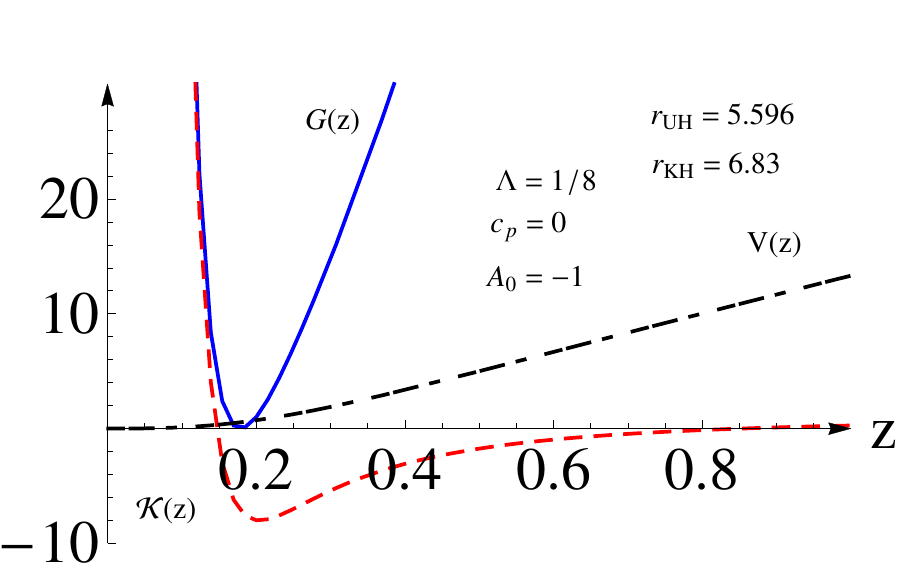}\includegraphics[width=5cm]{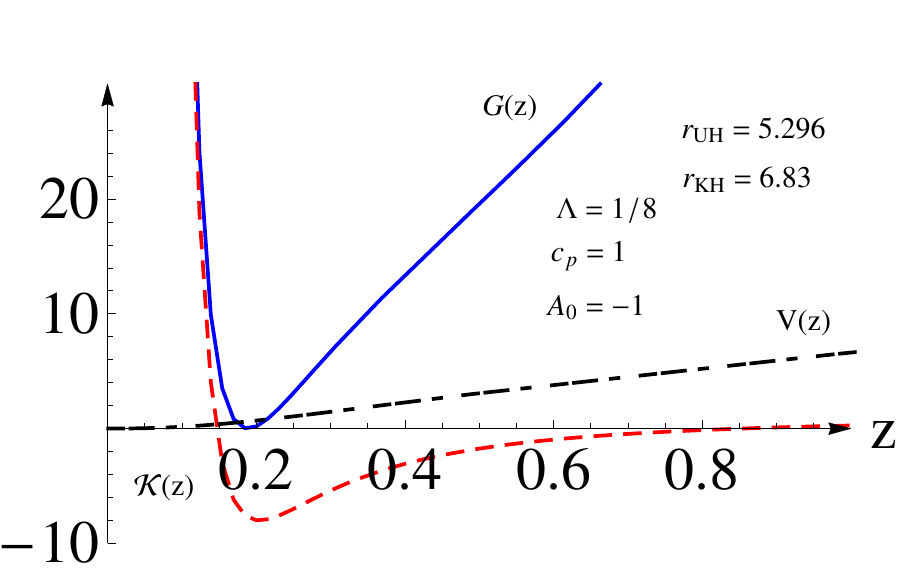}
\includegraphics[width=5cm]{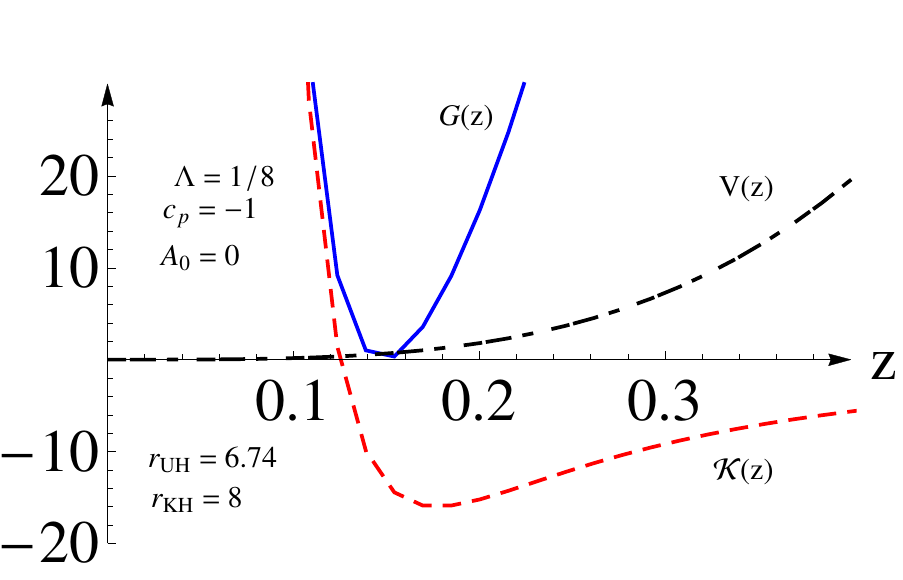}\includegraphics[width=5cm]{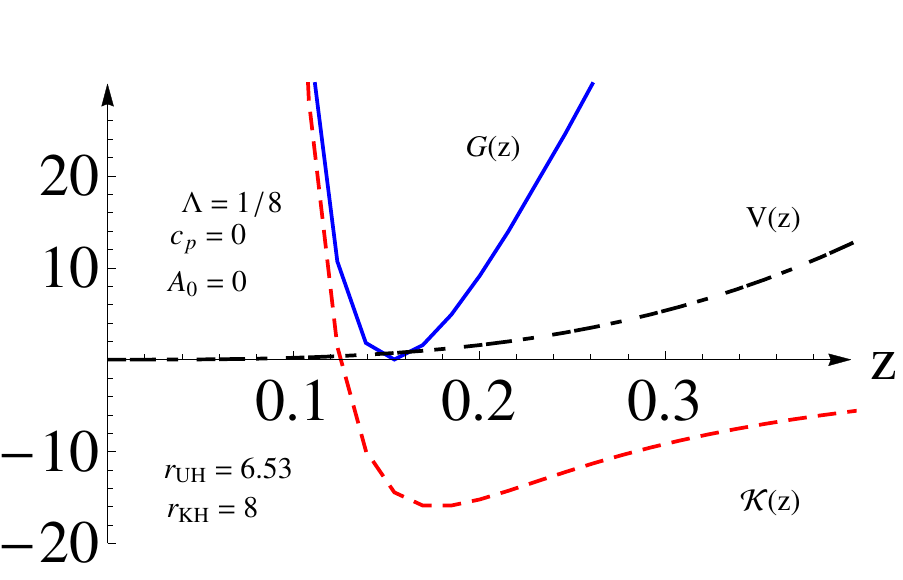}\includegraphics[width=5cm]{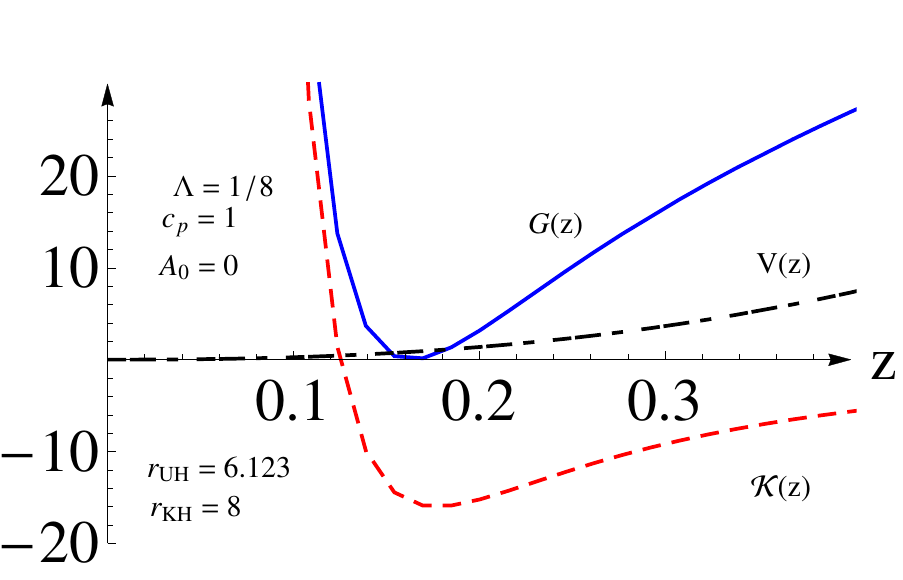}
\caption{The functions $G, V$ and ${\cal{K}}$  and the locations of the Killing and universal horizons for the solutions given by Eq.(\ref{4.22d}) with $C_3 = 0$ for various choices of the free parameters.} 
\label{Fig9}
\end{figure*}

When $C_3=0$, we assume that $C_2 \not= 0$. Otherwise, the metric will be singular across the Killing horizons.
Then, the  rescaling,
  \bqn
  \lb{4.22c}
  &&t\rightarrow C_2^{-1}t,~~A_0\rightarrow C_2A_0,~~\Lambda\rightarrow C_2\Lambda,
  \eqn
leads  the metric to the form,
  \bqn
  \lb{4.22d}
  ds^2&=&-\left(A_0-\Lambda
  r^2\right)^2dt^2+\left(dr+rdt\right)^2\nb\\
  && +r^2d\theta^2, \left(C_3 = 0\right),
  \eqn
from which we find that
  \bqn
  \lb{4.22e}
  R&=&2\frac{A_0(3-8\Lambda^2r^2)+4A_0^2\Lambda+\Lambda r^2+4\Lambda^3 r^4}{(A_0-\Lambda
  r^2)^3},\nb\\
  K&=&\frac{2}{A_0-\Lambda r^2}.
  \eqn
Thus, to avoid spacetime singularity at $A_0-\Lambda r^2 = 0$, we shall assume   $A_0\Lambda\le0$. On the other
hand, the Killing horizon is located at,
  \bqn
  \lb{4.22f}
  (A_0-\Lambda r_{KH}^2)^2-r_{KH}^2=0,
  \eqn
which has real and positive roots only when $1+4A_0\Lambda\ge0$. Moreover,  in the present case  Eq.(\ref{4.11a}) reduces to
  \bqn
  \lb{4.22f1}
  &&V''+\frac{A_0-3\Lambda r^2}{A_0r-\Lambda
  r^3}V'\nb\\
  &&-\frac{A_0^2+2A_0c_p\Lambda r^2+(3-2c_p)\Lambda^2r^4}{\left(A_0r-\Lambda r^3\right)^2}V=0.
  \eqn
In   Fig.~\ref{Fig9}, we show the functions $G, V$ and ${\cal{K}}$  and the locations of the Killing
and universal horizons for various choices of the free parameters, as specified in each of the panels of the figure.

When $C_3\not=0$, the rescaling of the timelike coordinate and the redefinitions of the parameters,  
  \bqn
  \lb{4.22g}
  &&t\rightarrow C_3^{-1}t,~~A_0\rightarrow C_3A_0,~~\Lambda\rightarrow
  C_3\Lambda,\nb\\
  &&C_1\rightarrow2C_A/C_3,
  \eqn
lead  the metric to the form,
  \bqn
  \lb{4.22h}
  ds^2&=&-\left(A_0+\frac{C_A}{r^2}-\Lambda r^2\right)^2dt^2\nb\\
  &&+\left[dr+\left(C_2r+\frac{1}{r}\right)dt\right]^2+r^2d\theta^2.
  \eqn

\begin{figure*}[h]
\includegraphics[width=5cm]{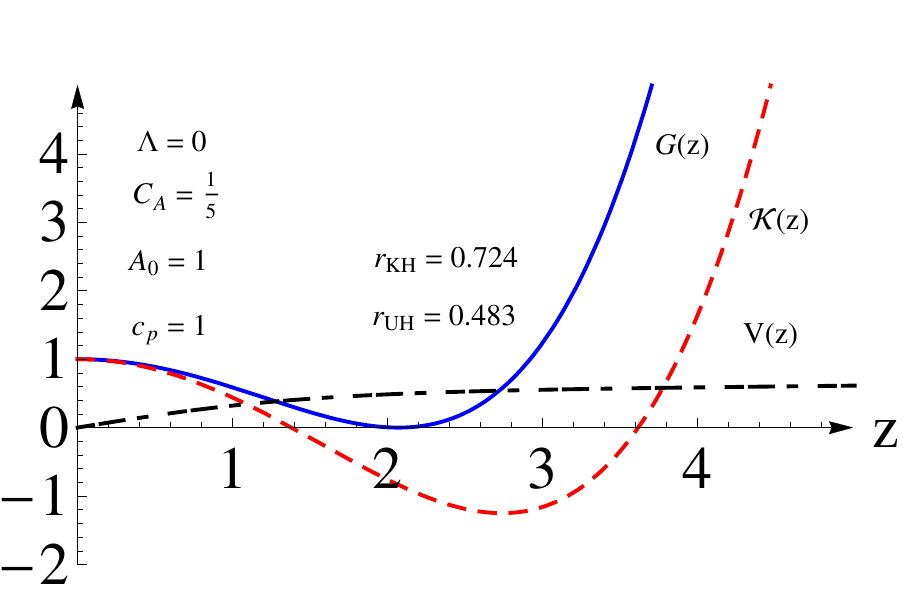}\includegraphics[width=5cm]{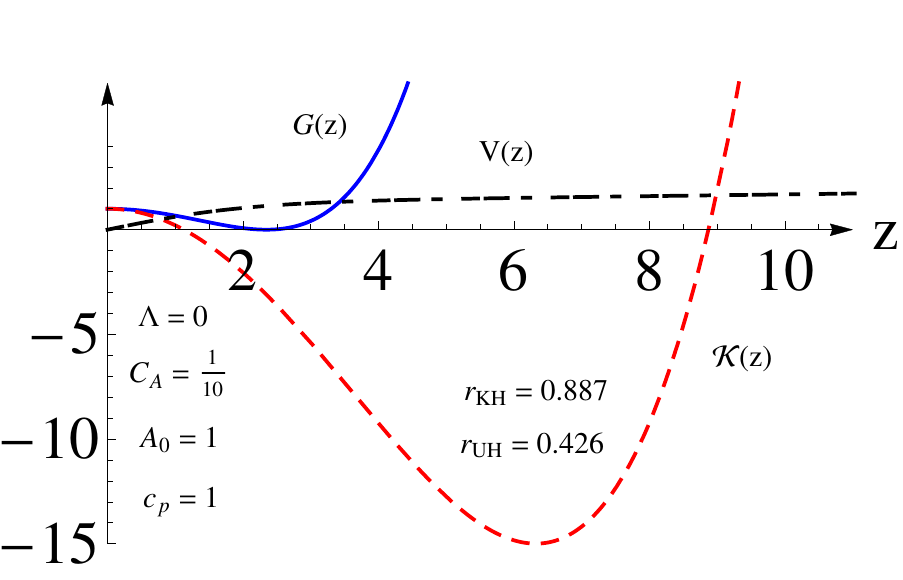}\includegraphics[width=5cm]{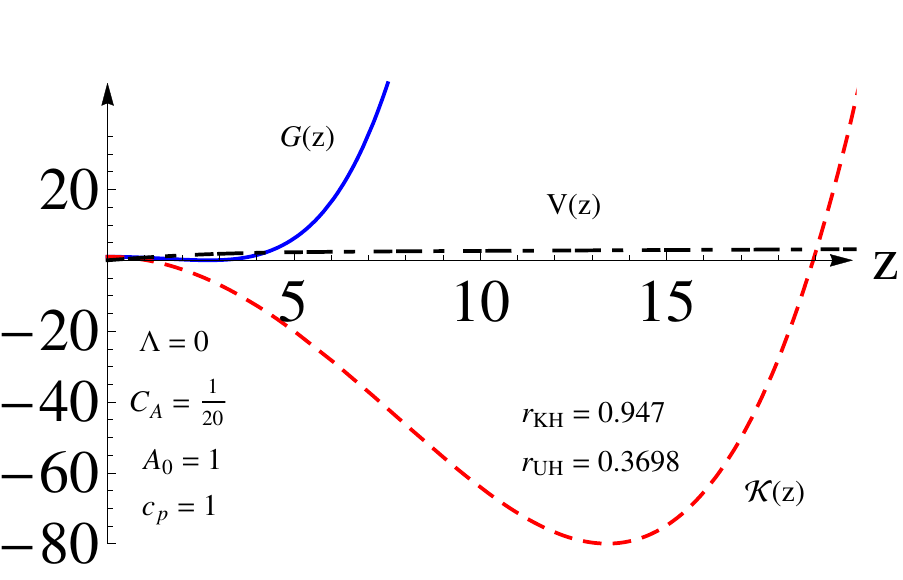}
\includegraphics[width=5cm]{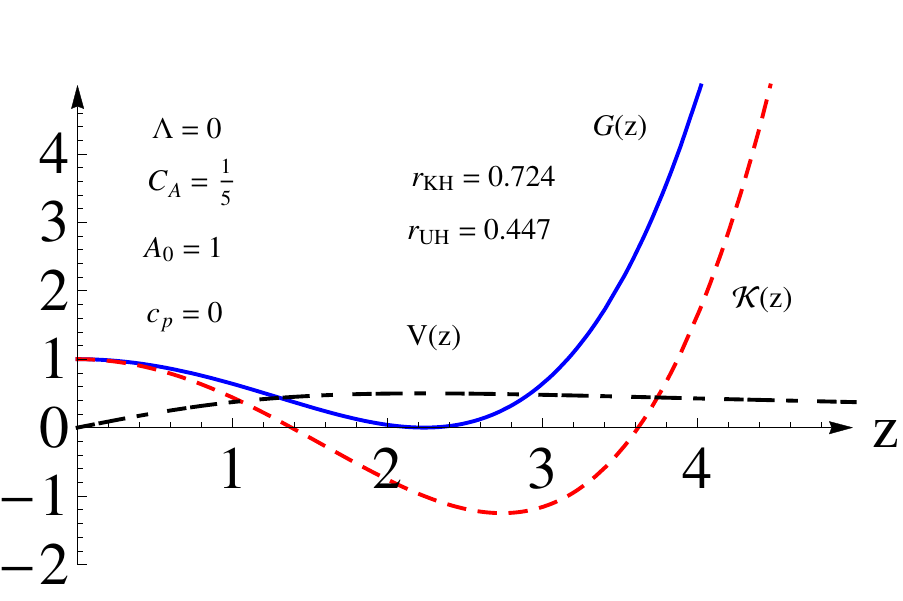}\includegraphics[width=5cm]{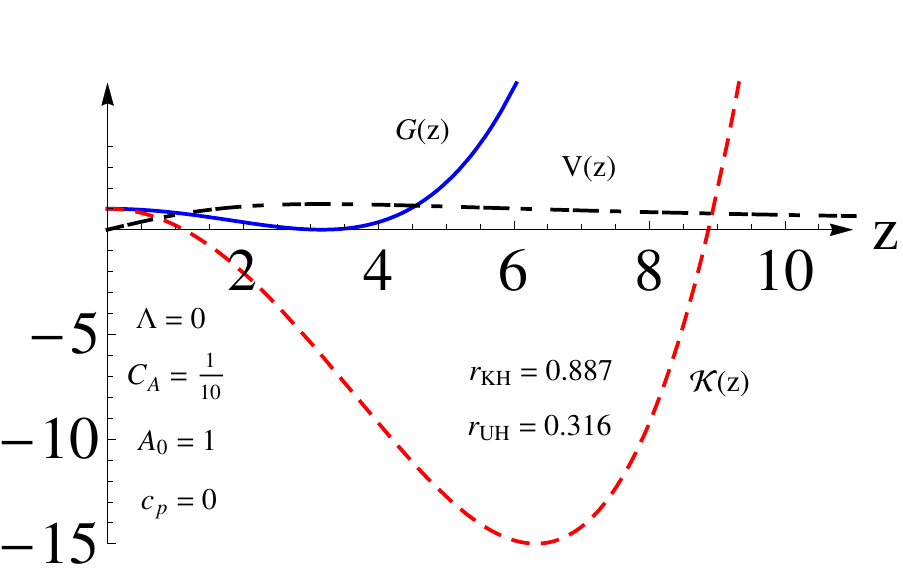}\includegraphics[width=5cm]{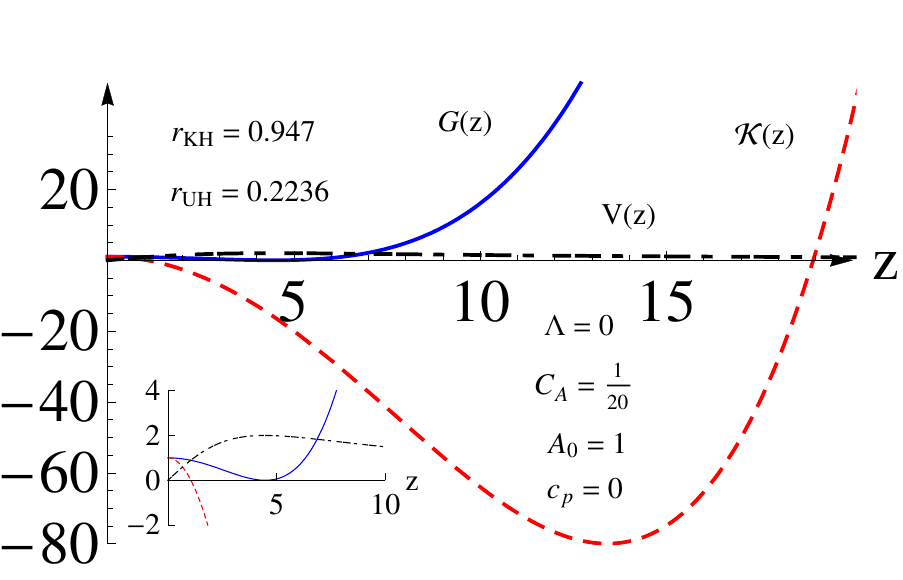}
\caption{The functions $G, V$ and ${\cal{K}}$  and the locations of the Killing
and universal horizons for the solutions given by Eq.(\ref{4.22h}) with $C_c = 0 = \Lambda$. } 
\label{Fig10}
\end{figure*}

If $C_2=0=\Lambda$, this metric becomes asymptotically flat at
spatial infinity,  and Eq.(\ref{4.11a}) is given by
  \bqn
  \lb{4.22i}
  &&\frac{C_A^2(1-2c_p)-2A_0C_A(c_p-2)r^2-A_0^2r^4}{r^2(A_0r^2+C_A)^2}V\nb\\
  &&+V''+\frac{A_0r^2-C_A}{A_0r^3+C_Ar}V'=0.
  \eqn
In Fig.~\ref{Fig10}, we show the locations of the Killing and universal horizons.

\begin{figure*}[h]
\includegraphics[width=5cm]{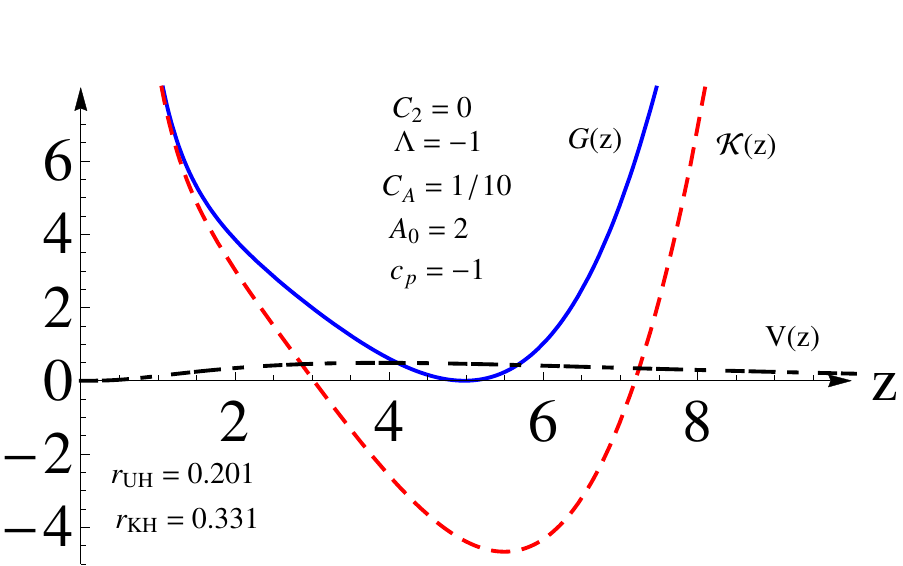}\includegraphics[width=5cm]{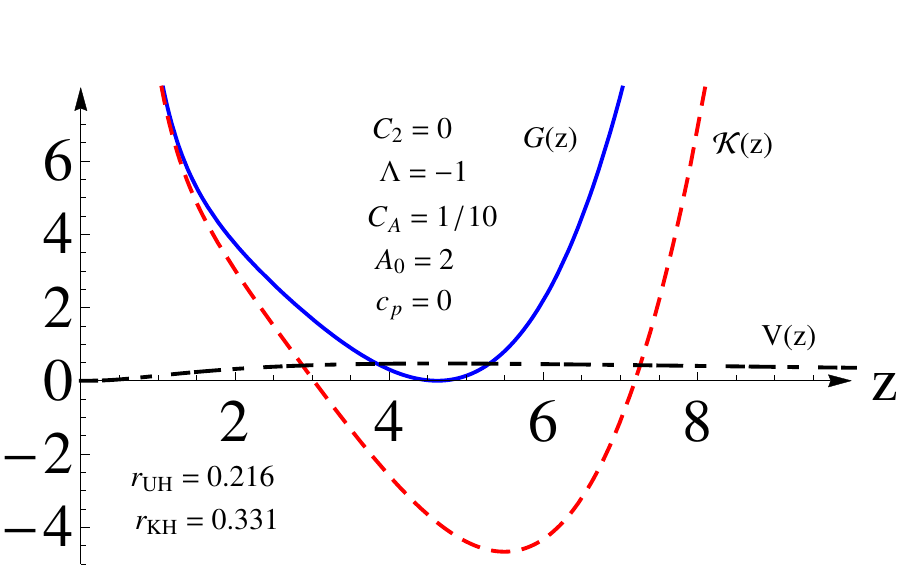}\includegraphics[width=5cm]{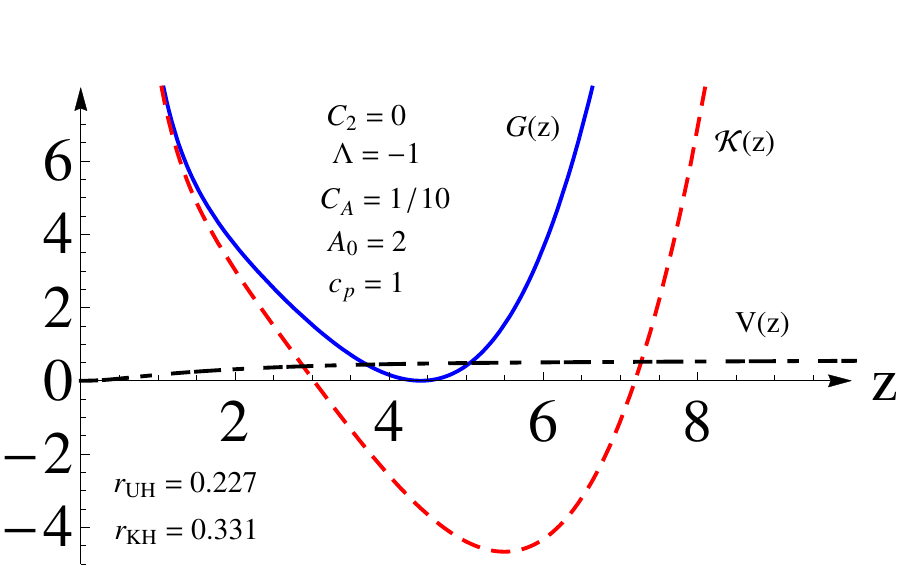}
\includegraphics[width=5cm]{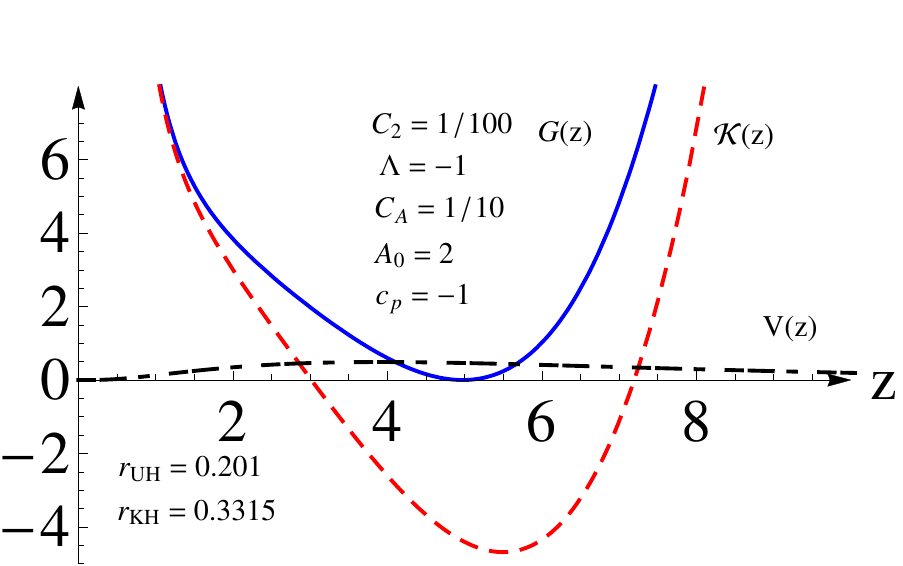}\includegraphics[width=5cm]{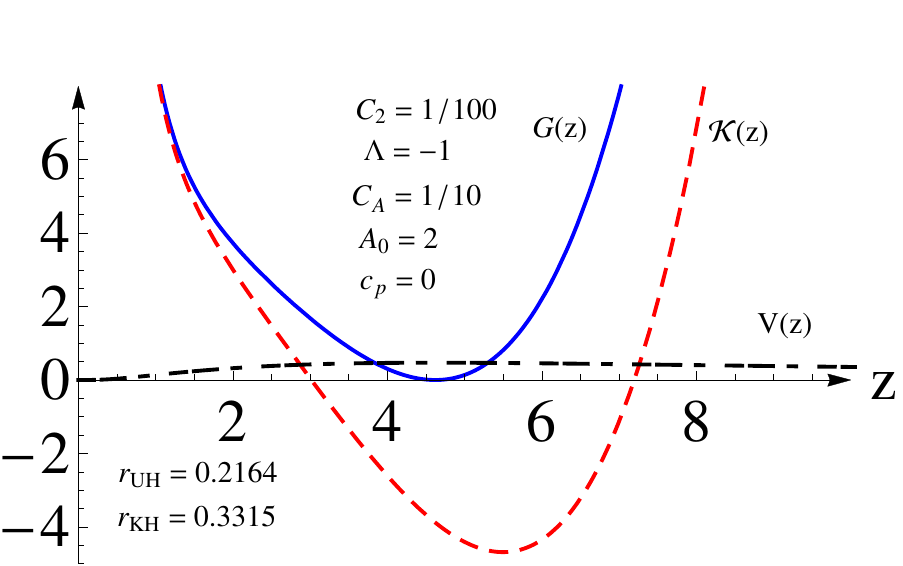}\includegraphics[width=5cm]{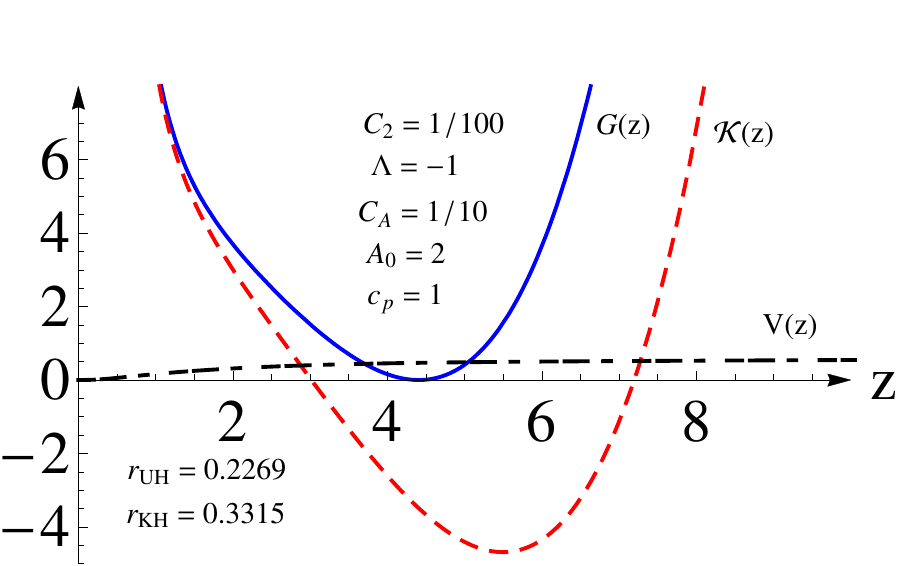}
\caption{The functions $G, V$ and ${\cal{K}}$  and the locations of the Killing
and universal horizons for the spacetimes given by Eq.(\ref{4.22i}) with $C_3 \not= 0, \Lambda = -1, C_2 = 0$.} 
\label{Fig11}
\end{figure*}

If $C_2$ and $\Lambda$ do not vanish at the same time, we find
  \bqn
  \lb{4.22j}
  R&=&\frac{2}{r^{2} {\cal{D}}(r)^3}\left[C_A^2(12\Lambda r^4-8A_0r^2)\right.\nb\\
  &&-4C_A^2+C_A[r^2-4(A_0^2-C_2)r^4-12\Lambda^2r^8\nb\\
  &&+(7C_2^2+16A_0\Lambda)r^6]+r^4[4A_0^2\Lambda r^4\nb\\
  &&+A_0(1+3C_2^2r^4-8\Lambda^2r^6)+\Lambda r^2(4\Lambda^2r^6+C_2^2r^4\nb\\
  &&\left.+4C_2r^2-1)]\right],\nb\\
  K&=&\frac{2C_2r^2}{ {\cal{D}}(r)},\;\;\;
  {\cal{D}}(r) \equiv C_A+A_0r^2-\Lambda r^4.
  \eqn
Again, to avoid spacetime singularities at finite but non-zero $r$, we must assume that $ {\cal{D}}(r) \not= 0$ for $r \in(0, \infty]$. In
Fig.~\ref{Fig11}, we show the functions $G, V$ and ${\cal{K}}$  and the locations of the Killing
and universal horizons.

\subsubsection{$\Lambda_g\not=0$}

When $\Lambda_g\not=0$, the solutions are mathematically much involved, and in this subsection we only consider the case where
$\lambda \not= 1$ but $C_1 = 0$. Then, the solutions are given by Eqs.(\ref{9}), (\ref{13})-(\ref{13.c}), for which
 the extrinsic curvature is given by
  \bqn
  \lb{4.23aA}
  K&=&\frac{r^{\frac{\zeta-\lambda_p}{2\zeta}}}{2\zeta}\left[C_-r^{\frac{\lambda-5}{\lambda_p}}\left(\lambda_p-\zeta\right)-C_+\left(\lambda_p+\zeta\right)\right]\nb\\
  &&\times\left[\Lambda_2-A_0r-C_+^2\frac{r^{\frac{5-\lambda}{\lambda_p}-1}}{4\Lambda_g(1+3\lambda)}\left(\zeta(13+3\lambda_p\right.\right.\nb\\
  &&\left.-3\lambda)+8\right)-C_-^2\frac{r^{\frac{-5+\lambda}{\lambda_p}-1}}{4\Lambda_g(1+3\lambda)}\left(\zeta(-13+3\lambda_p\right.\nb\\
  &&\left.\left.+3\lambda)-8\right)\right]^{-1},
  \eqn
where $\lambda_p=\sqrt{(\lambda-1)(\lambda-5)}$ and
$\zeta=\lambda-1$. Note that to have the metric real, as noticed in the last section, we must have either   $\lambda\le1$ or $\lambda\ge5$.
 When $\lambda\le1$, $\lambda$ should be larger than $-\frac13$ so
that $K$ remains finite at the spatial infinity.

In particular, for   $\Lambda_g<0, \; \lambda=\frac{1}{2}$ and $C_+=0$, we have
  \bqn
  \lb{4.23a}
  {\cal{N}}(r)&=&\Lambda_2-A_0r-\frac{9C_2^2}{20\Lambda_gr^4},\nb\\
  h(r) &=& \frac{C_2}{r},~~f(r)=-{\Lambda_gr^2},
  \eqn
where we had replaced $C_-$ by $C_2$.  Then, Eq.(\ref{4.11a}) can be rewritten as
  \bqn
  \lb{4.23b}
  &&\frac{V''}{V}+4\frac{5A_0\Lambda_gr^5-9C_2^2}{9C_2^2r+20r^5(A_0r-\Lambda_2)\Lambda_g}\frac{V'}{V}\nb\\
  &&-\left[9C_2^2r+20r^5(A_0r-\Lambda_2)\Lambda_g\right]^{-2}\left[81C_2^4(5c_p-4)\right.\nb\\
  &&+180C_2^2r^4\Lambda_g\left(A_0(5c_p-28)r+20\Lambda_2-6c_p\Lambda_2\right)\nb\\
  &&\left.+400r^8\Lambda_g^2\left(A_0^2r^2-A_0c_p\Lambda_2r+c_p\Lambda_2^2\right)\right].
  \eqn

\begin{figure*}[h]
\includegraphics[width=5cm]{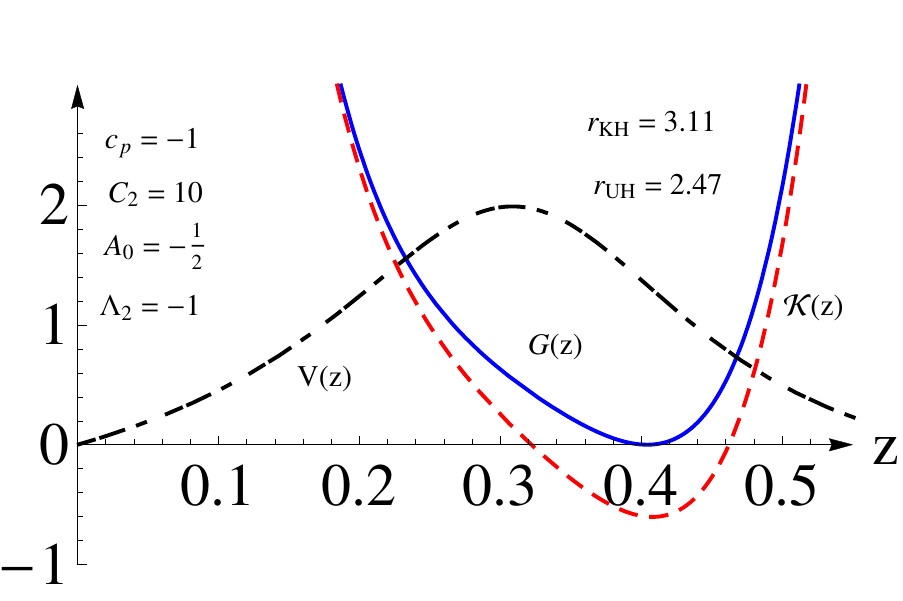}\includegraphics[width=5cm]{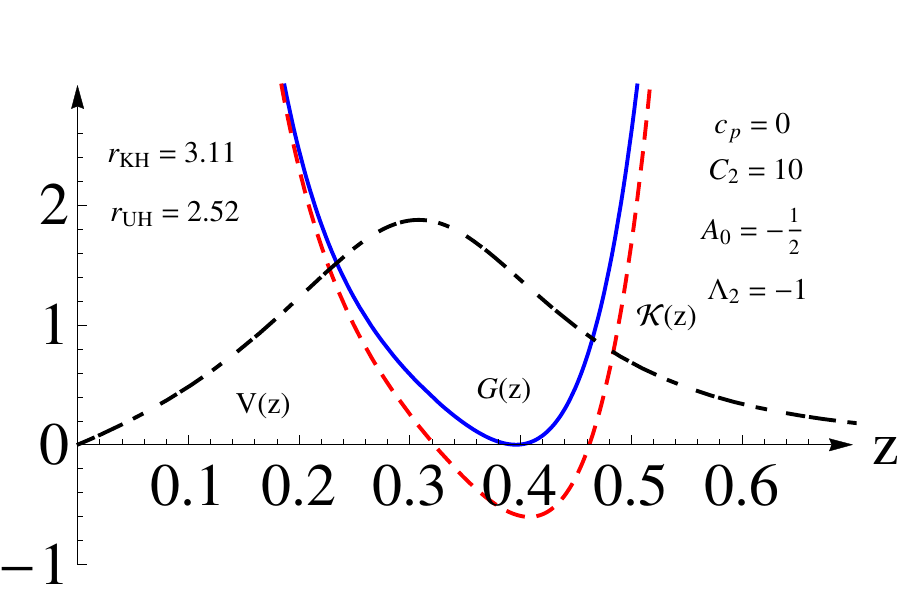}\includegraphics[width=5cm]{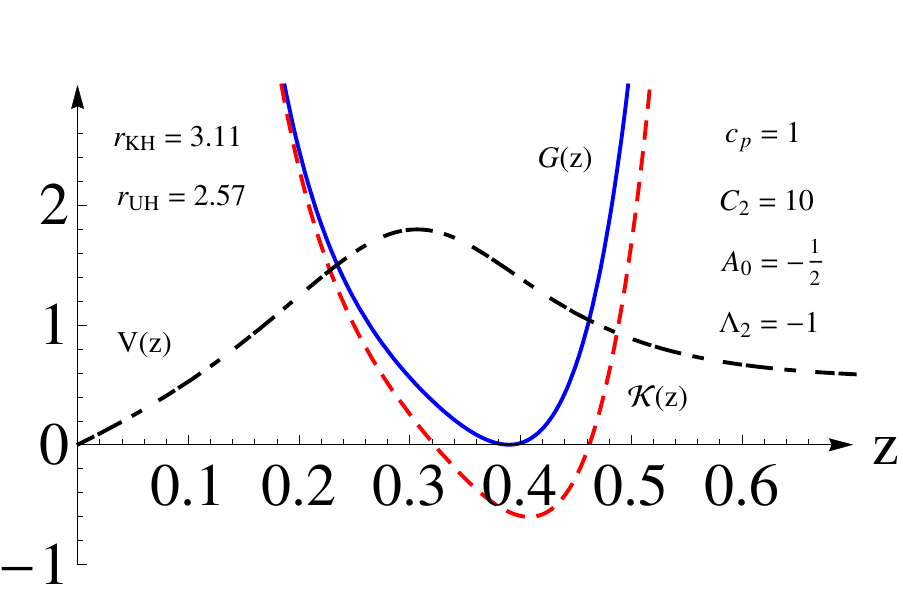}
\caption{The functions $G, V$ and ${\cal{K}}$  and the locations of the Killing and universal horizons for the spacetimes given by Eq.(\ref{4.23a}) with $\lambda\le1,\; \Lambda=-1$.} 
\label{Fig12}
\end{figure*}

In Fig.~\ref{Fig12}, we show the functions ${\cal{K}}$, $V$ and $G$, and numerically find the radii of the Killing and universal horizons for $\lambda \le 1$.  

\begin{figure*}[h]
\includegraphics[width=5cm]{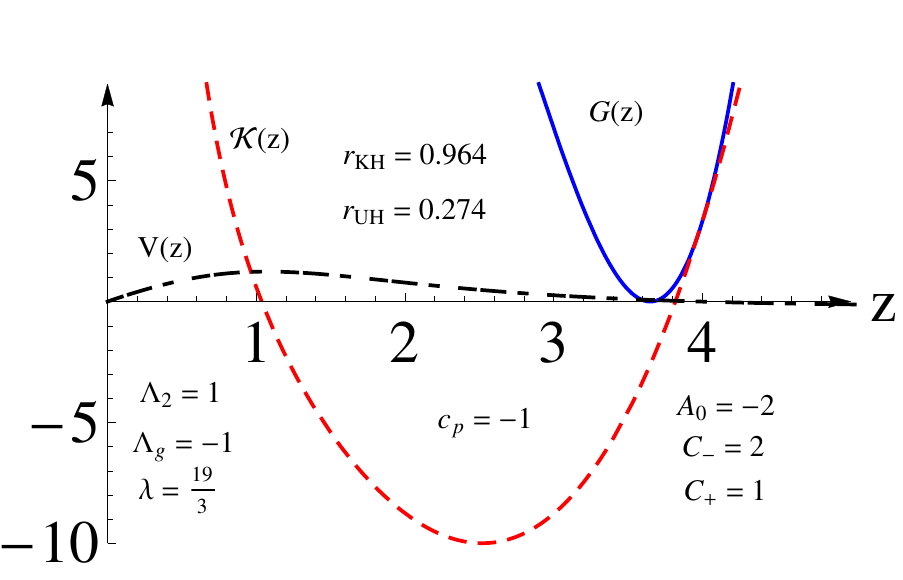}\includegraphics[width=5cm]{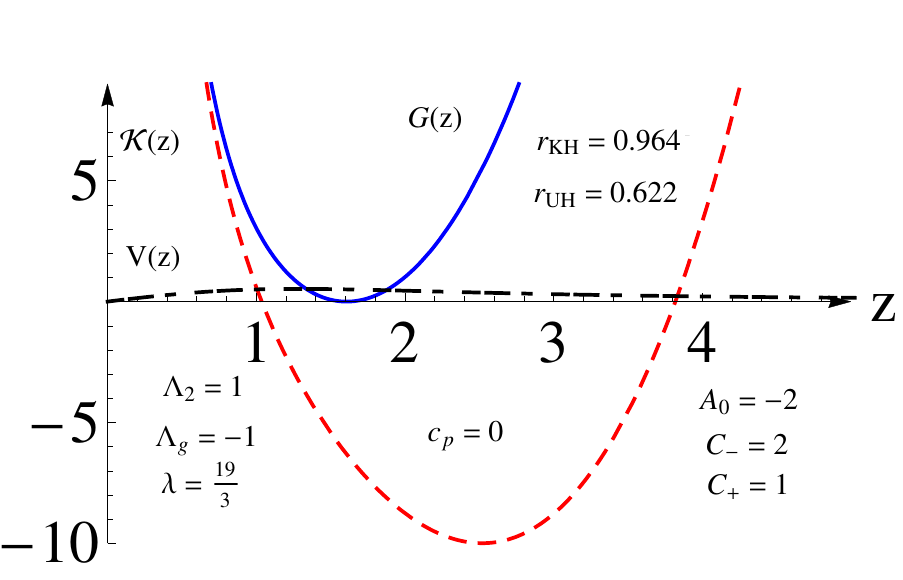}\includegraphics[width=5cm]{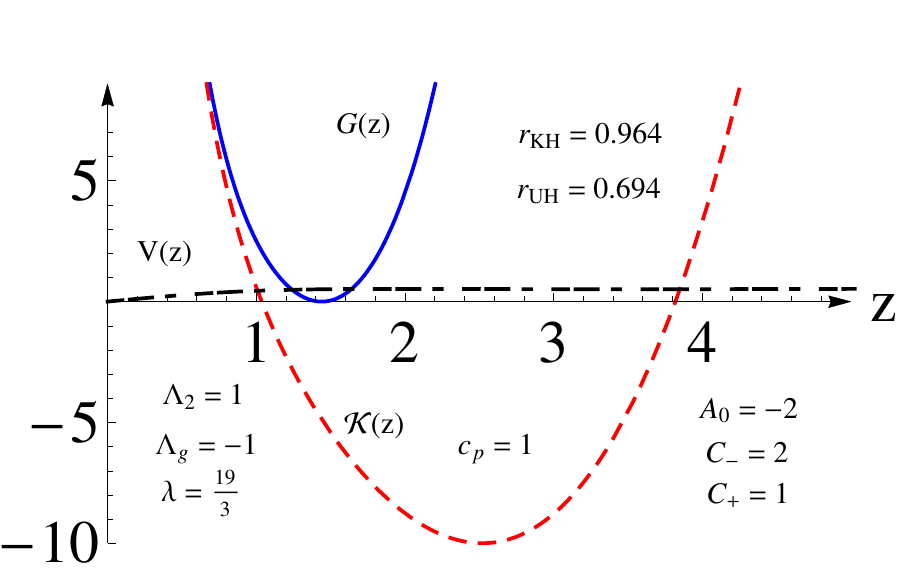}
\caption{The functions $G, V$ and ${\cal{K}}$  and the locations of the Killing and universal horizons for the spacetimes given by Eq.(\ref{4.23a}) with $\lambda = 19/3 > 5,\; \Lambda=-1$.} 
\label{Fig13}
\end{figure*}

A similar consideration is presented in Fig.~\ref{Fig13} for $\lambda \ge 5$. In particular, in this figure we have chosen   $\lambda=\frac{19}{3}$, $\Lambda_2=1$, $\Lambda_g=-1$, $A_0=-2$, $C_-=2$ and $C_+=1$. 

In all the cases considered above, a universal horizon always exists inside a Killing horizon. To assure that no coordinate singularities appear across the Killing horizons, 
we are forced to use the PG-like coordinates, in which $N^{i}$ does not vanishes identically. Although not all of the solutions of the theory can be written in this form, as we mentioned above, 
the solutions considered in this paper indeed  all possess  these properties.

It is also important to
note that the solutions presented above are the solutions of the full theory, that is, including the contribution of the higher-order derivative term specified by the coefficient $g_2$. Therefore,  our above
results show that {\em universal horizons and black holes exist  not only in the infrared limit, but also in the ultraviolet  limit.}

\subsection{Universal Horizons and Black Holes in Rotating Spacetimes}

When spacetimes have rotations, the Killing horizons are different from the event horizons. The former are often called ergosurfaces (or ergospheres when the topology is $S^n$, where $n$ denotes the dimensions of the
horizon), while  the latter is defined by the existence of a null surface of the normal vector, $r_\mu \equiv \partial_{\mu} r$ \cite{Visser07}, that is,
\bq
\lb{eq3.a}
\left.\gamma^{\mu\nu} r_{\mu}r_{\nu}\right|_{r= r_e} = \gamma^{rr}\left(r_e\right) = 0.
\eq
In the following, we shall continuously denote the locations of the Killing horizons (or ergosurfaces) by $r_{KH}$, while the locations of the event horizons by
$r_e$.

In the rotating spacetimes described by Eqs.(\ref{Eq5A1}), (\ref{Eq5A5}), (\ref{E1}) and (\ref{E2}),  it is reasonable to assume that the khronon field $\phi$ depends only on $t$ and $r$, as all the metric coefficients are  independent of
$\theta$. Then, we find that $u_{\mu} \left(\propto \phi_{,\mu}\right)$ must take the form $u_{\mu} = \left(u_t(r), u_r(r), 0\right)$, where $u_t = u_t\left(u_r\right)$, as now we have $\gamma^{\mu\nu}u_{\mu}u_{\nu} = -1$.
Note that all of the three components $u^{\mu} = \left(u^t, u^r, u^{\theta}\right)$ in general  do not vanish, due to the rotation, although  both of $u^{t}$ and $u^{\theta}$ are not independent, and can be expressed as functions
of $u_r$ (or equivalently as functions of $u^r$).    With these in mind, we find that  $V \left(\equiv u^r\right)$ satisfies the following differential equation,
  \bqn
  \lb{Eq5A4a}
  \frac{V''}{V}+\left(\frac{{\cal{N}}'}{{\cal{N}}}+\frac{1}{r}\right)\frac{V'}{V}+\frac{{\cal{N}}''}{{\cal{N}}}\nb\\
  -\left(\frac{{\cal{N}}'}{{\cal{N}}}\right)^2-\frac{1}{r^2}+\frac{c_p}{r}\frac{{\cal{N}}'}{{\cal{N}}}=0,
  \eqn
which just depends on the function ${\cal{N}}$. To proceed further, we consider the three classes of solutions, separately.

\subsubsection{$R_{ij} = 0, \; \lambda=1$}

 In this case,  the rescaling
 \bqn
 \lb{Eq5A5a}
&& r\rightarrow C_1^{1/2}r,~~~\theta\rightarrow C_1^{-1/2}\theta,~~~h_r\rightarrow
 C_1^{1/2}h_r,\nb\\
&& h_a\rightarrow C_1^{1/2}h_a,~~~h_b\rightarrow C_1^{-1/2}h_b,
 \eqn
brings the metric into the form, 
 \bqn
 \lb{Eq5A5b}
  ds^2&=&-\left(A_0+\frac{H_R^2-\Lambda
 r^2}{2}\right)^2dt^2\nb\\
 &&+\left[dr\pm\sqrt{H_R^2-\frac{h_a^2}{r^2}}dt\right]^2\nb\\
 &&+r^2\left[d\theta+\left(\frac{h_a}{r^2}+h_b\right)dt\right]^2,
 \eqn
where $H_R^2 \equiv h_r(r)^2+\frac{h_a^2}{r^2}$.  Recall that $h(r)$ is an arbitrary function of $r$, and $h_a$ and $h_b$ are the integration constants.
Then,  the killing horizon satisfies the equation,
 \bqn
 \lb{Eq5A5c}
  \left(2A_0+{H_R^2-\Lambda
 r^2}\right)^2- 4\left(H_R^2+2h_ah_b+h_b^2r^2\right) = 0.\nb\\
 \eqn

\begin{figure*}[h]
\includegraphics[width=5cm]{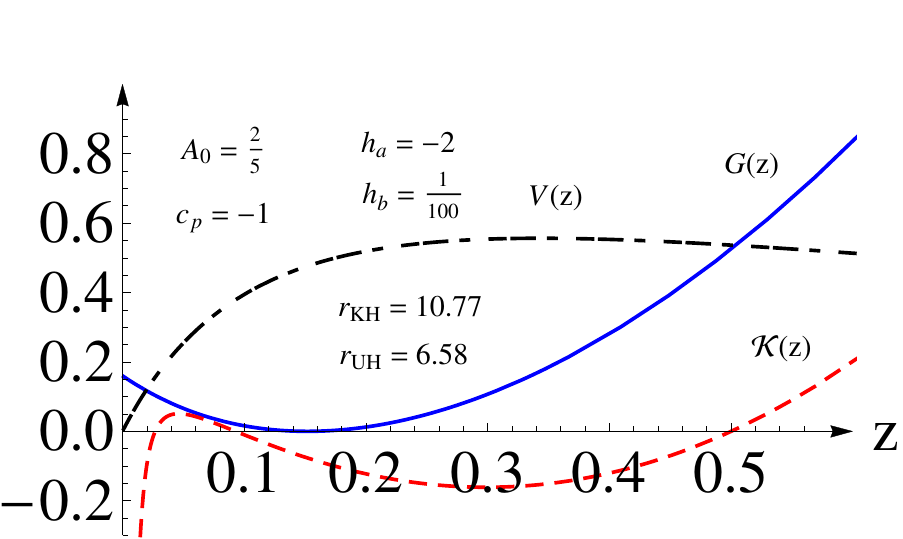}\includegraphics[width=5cm]{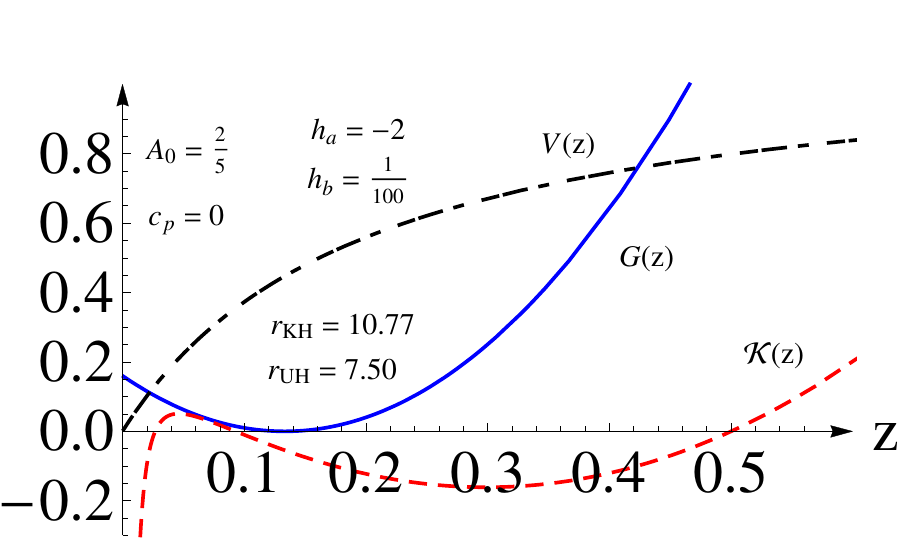}\includegraphics[width=5cm]{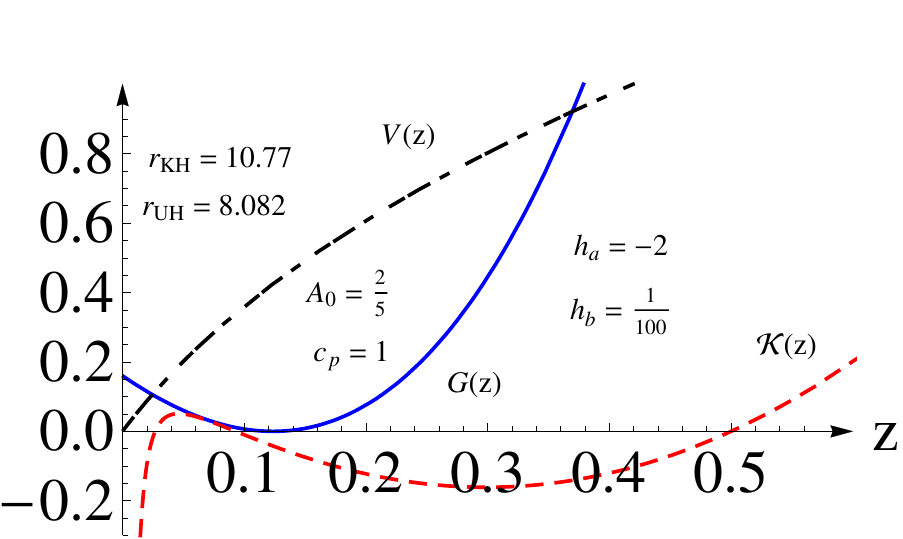}
\caption{The functions $V, G, {\cal{K}}$ and the locations of the Killing, event and universal horizons in the case $\lambda=1$ of the rotating spacetimes given by Eq.(\ref{Eq5A5b})  with $\Lambda=0$.} 
\label{Fig14}
\end{figure*}

\begin{figure*}[h]
\includegraphics[width=5cm]{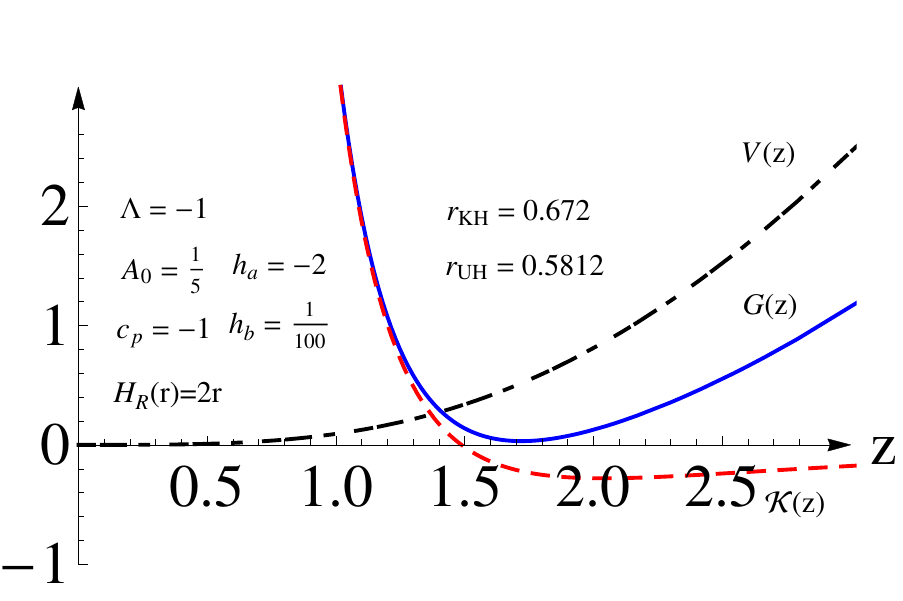}\includegraphics[width=5cm]{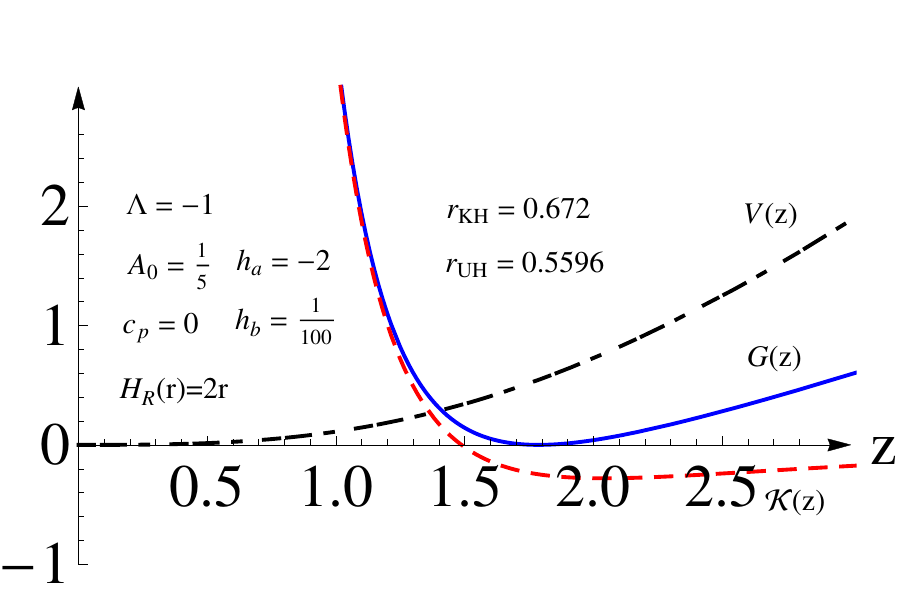}\includegraphics[width=5cm]{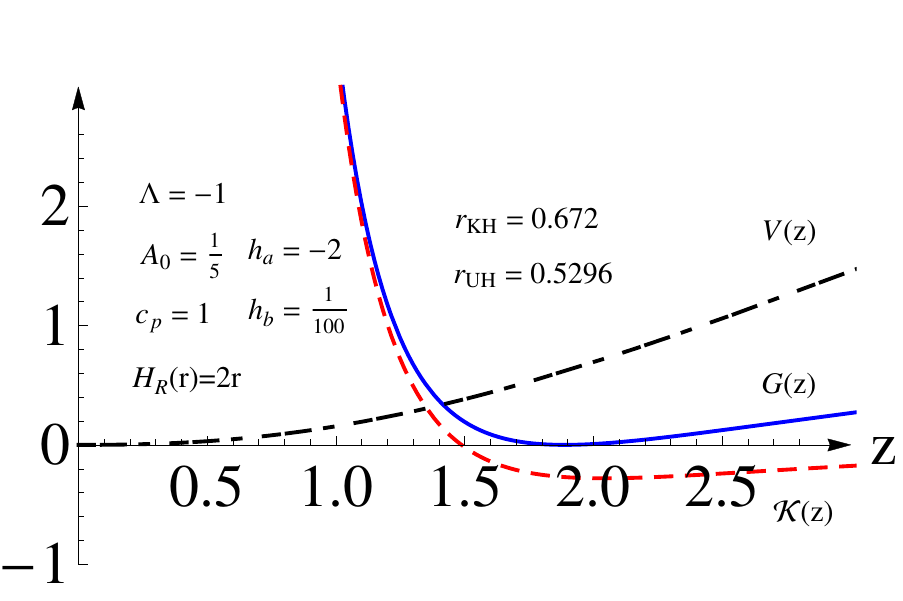}
\caption{The functions $V, G, {\cal{K}}$ and the locations of the Killing, event and universal horizons in the case $\lambda=1$ of the rotating spacetimes given by Eq.(\ref{Eq5A5b})  with $\Lambda\not=0$ and $\beta=-1$.} 
\label{Fig15}
\end{figure*}

\begin{figure*}[h]
\includegraphics[width=5cm]{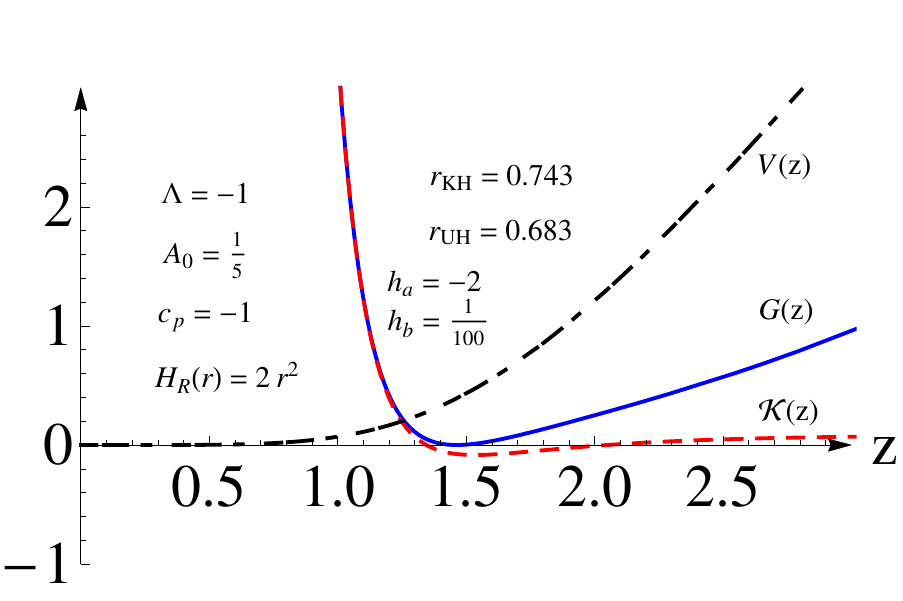}\includegraphics[width=5cm]{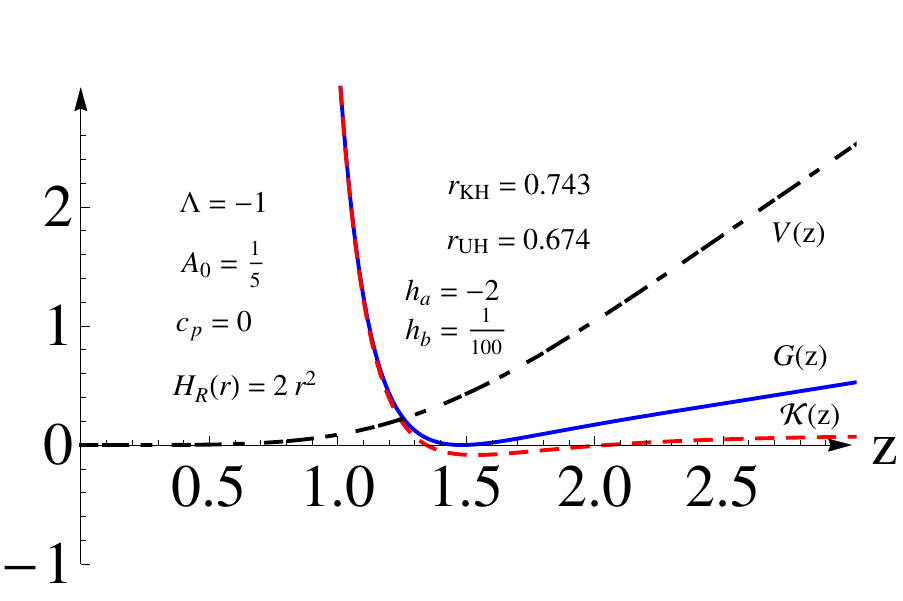}\includegraphics[width=5cm]{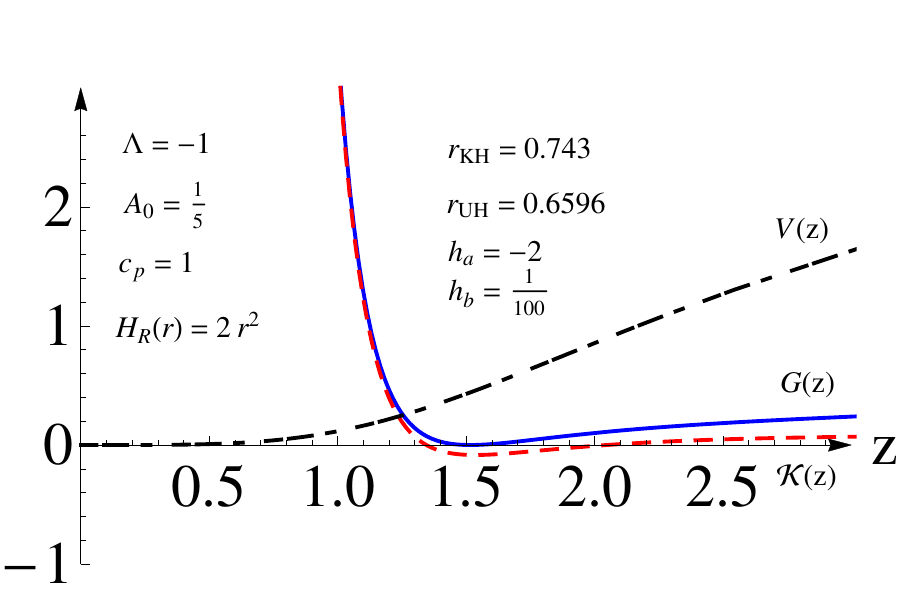}
\caption{The functions $V, G, {\cal{K}}$ and the locations of the Killing, event and universal horizons in the case $\lambda=1$ of the rotating spacetimes given by Eq.(\ref{Eq5A5b})  with $\Lambda\not=0$ and $\beta<-1$.} 
\label{Fig16}
\end{figure*}

\begin{figure*}[h]
\includegraphics[width=5cm]{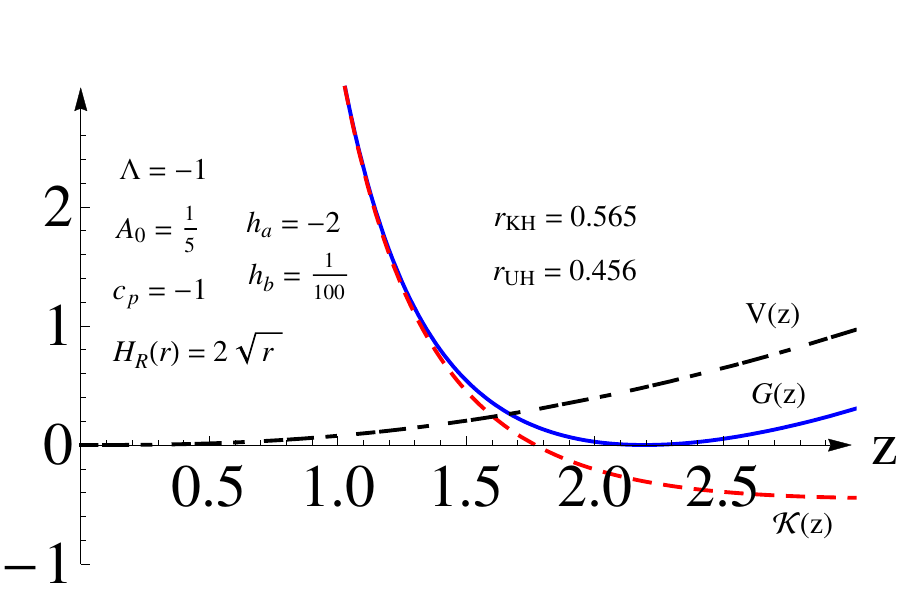}\includegraphics[width=5cm]{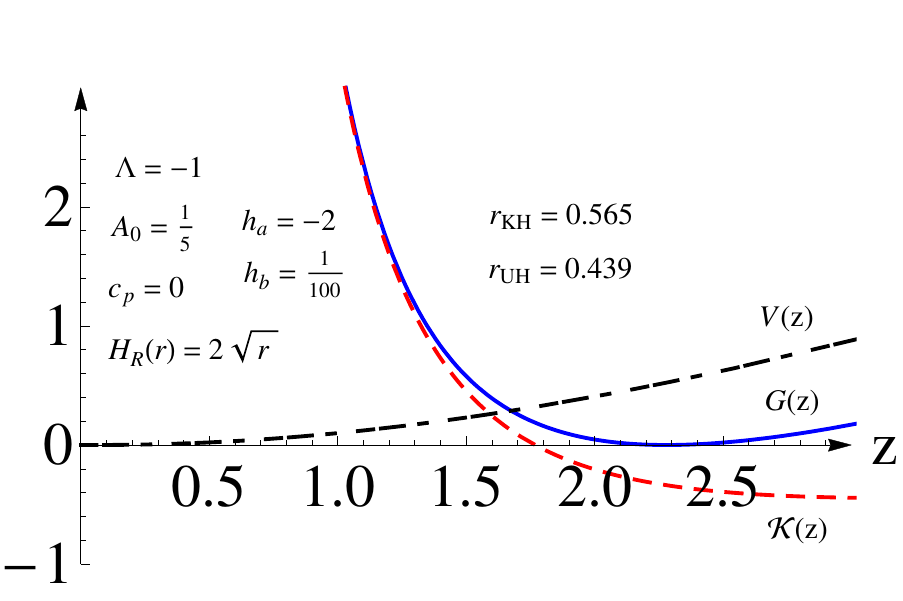}\includegraphics[width=5cm]{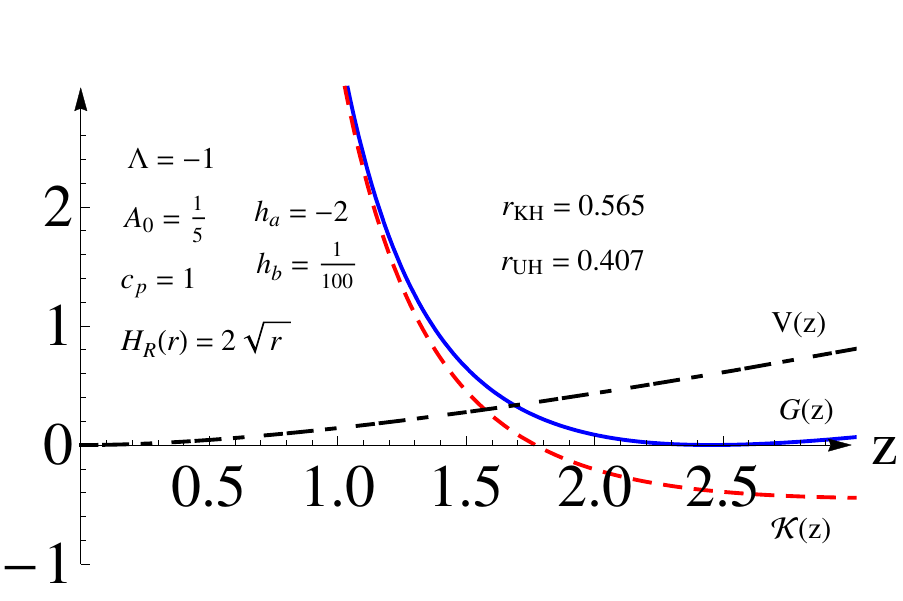}
\caption{The functions $V, G, {\cal{K}}$ and the locations of the Killing, event and universal horizons in the case $\lambda=1$ of the rotating spacetimes given by Eq.(\ref{Eq5A5b})  with $\Lambda\not=0$ and $\beta>-1$. } 
\label{Fig17}
\end{figure*}

\begin{figure*}[h]
\includegraphics[width=5cm]{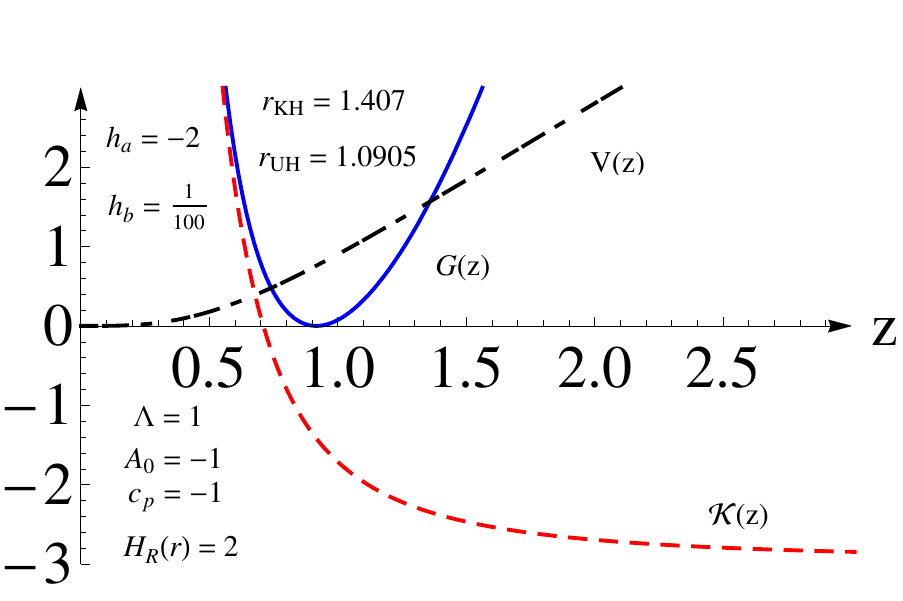}\includegraphics[width=5cm]{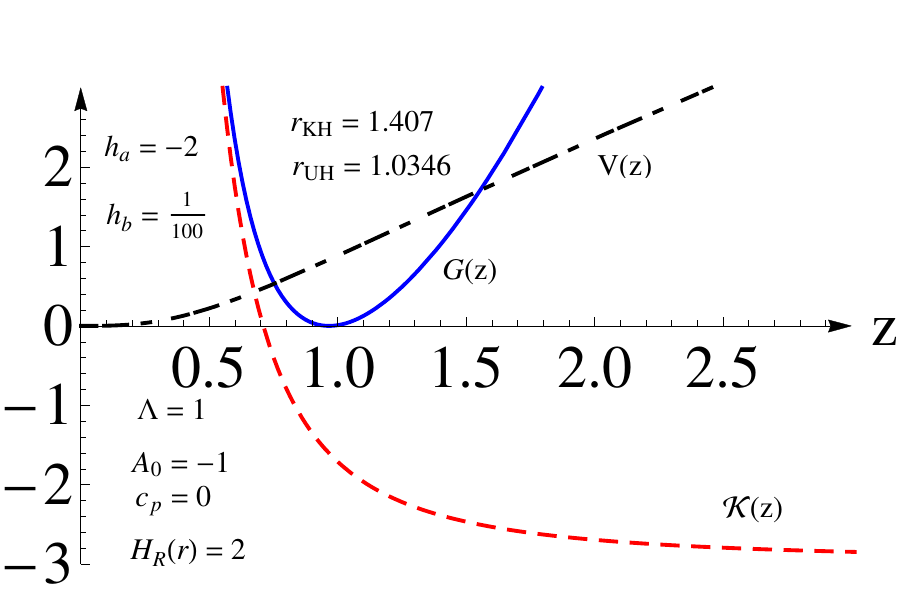}\includegraphics[width=5cm]{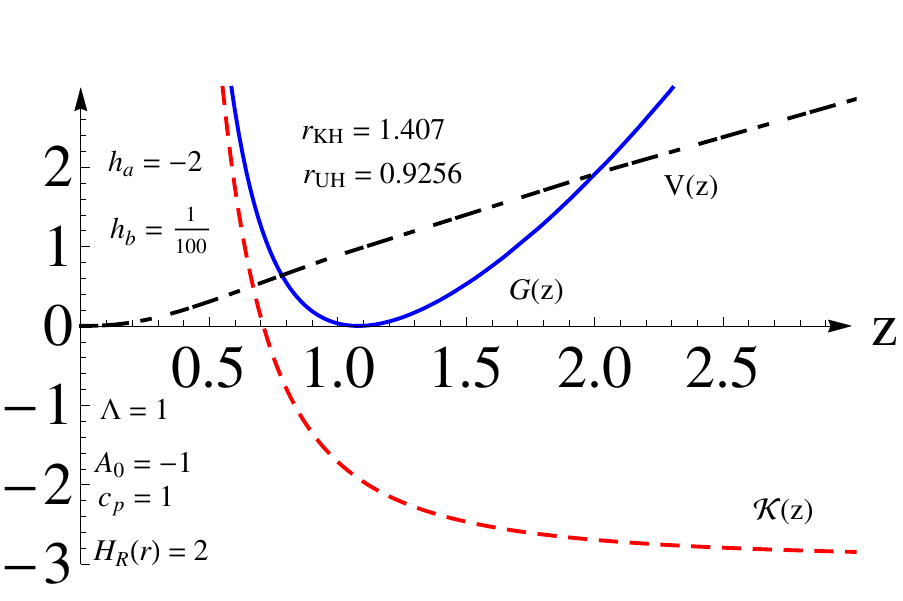}
\caption{The functions $V, G, {\cal{K}}$ and the locations of the Killing, event and universal horizons in the case $\lambda=1$ of the rotating spacetimes given by Eq.(\ref{Eq5A5b})  with $\Lambda\not=0$ and $\beta=0$. } 
\label{Fig18}
\end{figure*}

\begin{figure*}[h]
\includegraphics[width=5cm]{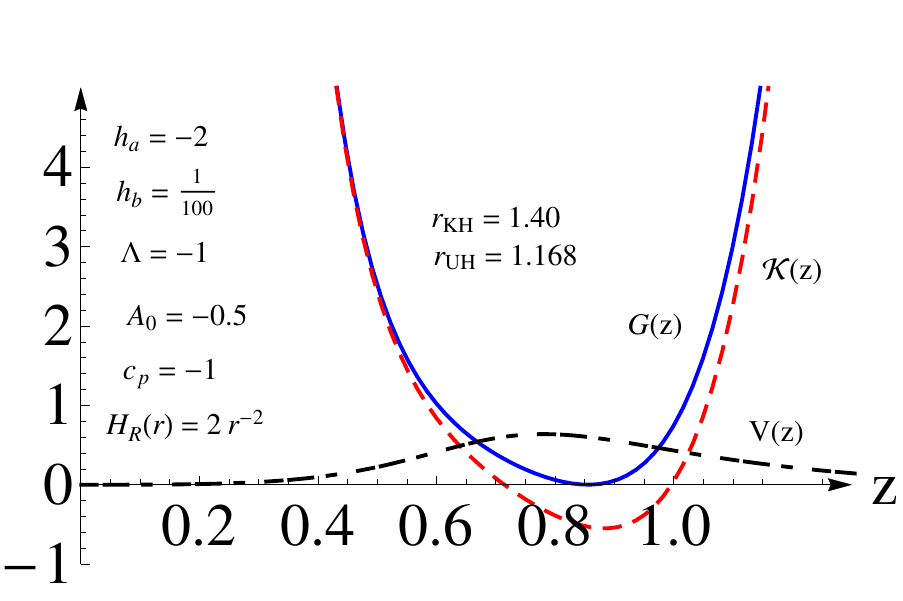}\includegraphics[width=5cm]{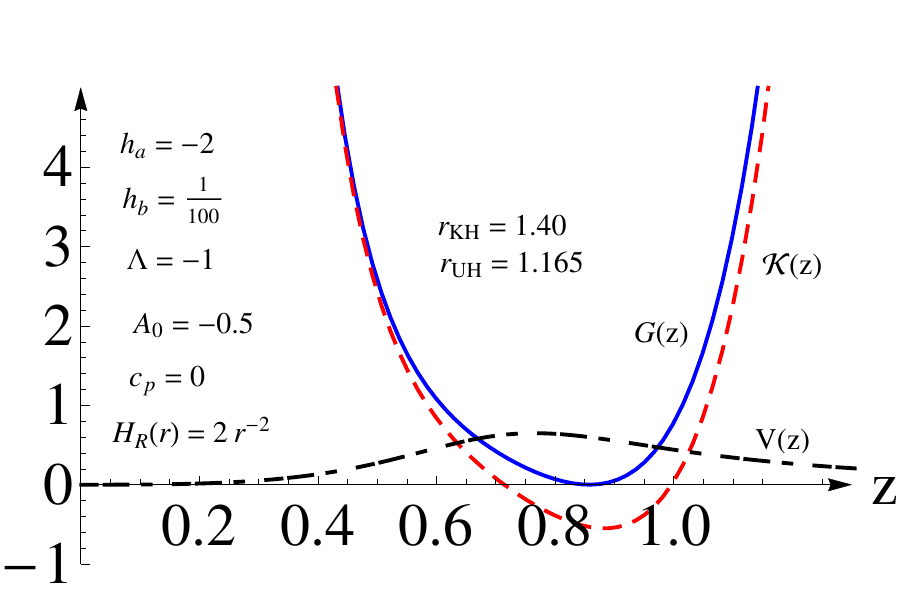}\includegraphics[width=5cm]{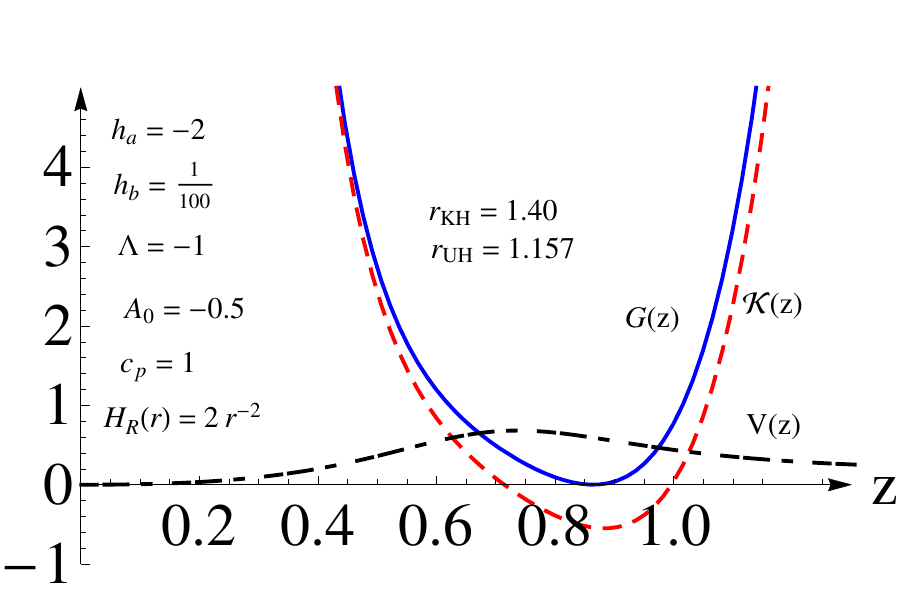}
\caption{The functions $V$, $G$, $\cal K$ and the locations of the Killing, event and universal horizons in the case $\lambda=1$ of the rotating spacetimes given by Eq.(4.51) with $\Lambda\not=0$ and $\beta>0$.}
\label{Fig19}
\end{figure*}

Choosing $H_R=\frac{h_0}{r^\beta}$ where
$h_0$ is a constant, we have
\bq
\lb{Eq5A5cA}
{\cal{N}}=A_0+\frac{h_0^2}{2r^{2\beta}}-\frac{\Lambda
 r^2}{2},
 \eq
for which we find,
  \bqn
  \lb{Eq5A6}
  R&=&\frac{16h_0^2(1-\beta)r^{2\beta-2}}{D^3}\left[2A_0r^{2\beta}+h_0^2(1+\beta)\right]\nb\\
  &&+\frac{8}{R_t^2}\left[(h_a-1)h_ar^{2\beta}+h_0^2(2\beta^2+\beta-2)r^2\right]r^{2\beta-4}\nb\\
  &&+\frac{8}{r^2R_t}\left[2A_0r^{2\beta}+h_0^2(1-\beta^2)\right]-\frac{8}{r^2},\nb\\
  K&=&\pm\frac{2h_0^2(1-\beta)}{\sqrt{h_0^2r^{2-2\beta}-h_a}D},
  \eqn
but now with $D(r) \equiv h_0^2+r^{2\beta}(\Lambda r^2-2A_0)$. Again, to avoid spacetime singularities occurring at a finite and non-zero radius, we shall choose the free parameters so that $D(r) \not= 0$ for $r\in (0, \infty]$.  In Figs. \ref{Fig14} - \ref{Fig19}, we show the functions $V, G, {\cal{K}}$ and the locations of the Killing and universal horizons  for various choices of the free parameters, as indicated in each of the panels of the figures. From these figures one can see that universal horizons always exist.

\subsubsection{$R_{ij} = 0, \; \lambda\not=1$}

In this case, the solutions are given by Eq.(\ref{Eq5A8}), and the rescaling,
 \bqn
 \lb{Eq5A8a}
 r\rightarrow C_1^{1/2}r,~~~\theta\rightarrow
 C_1^{-1/2}\theta,~~~h_a\rightarrow C_1^{1/2}h_a,\nb\\
 h_b\rightarrow
 C_1^{-1/2}h_b,~~~H_A\rightarrow C_1H_A,
 \eqn
leads the metric to the form,  
 \bqn
 \lb{Eq5A8b}
 ds^2&=&-\left(A_0+\frac{H_A^2+h_a^2}{2r^2}-\frac{\Lambda_C}{2}r^2\right)^2dt^2\nb\\
 &&+\left[dr+\left(\frac{H_A}{r}+H_Br\right)dt\right]^2\nb\\
 &&+r^2\left[d\theta+\left(\frac{h_a}{r^2}+h_b\right)dt\right]^2.
 \eqn
The corresponding scalar and extrinsic curvatures   are given
by
 \bqn
 \lb{Eq5A9}
 R&=&-\frac{8}{r^2}+64R_C^3H_Br^2(H_A+H_Br^2)(h_a^2+H_A^2+A_0r^2)\nb\\
 &&+8R_C^2(h_a^2+H_A^2-4H_AH_Br^2-H_B^2r^4)+16A_0R_C,\nb\\
 K&=&4H_BR_Cr^2,
 \eqn
where $R_C\equiv (h_a^2+H_A^2+2A_0r^2-\Lambda_0 r^4)^{-1}$ and
$\Lambda_0\equiv \Lambda+H_B^2(1-2\lambda)$.

\begin{figure*}[tbp]
\includegraphics[width=5cm]{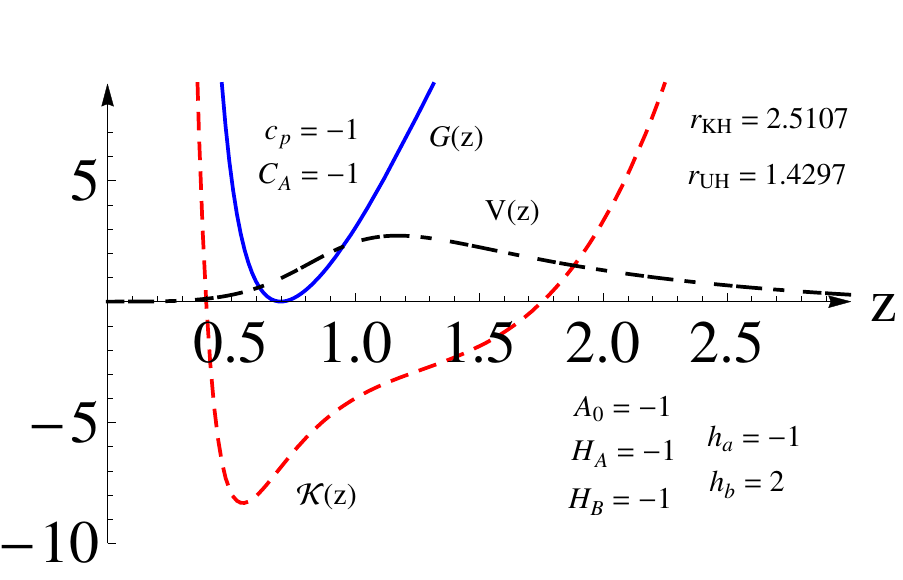}\includegraphics[width=5cm]{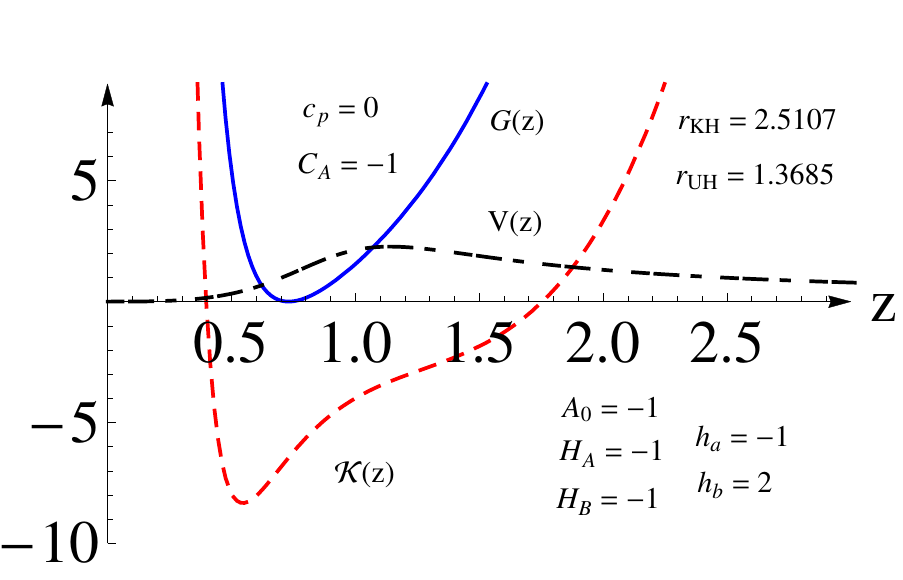}\includegraphics[width=5cm]{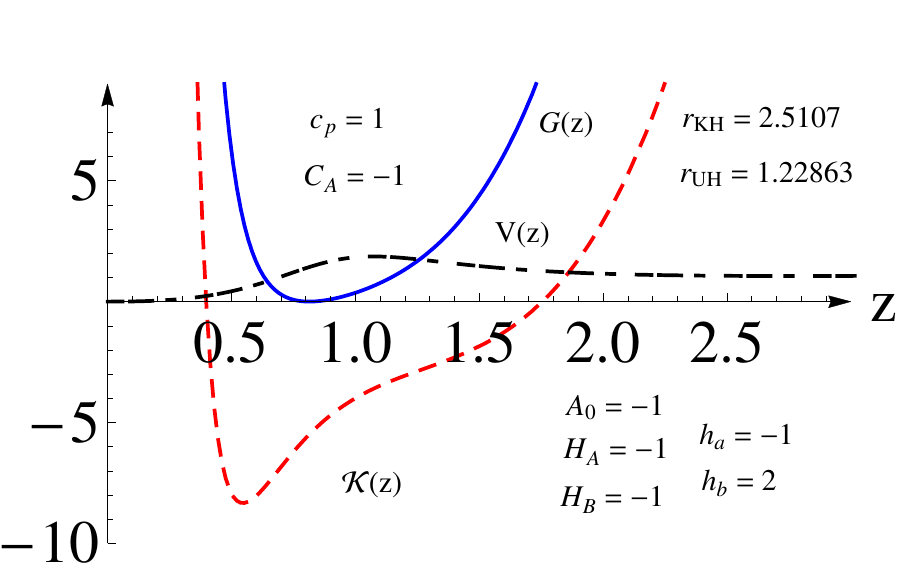}
\caption{The functions $V, G, {\cal{K}}$ and the locations of the Killing, event and universal horizons in the case $\lambda\not=1$ of the rotating spacetimes given by Eq.(\ref{Eq5A11})  for $\Lambda_C\not=0$.} 
\label{Fig20}
\includegraphics[width=5cm]{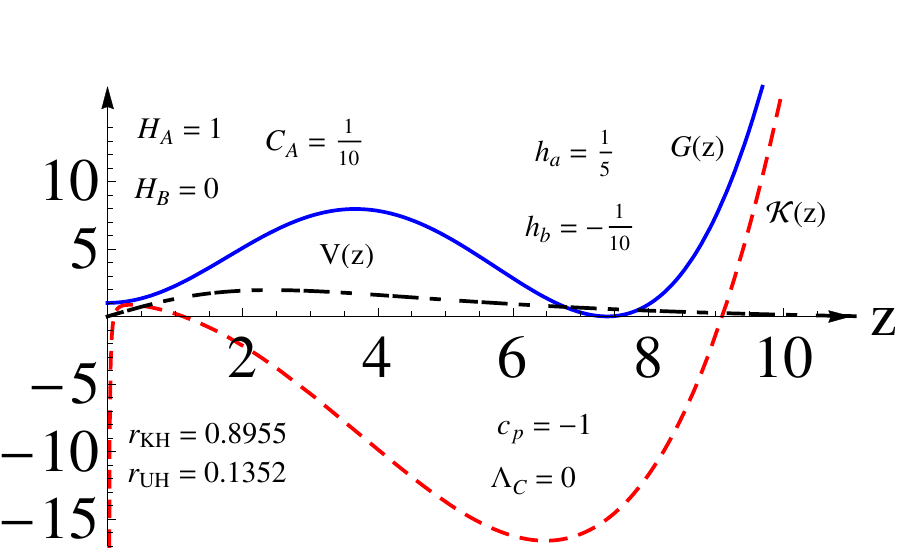}\includegraphics[width=5cm]{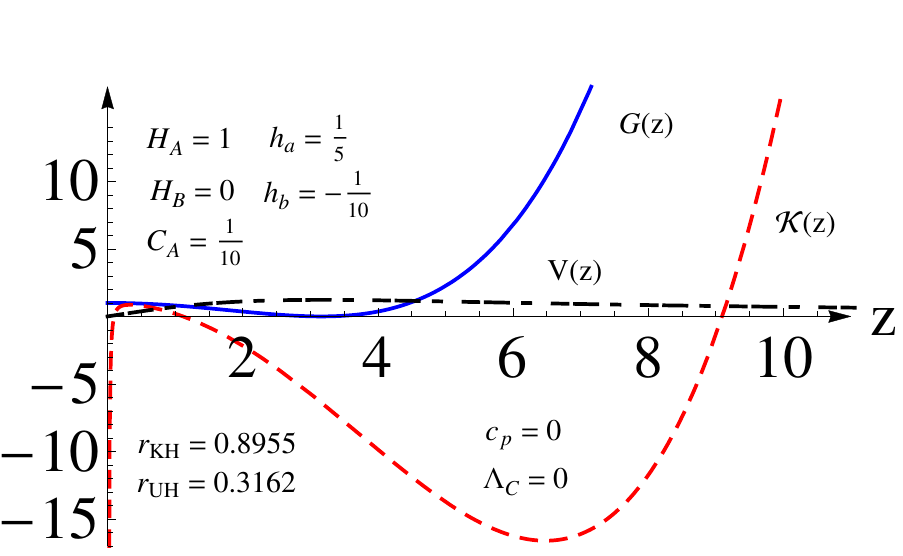}\includegraphics[width=5cm]{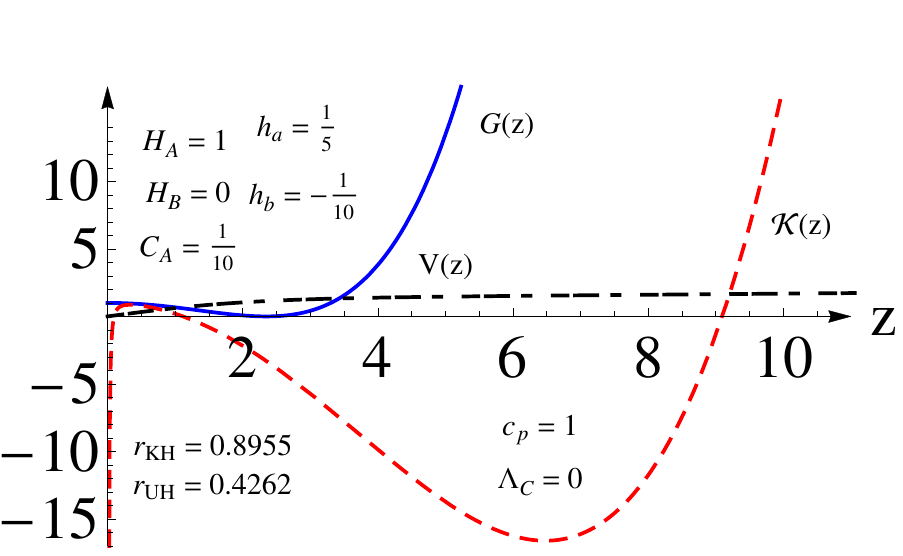}
\caption{The functions $V, G, {\cal{K}}$ and the locations of the Killing, event and universal horizons in the case $\lambda=1$ of the rotating spacetimes given by Eq.(\ref{Eq5A14})  with  $\Lambda_C=0$ but $A_0\not=0$.} 
\label{Fig21}
\end{figure*}

When $\Lambda_C\not=0$, we can simplify the above metric further by,  
 \bqn
 \lb{Eq5A10}
 t\rightarrow -\frac{2}{\Lambda_C}t,~~~A_0\rightarrow -\frac{2}{\Lambda_C}A_0,~~~H_A\rightarrow -\frac{2}{\Lambda_C}H_A,\nb\\
 H_B\rightarrow -\frac{2}{\Lambda_C}H_B,~h_a\rightarrow
 -\frac{2}{\Lambda_C}h_a,~h_b\rightarrow -\frac{2}{\Lambda_C}h_b,  ~~~~~~~
 \eqn
which leads Eq.(\ref{Eq5A8b}) to,  
 \bqn
 \lb{Eq5A11}
 ds^2&=&-\left(1+\frac{C_A}{r^2}+r^2\right)^2dt^2\nb\\
 &&+\left[dr+\left(\frac{H_A}{r}+H_Br\right)dt\right]^2\nb\\
 &&+r^2\left[d\theta+\left(\frac{h_a}{r^2}+h_b\right)dt\right]^2,
 \eqn
where $C_A=-\frac{\Lambda_C}{4}\left(H_A^2+h_a^2\right)$. In Fig.~\ref{Fig20} we show the locations of the Killing, event and universal horizons in this case.

When $\Lambda_C=0$, we assume $H_{B} \not= 0$. Then, we consider the cases
$A_0\not=0$ and $A_0=0$, separately.  In particular, when $A_0\not=0$, the rescaling
 \bqn
 \lb{Eq5A13}
 t\rightarrow A_0^{-1}t,~~~H_A\rightarrow A_0H_A,\nb\\
 h_b\rightarrow A_0h_b,~~~h_a\rightarrow A_0h_a,
 \eqn
leads to,  
 \bqn
 \lb{Eq5A14}
 ds^2&=&-\left(1+\frac{C_A}{r^2}\right)^2dt^2+\left(dr+\frac{H_A}{r}dt\right)^2\nb\\
 &&+r^2\left[d\theta+\left(\frac{h_a}{r^2}+h_b\right)dt\right]^2,
 \eqn
where $C_A=-\frac{A_0}{2}\left(H_A^2+h_a^2\right)$. Fig.~\ref{Fig21} shows the locations of the Killing and universal horizons.

When $A_0=0$, the rescaling,  
 \bqn
 \lb{Eq5A15}
 t\rightarrow C_H t,~~~H_A\rightarrow C_H^{-1}H_A,\nb\\
 h_b\rightarrow C_H^{-1}h_b,~~~h_a\rightarrow C_H^{-1}h_a,
 \eqn
leads to,  
 \bqn
 \lb{Eq5A16}
 ds^2&=&-\frac{dt^2}{r^4}+\left[dr+\frac{H_A}{r}dt\right]^2\nb\\
 &&+r^2\left[d\theta+\left(\frac{h_a}{r^2}+h_b\right)dt\right]^2,
 \eqn
 where $C_H=\frac{2}{H_A^2+h_a^2}$. In this case,
the equation of $V$ takes the simple form,  
 \bqn
 \lb{Eq5A17}
 V''-\frac{V'}{r}+(1-2c_p)\frac{V}{r^2}=0,
 \eqn
and has the solution,
 \bqn
 \lb{Eq5A18}
 V=r_Ar^{1+\sqrt{2c_p}}+r_Br^{1-\sqrt{2c_p}}.
 \eqn
Consider the boundary condition $V(\infty)\rightarrow 0$, we have
$r_A=0$, so $G$ is given by
 \bqn
 \lb{Eq5A19}
 G=\frac{1+r_B^2r^{2-2\sqrt{2c_p}}}{r^4}-\frac{H_A^2}{r^2}.
 \eqn
Clearly,  $G<0$ as $r\rightarrow \infty$, which is not allowed by the existence of the khronon field in the whole spacetime.
Therefore, in this case the solution must be discarded.

\subsubsection{$R_{ij}\not=0$}

In this case, the solutions are given by Eqs.(\ref{E1}) and (\ref{E2}), and the  corresponding spacetimes with  $a_i=(1,0)$ describes a spacetime, but the metric at the  killing horizon located at  $f(r_{KH})=0$ becomes singular. Therefore, to study the location of the universal horizon inside it, an extension of the solution into the internal of the Killing horizon is needed. Such extension is standard \cite{Sotiriou:2014gna}, so in the following we shall not consider it further.  

\section{Conclusions}
\renewcommand{\theequation}{5.\arabic{equation}} \setcounter{equation}{0}

In this paper, we have studied the existence of universal horizons and black holes in gravitational theories with broken Lorentz invariance. We have paid particular attention to the case where the gravitational field is so strong that  the infrared limit does not exist, and the higher-order derivative terms must be included, in order for the theory to be UV complete. We have shown that even in the UV regime both static and rotating 
universal horizons and black holes exist. Therefore, universal horizons and black holes are not only the low energy phenomena but also phenomena existing in the UV regime. To reach this conclusion, we have first constructed exact solutions of the full theory of Ho\v{r}ava gravity with the projectability and U(1) symmetry in (2+1)-dimensions, which is power-counting renormalizable \cite{HMT,WW,Silva,HW}.  To avoid coordinate  singularities across the Killing horizons, we have chosen to work with the PG-like coordinates \cite{PG}. Although this normally makes the field equations very complicated, we are still 
 able to find analytical solutions in both static and stationary spacetimes. Then, we have numerically solved the khronon field equations, and  identified  the locations of the  universal horizons. In all the cases considered, universal horizons exist, and are always located inside the Killing horizons.

With these exact solutions, we hope that the study of black hole thermodynamics at the universal horizons, specially the ones with rotations, can be made more accessible.  In the spherically symmetric and neutral case, the first law of black hole thermodynamics
at universal horizons holds \cite{Berglund:2012bu}, provided that the entropy is still proportional to the area of the horizon, and the surface gravity is defined by, 
\bq
\lb{eqa}
\kappa \equiv  \frac{1}{2} u^{\alpha} D_{\alpha} \left(u_{\lambda} \zeta^{\lambda}\right),
\eq
which is identical to the one obtained from  the peeling behavior of the khronon field  \cite{Cropp:2013sea},
 as shown explicitly in \cite{Lin:2014eaa}. In the neutral case, the temperature of the black hole takes its standard form, $T = \kappa/2\pi$ \cite{Berglund:2012fk,Cropp:2013sea}. However, when the black hole is charged, such a first law does not exist \cite{Ding:2015kba}, if we insist that the 
temperature of the black hole still takes its standard form with the surface gravity given as above, and that the entropy of the black hole is proportional to its area.  In addition, when high-order powers of momentum appear in the dispersion relation of the particles emitted through the Hawking radiation process, which generically is always the case,  the temperature of the black hole at the universal horizon depends on the order of the powers, although it is still proportional to the surface gravity $\kappa$ defined above \cite{JB13,Ding:2015fyx}.

We also hope that  these exact solutions  will help us to get deeper insights into  the problem of quantization of the theory  \cite{Carlip,Kiefer}, and the non-relativistic AdS/CFT correspondence \cite{Kachru:2008yh,LSWW,Janiszewski:2014ewa,Basu:2016vyz}. The studies of these important issues are out of scope of this paper, and we wish to come back to them in a different occasion.

Finally, we would like to note  that our main  conclusion regarding the existence of universal horizons and black holes at all energy scales (including the UV regime) should be easily generalized to other versions of Ho\v{r}ava gravity \cite{Horava:2011gd,Wang:2012nv}, although in this paper we have considered  it   only in the projectable Ho\v{r}ava theory with an extra U(1) symmetry \cite{HMT,WW,Silva,HW}. It is also quite reasonable to expect that this is also the generic case in other theories of gravity without Lorentz symmetry.

\begin{acknowledgments}

We would like to thank  Ahmad Borzou and Raziyeh Yousefi for the involvement in the early stage of this project. We would also like to thank M.F. da Silva for valuable discussions and comments. The work was done partly when K.L. was visiting Baylor University (BU), and A.W. was visiting the State University of Rio de Janeiro (UERJ). They would like to express their gratitude to BU and UERJ  for hospitality.  A.W. and V.H.S. are supported  in part by Ci\^{e}ncia Sem Fronteiras, No. A045/2013 CAPES, Brazil. A.W. is also supported  in part by   NNSFC No. 11375153, China.  K.L. is supported  in part by  CAPES,   FAPESP No. 2012/08934-0, Brazil,  and  NNSFC No.11573022  and No.11375279, China.

\end{acknowledgments}

\section*{Appendix A. Projectable Ho\v{r}ava theory with U(1) symmetry in (d+1) dimensions }
\renewcommand{\theequation}{A.\arabic{equation}} \setcounter{equation}{0}

In this Appendix, we give a brief introduction to the projectable Ho\v{r}ava theory with U(1) symmetry. for detail, we refer readers to
\cite{HMT,WW,Silva,HW}.

\subsection{The Gauge Symmetries}

The  Ho\v{r}ava  theory  is based on the perspective that Lorentz symmetry should appear as an emergent symmetry at long
distances, but can be fundamentally  absent at short ones \cite{Pav}. In the latter regime,  the system
 exhibits a strong anisotropic scaling between space and time,
\bq
\lb{1.1}
{\bf x} \rightarrow \ell {\bf x}, \;\;\;  t \rightarrow \ell^{z} t,
\eq
where $z \ge 3$ in the $(3+1)$-dimensional spacetime \cite{Horava,Visser}. At long distances, higher-order curvature corrections become
negligible, and  the lowest order terms   $R$ and $\Lambda$ take over,  whereby the Lorentz invariance is expected to be
 ``accidentally restored," where $R$ denotes the 3-dimensional Ricci scalar, and $\Lambda$ the cosmological constant.
Because of the anisotropic scaling,   the gauge symmetry of the theory is broken down to the  foliation-preserving
diffeomorphism, Diff($M, \; {\cal{F}}$),
\bq
\lb{1.4}
\tilde{t} = t - f(t),\; \;\; \tilde{x}^{i}  =  {x}^{i}  - \zeta^{i}(t, {\bf x}),
\eq
for which the lapse function $N$, shift vector  $N^{i}$, and 3-spatial metric $g_{ij}$   transform as
\bqn
\lb{1.5}
\delta{N} &=& \zeta^{k}\nabla_{k}N + \dot{N}f + N\dot{f},\nb\\
\delta{N}_{i} &=& N_{k}\nabla_{i}\zeta^{k} + \zeta^{k}\nabla_{k}N_{i}  + g_{ik}\dot{\zeta}^{k}
+ \dot{N}_{i}f + N_{i}\dot{f}, \nb\\
\delta{g}_{ij} &=& \nabla_{i}\zeta_{j} + \nabla_{j}\zeta_{i} + f\dot{g}_{ij},
\eqn
where $\dot{f} \equiv df/dt,\;  \nabla_{i}$ denotes the covariant
derivative with respect to   $g_{ij}$,  $N_{i} = g_{ik}N^{k}$, and $\delta{g}_{ij}
\equiv {\gamma}_{ij}\left(t, x^k\right) - {g}_{ij}\left(t, x^k\right)$,
 etc. From these expressions one can see that   $N$ and   $N^{i}$ play the role of gauge fields of the Diff($M, \; {\cal{F}}$).
 Therefore, it is natural to assume that $N$ and $N^{i}$ inherit the same dependence on
space and time as the corresponding generators \cite{Horava},
\bq
\lb{1.6}
N = N(t), \;\;\; N^{i} = N^{i}(t, x),
\eq
which is  often referred to as the projectability condition.

Due to the  Diff($M, \; {\cal{F}}$) diffeomorphisms (\ref{1.4}), one more degree of freedom appears
in the gravitational sector - a spin-0 graviton. This is potentially dangerous, and needs to decouple
in the IR regime, in order to be consistent with observations.    A very promising approach along this direction  is to eliminate the spin-0 graviton by introducing two auxiliary fields,
the $U(1)$ gauge field $A$ and the Newtonian prepotential $\varphi$, by   extending
the  Diff($M, \; {\cal{F}}$) symmetry (\ref{1.4}) to include  a local $U(1)$ symmetry \cite{HMT},
\bq
\lb{symmetry}
 U(1) \ltimes {\mbox{Diff}}(M, \; {\cal{F}}).
 \eq
Under this extended symmetry,   the special status of time  maintains,  so that the anisotropic scaling (\ref{1.1})
can  still be  realized,  and the theory is UV complete. Meanwhile, because of the elimination of the spin-0 graviton,  its IR  behavior can be
 significantly improved.  Under the Diff($M, \; {\cal{F}}$), $A$ and $\varphi$ transform as,
\bqn
\lb{2.2}
\delta{A} &=& \zeta^{i}\partial_{i}A + \dot{f}A  + f\dot{A},\nb\\
\delta \varphi &=&  f \dot{\varphi} + \zeta^{i}\partial_{i}\varphi.
\eqn
Under the local $U(1)$ symmetry,  the fields
 transform as
\bqn
\lb{2.3}
\delta_{\alpha}A &=&\dot{\alpha} - N^{i}\nabla_{i}\alpha,\;\;\;
\delta_{\alpha}\varphi = - \alpha,\nb\\
\delta_{\alpha}N_{i} &=& N\nabla_{i}\alpha,\;\;\;
\delta_{\alpha}g_{ij} = 0,\;\;\; \delta_{\alpha}{N} = 0,
\eqn
where $\alpha$ is   the generator  of the local $U(1)$ gauge symmetry. For the detail, we refer readers to \cite{HMT,WW}.

The elimination of the spin-0 graviton was done  initially in the   case $\lambda = 1$ \cite{HMT,WW}, but soon generalized
to the case with any $\lambda$ \cite{Silva,HW,LWWZ}, where $\lambda$ denotes a coupling constant that
characterizes the deviation of  the kinetic part of action from the corresponding one given in GR with $\lambda_{GR} =1$.
For the analysis of Hamiltonian consistency, see \cite{HMT,Kluson}.

\subsection{Universal Coupling with Matter and Field Equations}

The basic variables in the HMT setup are
\bq
\left(A, \; \varphi, \; N, \; N^{i},\; g_{ij}\right),\; (i, j = 1, 2, ..., d),
\eq
and the total action of the theory in ($d+1$)-diemnsions can be written in the form,
\bqn
\lb{2.4}
S &=& \zeta^2\int dt d^{d}x N \sqrt{g} \Big({\cal{L}}_{K} -
{\cal{L}}_{{V}} +  {\cal{L}}_{{\varphi}} +  {\cal{L}}_{{A}} +{\zeta^{-2}} {\cal{L}}_{M} \Big),\nb\\
\eqn
where $g={\rm det}(g_{ij})$, $\zeta^2 = {1}/{(16\pi G)}$, and
\bqn
\lb{2.5}
{\cal{L}}_{K} &=& K_{ij}K^{ij} -   \lambda K^{2},\nb\\
{\cal{L}}_{\varphi} &=&\varphi {\cal{G}}^{ij} \Big(2K_{ij} + \nabla_{i}\nabla_{j}\varphi\Big)\nb\\
&& + \big(1-\lambda\big)\Big[\big(\Delta\varphi\big)^{2} + 2 K \Delta\varphi\Big],\nb\\
{\cal{L}}_{A} &=&\frac{A}{N}\Big(2\Lambda_{g} - R\Big).
\eqn
Here $\Delta \equiv g^{ij}\nabla_{i}\nabla_{j}$, $\Lambda_{g}$ is a    coupling constant with dimension of $({\mbox{length}})^{-2}$,  the
Ricci and Riemann tensors $R_{ij}$ and $R^{i}_{jkl}$  all refer to the d-dimensional metric $g_{ij}$, and
\bqn \lb{2.6}
K_{ij} &=& \frac{1}{2N}\left(- \dot{g}_{ij} + \nabla_{i}N_{j} +
\nabla_{j}N_{i}\right),\nb\\
{\cal{G}}_{ij} &=& R_{ij} - \frac{1}{2}g_{ij}R + \Lambda_{g} g_{ij}.
\eqn
${\cal{L}}_{{V}}$ is an arbitrary Diff($\Sigma$)-invariant local scalar functional
built out of the spatial metric, its Riemann tensor and spatial covariant derivatives, without the use of time derivatives.

${\cal{L}}_{{M}}$ is the Lagrangian of matter fields, which is a scalar not only with respect to the ${\mbox{Diff}}(M,{\cal{F}})$
symmetry (\ref{1.4}), but also to  the $U(1)$ symmetry (\ref{2.3}). When the gravity is  universally coupled with matter, it is given by \cite{LMWZ}
\bqn
\lb{ac tionM}
S_{M} &=& \int dt d^{d}x N \sqrt{g}{\cal{L}}_{M}\left(A, \; \varphi, \; N, \; N^{i},\; g_{ij}; \psi_{n}\right)\nb\\
&=& \int d^{d+1}x \sqrt{|\gamma|}\; \tilde{{\cal{L}}}_{M}\left( \gamma_{\mu\nu}; \psi_{n}\right),
\eqn
where $\gamma \equiv det(\gamma_{\mu\nu})\; (\mu, \nu = 0, 1, ..., d)$, and $\psi_n$ collectively  stands for matter fields,  minimally coupled to the
(d+1)-dimensional metric $\gamma_{\mu\nu}$, defined as
\bqn
\lb{Pmetric}
\left(\gamma_{\mu\nu}\right) &\equiv& \left(\matrix{-{\cal{N}}^2 +{\cal{N}}^{i}{\cal{N}}_{i} &{\cal{N}}_{i}\cr
{\cal{N}}_{i} & {\gamma}_{ij}\cr}\right),\nb\\
\left(\gamma^{\mu\nu}\right) &=&
\left(\matrix{-\frac{1}{{\cal{N}}^2}   &
\frac{{\cal{N}}^{i}}{{\cal{N}}^2}\cr
 \frac{{\cal{N}}^{i}}{{\cal{N}}^2} & {\gamma}^{ij} - \frac{{\cal{N}}^{i}{\cal{N}}^{j}}{{\cal{N}}^2}\cr}\right),
\eqn
where ${\gamma}^{ij}{\gamma}_{ik} = \delta^{j}_{k},\;{\cal{N}}_{i} \equiv {\gamma}_{ij}{\cal{N}}^{j}$, and
\bqn
\lb{eq8-1}
& &{\cal{N}} = \left(1 - a_1\sigma\right)N,\;\;\;
{\cal{N}}^i = N^i + Ng^{ij} \nabla_j\varphi, \nb\\
&&  {\gamma}_{ij} = \left(1 - a_2\sigma\right)^2g_{ij},\;\;\;
\sigma  = \frac{A - {\cal{A}}}{N}, \nb\\
&&  {\cal{A}} \equiv - \dot{\varphi}  + N^i\nabla_i\varphi
+\frac{1}{2}N\left(\nabla^i\varphi\right)\left(\nabla_i\varphi\right).
\eqn
Here $a_1$ and $a_2$ are two arbitrary constants.
It is should be noted that the line element
\bqn
\lb{eq8-2b{gauge}}
ds^{2} &=& \gamma_{\mu\nu}dx^{\mu}dx^{\nu} =  -{\cal{N}}^2dt^2 \nb\\
&& + {\gamma}_{ij}\left(dx^i +{\cal{N}}^i dt\right) \left(dx^j +{\cal{N}}^j dt\right), ~~~~
\eqn
is invariant not only under the gauge transformations (\ref{1.4}), but also under the U(1) transformations (\ref{2.3}).
In terms of $ \tilde{{\cal{L}}}_{M}$, the (d+1)-dimensional energy-momentum tensor $T_{\mu\nu}$ is given by
\bq
\lb{4.7}
T^{\mu\nu} \equiv \frac{1}{\sqrt{|\gamma|}} \frac{\delta\left(\sqrt{|\gamma|}{\tilde{\cal{L}}}_{M}(\gamma_{\alpha\beta}; \psi_{n})\right)}{\delta \gamma_{\mu\nu}}.
\eq

The  variations of the total action with respect to $N$ and $N^i$ yield the Hamiltonian and momentum constraints,  given, respectively, by,
 \bqn
 \lb{eq1}
& & \int{ d^{d}x\sqrt{g}\left[{\cal{L}}_{K} + {\cal{L}}_{{V}} - \varphi {\cal{G}}^{ij}\nabla_{i}\nabla_{j}\varphi
- \big(1-\lambda\big)\big(\Delta\varphi\big)^{2}\right]}\nb\\
& & ~~~~~~~~~~~~~~~~~~~~~~~~~~~~~
= 8\pi G \int d^{d}x {\sqrt{g}\, J^{t}},\\
\lb{eq2}
& & \nabla^{j}\Big[\pi_{ij} - \varphi  {\cal{G}}_{ij} - \big(1-\lambda\big)g_{ij}\nabla^{2}\varphi \Big] = 8\pi G J_{i},
 \eqn
where
 \bqn
  \lb{eq2b}
  J^{t} &\equiv& 2 \frac{\delta\left(N{\cal{L}}_{M}\right)}{\delta N},\;\;\; J_{i} \equiv - N\frac{\delta{\cal{L}}_{M}}{\delta N^{i}}, \nb\\
   \pi^{ij} &\equiv&   \frac{\delta(N{\cal{L}}_{K})}{\delta\dot{g}_{ij}} =
   - K^{ij} +  \lambda K g^{ij}.
 \eqn
 Variation of the action (\ref{2.4}) with respect to   $\varphi$ and $A$ yield,
\bqn
\lb{eq4a}
& & {\cal{G}}^{ij} \Big(K_{ij} + \nabla_{i}\nabla_{j}\varphi\Big) + \big(1-\lambda\big)\Delta \Big(K + \Delta \varphi\Big) \nb\\
& & ~~~~~~~~~~~~~~~~~~~~~~~~~~~~~~~~ = 8\pi G J_{\varphi}, \\
\lb{eq4b}
& & R - 2\Lambda_{g} =   8\pi G J_{A},
\eqn
which will be referred, respectively, to as the $\varphi$- and $A$- constraint, where
\bq
\lb{eq5}
J_{\varphi} \equiv - \frac{\delta{\cal{L}}_{M}}{\delta\varphi},\;\;\;
J_{A} \equiv 2 \frac{\delta\left(N{\cal{L}}_{M}\right)}{\delta{A}}.
\eq
On the other hand,  the dynamical equations now read\footnote{Note that the dynamical equations given here differ from
those given in \cite{HW} because here we took $N^i$ as the fundamental variable instead of $N_i $ as what we did in \cite{HW}.
The subtle is that $N_i$ now are  functions of $g_{ij}$ via the relations $N_i = g_{ij}N^j$, once $N^i$ are considered as the fundamental variables, or vice versa.
Of course, they are equivalent,  if  one consistently uses either $N^i$ or $N_i$ to carry out the derivation of all the field equations.},
\bqn \lb{eq3}
&&
\frac{1}{N\sqrt{g}}\Bigg\{\sqrt{g}\Big[\pi^{ij} - \varphi {\cal{G}}^{ij} - \big(1-\lambda\big) g^{ij} \Delta \varphi\Big]\Bigg\}_{,t}
\nb\\
& &~~~ = -2\left(K^{2}\right)^{ij}+2\lambda K K^{ij} - \frac{2}{N}\pi^{k(i}\nabla_k N^{j)}\nb\\
& &  ~~~~~ + \nabla_{k}\Big[\frac{N^k}{N} \pi^{ij}-(1-\lambda) F^k_{\varphi} g^{ij}\Big]\nb\\
& &  ~~~~~ - 2\big(1-\lambda\big) \Big[\big(K + \Delta \varphi\big)\nabla^{i}\nabla^{j}\varphi + K^{ij}\Delta \varphi\Big]\nb\\
& &  ~~~~~ + 2 (1-\lambda)\nabla^{(i}\left[\nabla^{j)} \varphi \left( K + \Delta\varphi\right)\right] \nb\\
&& ~~~~~ +  \frac{1-\lambda}{N}\Delta\varphi \nabla^{(i}N^{j)} \nb\\
& & ~~~~~ +  \frac{1}{2} \Big({\cal{L}}_{K} + {\cal{L}}_{\varphi} + {\cal{L}}_{A}\Big) g^{ij} \nb\\
& &  ~~~~~    + F^{ij} + F_{\varphi}^{ij} +  F_{A}^{ij} + 8\pi G {\cal{N}}^{ij},
 \eqn
where $\left(K^{2}\right)^{ij} \equiv K^{il}K_{l}^{j},\; f_{(ij)}
\equiv \left(f_{ij} + f_{ji}\right)/2$, and
 \bqn
\lb{eq3a}
F^{ij} &\equiv& -
\frac{1}{\sqrt{g}}\frac{\delta\left(\sqrt{g}
{\cal{L}}_{V}\right)}{\delta{g}_{ij}},\nb\\
F_{\varphi}^{ij} &=&  \sum^{3}_{n=1}{F_{(\varphi, n)}^{ij}},\nb\\
F_{\varphi}^{i} &=&  \Big(K + \nabla^{2}\varphi\Big)\nabla^{i}\varphi + \frac{N^{i}}{N} \Delta \varphi, \nb\\
F_{A}^{ij} &=& \frac{1}{N}\left[AR^{ij} - \Big(\nabla^{i}\nabla^{j} - g^{ij}\Delta \Big)A\right],\nb\\
 \eqn
with
$n_{{\scriptscriptstyle S}} =(2, 0, -2, -2)$, and $F_{(\varphi, n)}^{ij}$ are given by
 \cite{WW},
\bqn
  \lb{eq3c}
F_{(\varphi, 1)}^{ij} &=& \frac{1}{2}\varphi\left\{\Big(2K + \nabla^{2}\varphi\Big) R^{ij}
- 2 \Big(2K^{j}_{k} + \nabla^{j} \nabla_{k}\varphi\Big) R^{ik} \right.\nb\\
& & ~~~~~ - 2 \Big(2K^{i}_{k} + \nabla^{i} \nabla_{k}\varphi\Big) R^{jk}\nb\\
& &~~~~~\left.
- \Big(2\Lambda_{g} - R\Big) \Big(2K^{ij} + \nabla^{i} \nabla^{j}\varphi\Big)\right\},\nb\\
F_{(\varphi, 2)}^{ij} &=& \frac{1}{2}\nabla_{k}\left\{\varphi{\cal{G}}^{ik}
\Big(\frac{2N^{j}}{N} + \nabla^{j}\varphi\Big) \right. \nb\\
& & \left.
+ \varphi{\cal{G}}^{jk}  \Big(\frac{2N^{i}}{N} + \nabla^{i}\varphi\Big)
-  \varphi{\cal{G}}^{ij}  \Big(\frac{2N^{k}}{N} + \nabla^{k}\varphi\Big)\right\}, \nb\\
F_{(\varphi, 3)}^{ij} &=& \frac{1}{2}\left\{2\nabla_{k} \nabla^{(i}f^{j) k}_{\varphi}
- \nabla^{2}f_{\varphi}^{ij}   - \left(\nabla_{k}\nabla_{l}f^{kl}_{\varphi}\right)g^{ij}\right\},\nb\\
\eqn
where
\bqn
\lb{eq3d}
f_{\varphi}^{ij} &=& \varphi\left\{\Big(2K^{ij} + \nabla^{i}\nabla^{j}\varphi\Big) 
- \frac{1}{2} \Big(2K + \nabla^{2}\varphi\Big)g^{ij}\right\}.\nb\\
\eqn

  The tensor ${\cal{N}}^{ij}$ is defined as
 \bq
 \label{tau}
{\cal{N}}^{ij} = {2\over \sqrt{g}}{\delta \left(\sqrt{g}
{\cal{L}}_{M}\right)\over \delta{g}_{ij}}.
 \eq

The  conservation laws of energy and momentum of  matter fields read, respectively,
 \bqn \lb{eq5a} & &
 \int d^{d}x \sqrt{g} { \left[ \dot{g}_{kl}{\cal{N}}^{kl} -
 \frac{1}{\sqrt{g}}\left(\sqrt{g}J^{t}\right)_{, t}
 +   \frac{2N_{k}}  {N\sqrt{g}}\left(\sqrt{g}J^{k}\right)_{,t}
  \right.  }   \nb\\
 & &  ~~~~~~~~~~~~~~ \left.   - 2\dot{\varphi}J_{\varphi} -  \frac{A} {N\sqrt{g}}\left(\sqrt{g}J_{A}\right)_{,t}
 \right] = 0,\\
\lb{eq5b} & & \nabla^{k}{\cal{N}}_{ik} -
\frac{1}{N\sqrt{g}}\left(\sqrt{g}J_{i}\right)_{,t}  - \frac{J^{k}}{N}\left(\nabla_{k}N_{i}
- \nabla_{i}N_{k}\right)   \nb\\
& & \;\;\;\;\;\;\;\;\;\;\;- \frac{N_{i}}{N}\nabla_{k}J^{k} + J_{\varphi} \nabla_{i}\varphi - \frac{J_{A}}{2N} \nabla_{i}A
 = 0.
 \eqn

Introducing the normal vector $n_{\nu}$ to the hypersurface $t=$ constant,
\bq
\lb{4.8}
n_{\mu} = -{\cal{N}}\delta^{t}_{\mu}, \;\;\; n^{\mu} =
\frac{1}{{\cal{N}}} \left(1,  -{\cal{N}}^{i} \right),
 \eq
one can decompose $T_{\mu\nu}$ as \cite{Anninos:2001yc},
\bqn
\lb{4.9}
\rho_H  &\equiv& T_{\mu\nu} n^{\mu} n^{\nu},\nb\\
s_{i}  &\equiv&  -  T_{\mu\nu} \left(\delta^{\mu}_{i} + n^{\mu}n_{i}\right)   n^{\nu},\nb\\
s_{ij}  &\equiv&  T_{\mu\nu}  \left(\delta^{\mu}_{i} +
n^{\mu}n_{i}\right)   \left(\delta^{\nu}_{j} +
n^{\nu}n_{j}\right),
 \eqn
  in terms of which, the quantities
$J^t, \; J_i, \; J_A, \; J_{\varphi}$ and ${\cal{N}}_{ij}$ are given by  \cite{LMWZ},
 \bqn
\lb{B.1}
J^{t} &=& 2\Omega^{3}(\sigma)\Bigg\{- \rho_H  \frac{\delta{\cal{N}}}{\delta{N}}
+  \frac{\delta{\cal{N}}_i}{\delta{N}}s^i   + \frac{1}{2}{\cal{N}}  \frac{\delta{\gamma}_{ij}}{\delta{N}}s^{ij}\Bigg\},\nb\\
J^{i} &=& - \Omega^{3}(\sigma)\Bigg\{- \rho_H  \frac{\delta{\cal{N}}}{\delta{N}_i}
+  \frac{\delta{\cal{N}}_k}{\delta{N}_i}s^k   + \frac{1}{2}{\cal{N}}  \frac{\delta{\gamma}_{kl}}{\delta{N}_i}s^{kl}\Bigg\},\nb\\
{\cal{N}}^{ij} &=& \frac{2 \Omega^{3}(\sigma)}{N}\Bigg\{- \rho_H
\frac{\delta{\cal{N}}}{\delta{g}_{ij}}
+  \frac{\delta{\cal{N}}_k}{\delta{g}_{ij}}s^k   + \frac{1}{2}{\cal{N}}  \frac{\delta{\gamma}_{kl}}{\delta{g}_{ij}}s^{kl}\Bigg\},\nb\\
J_{A} &=& 2 \Omega^{3}(\sigma)\Bigg\{- \rho_H
\frac{\delta{\cal{N}}}{\delta{A}}
+  \frac{\delta{\cal{N}}_k}{\delta{A}}s^k   + \frac{1}{2}{\cal{N}}  \frac{\delta{\gamma}_{kl}}{\delta{A}}s^{kl}\Bigg\},\nb\\
J_{\varphi} &=& - \frac{1}{N}\Bigg\{\frac{1}{\sqrt{g}}\left(B \sqrt{g}\right)_{,t} - \nabla_{i}\Big[B\left(N^i + N \nabla^i\varphi\right)\Big]\nb\\
&& ~~~~~~~~~ - \nabla_i\left(N\Omega^5 s^i\right)\Bigg\},
\eqn
where $\Omega \equiv 1 - a_2 \sigma$, and
\bqn
\lb{B.2}
B &\equiv& - \Omega^{3}(\sigma)\Bigg\{a_1\rho_H  - \frac{2a_2\left(1- a_2\sigma\right)}{N}s^k\left(N_k + N\nabla_k\varphi\right)\nb\\
&& -  a_2\left(1- a_1\sigma\right)\left(1- a_2\sigma\right)g_{ij}s^{ij}\Bigg\},
\eqn
and
\bqn
\lb{B.3}
 \frac{\delta{\cal{N}}}{\delta{N}} &=& 1 + \frac{1}{2}a_1 \left(\nabla_k\varphi\right)^2, \nb\\
 \frac{\delta{\cal{N}}_i}{\delta{N}} &=& \frac{\Omega}{N}\Bigg\{N\Omega\nabla_{i}\varphi \nb\\
 && ~~~~~~ + 2a_2\left(N_i + N\nabla_i\varphi\right)
 \left[\sigma + \frac{1}{2}\left(\nabla_k\varphi\right)^2\right]\Bigg\},\nb\\
  \frac{\delta{\gamma}_{ij}}{\delta{N}} &=& \frac{2a_2\Omega}{N} \left[\sigma + \frac{1}{2}\left(\nabla_k\varphi\right)^2\right]g_{ij},\nb\\
   \frac{\delta{\cal{N}}}{\delta{N}_i}  &=& a_1\nabla^i\varphi,\nb\\
   \frac{\delta{\cal{N}}_k}{\delta{N}_i}  &=& \frac{\Omega}{N}\Big\{N\Omega\delta^{i}_{k}
   + 2a_2\left(N_k + N\nabla_k\varphi\right)\nabla^i\varphi\Big\},\nb\\
\frac{\delta{\gamma}_{kl}}{\delta{N}_i}    &=& \frac{2a_2\Omega}{N}g_{kl}\nabla^i\varphi,\nb\\
\frac{\delta{\cal{N}}}{\delta{g}_{ij}} &=& -a_1\Bigg[N^{(i}\nabla^{j)}\varphi + \frac{1}{2}N \left(\nabla^i\varphi\right)\left(\nabla^j\varphi\right)\Bigg],\nb\\
\frac{\delta{\cal{N}}_k}{\delta{g}_{ij}} &=&-\frac{2a_2\Omega}{N}\left(N_k + N\nabla_k\varphi\right)\nb\\
&& \times \Bigg[N^{(i}\nabla^{j)}\varphi + \frac{1}{2}N \left(\nabla^i\varphi\right)\left(\nabla^j\varphi\right)\Bigg],\nb\\
 \frac{\delta{\gamma}_{kl}}{\delta{g}_{ij}} &=& \frac{1}{2}\Omega^2\left(\delta^i_k\delta^j_l  + \delta^i_l\delta^j_k\right)\nb\\
 && - \frac{2a_2\Omega g_{kl}}{N} \Bigg[N^{(i}\nabla^{j)}\varphi + \frac{1}{2}N \left(\nabla^i\varphi\right)\left(\nabla^j\varphi\right)\Bigg],\nb\\
  \frac{\delta{\cal{N}}}{\delta{A}} &=& -a_1,\nb\\
  \frac{\delta{\cal{N}}_i}{\delta{A}} &=& - \frac{2a_2\Omega}{N}\left(N_i + N\nabla_i\varphi\right),\nb\\
    \frac{\delta{\gamma}_{ij}}{\delta{A}} &=& - \frac{2a_2\Omega}{N}g_{ij}.
\eqn

For the gauge $\varphi = 0$, the above expressions reduce to
\bqn
\lb{B.4}
 \frac{\delta{\cal{N}}}{\delta{N}} &=& 1, \;\;\;\;
 \frac{\delta{\cal{N}}_i}{\delta{N}} = \frac{2a_2\sigma \Omega}{N} N_i,\;\;\;
  \frac{\delta{\gamma}_{ij}}{\delta{N}} = \frac{2a_2\sigma \Omega}{N}  g_{ij},\nb\\
   \frac{\delta{\cal{N}}}{\delta{N}_i}  &=& 0,\;\;\;
   \frac{\delta{\cal{N}}_k}{\delta{N}_i}  =  \Omega^2\delta^{i}_{k},\;\;\;
\frac{\delta{\gamma}_{kl}}{\delta{N}_i}    = 0,\;\;\;
\frac{\delta{\cal{N}}}{\delta{g}_{ij}} = 0,\nb\\
\frac{\delta{\cal{N}}_k}{\delta{g}_{ij}} &=&0,\;\;\;
 \frac{\delta{\gamma}_{kl}}{\delta{g}_{ij}} = \frac{1}{2}\Omega^2\left(\delta^i_k\delta^j_l  + \delta^i_l\delta^j_k\right),\nb\\
  \frac{\delta{\cal{N}}}{\delta{A}} &=& -a_1,\;\;\;
  \frac{\delta{\cal{N}}_i}{\delta{A}} = - \frac{2a_2\Omega}{N}N_i,\nb\\
    \frac{\delta{\gamma}_{ij}}{\delta{A}} &=& - \frac{2a_2\Omega}{N}g_{ij},\; \; (\varphi = 0).
\eqn
Inserting the above expressions into Eq.(\ref{B.1}), we find that
\bqn
\lb{B.5}
J^{t} &=& 2\Omega^{3}\Bigg\{- \rho_H
+  \frac{2a_2\sigma\Omega}{N}N_is^i   + a_2\sigma\Omega(1-a_1\sigma)g_{ij}s^{ij}\Bigg\},\nb\\
J^{i} &=& - \Omega^{5}s^i,\nb\\
{\cal{N}}^{ij} &=& (1-a_1\sigma)\Omega^5s^{ij},\nb\\
J_{A} &=& 2 \Omega^{3}\Bigg\{a_1 \rho_H  -\frac{2a_2\Omega}{N}N_ks^k  -a_2\Omega(1-a_1\sigma)g_{ij}s^{ij}\Bigg\},\nb\\
J_{\varphi} &=& - \frac{1}{N}\Bigg\{\frac{1}{\sqrt{g}}\left(B \sqrt{g}\right)_{,t} - \nabla_{i}\Big[B\left(N^i + N \nabla^i\varphi\right)\Big]\nb\\
&& ~~~~~~~~~ - \nabla_i\left(N\Omega^5 s^i\right)\Bigg\},\; (\varphi = 0).
\eqn

Note that the solar system tests (with $d =3$)  lead to the constraints \cite{LMWZ},
\bq
\lb{eq8-2}
\left|a_1 - 1\right| < 10^{-5},\;\;\; |a_2| < 10^{-5}.
\eq
In particular, for
\bq
\lb{PPNGR}
(a_1, a_2) = (1, 0),
\eq
 the corresponding parameterized post-Newtonian (PPN) parameters can take the same values as those given in GR.

%
%
%
%
%
%
%
%

\section*{Appendix B:  Some quantities  in $2+1$ dimensional static spacetimes}
\renewcommand{\theequation}{B.\arabic{equation}} \setcounter{equation}{0}

 The quantities $K_{ij},\; R_{ij}, \pi_{ij},\; F_{ij}^A, \; F_{ij}$
and ${\cal{L}}_{i}$ for the static spacetimes (\ref{8}) are given by,
 \bqn
\lb{2.1a}
K_{ij} &=& \frac{2h'f-f'h}{2f^2}\delta{^r _i}\delta{^r _j}+{rh}\delta{^\theta _i}\delta{^\theta _j},\nb\\
\lb{2.1b}
K &=& \frac{2rh'f-rhf'+2fh}{2rf},\nb\\
\lb{2.1c}
R_{ij} &=& - \frac{f'}{2r f}\delta{^r _i}\delta{^r _j}-\frac{1}{2}r f'\delta{^\theta _i}\delta{^\theta _j},\nb\\
\lb{2.1d}
R &=& -\frac{f'}{r},\nb\\
\lb{2.1e}
\pi_{ij} &=& \frac{1}{2rf^2}\Big[(\lambda-1)r(2fh' + f'h)+2\lambda hf\Big]\delta{^r _i}\delta{^r _j}\nb\\
&&+\frac{1}{2f}\Big[2r(\lambda-1)hf +\lambda r(2 h'f- hf')\Big]\delta{^\theta _i}\delta{^\theta _j},\nb\\
\lb{2.1f}
F_{ij}^{A} &=& \frac{2fA' -Af'}{2r f}\delta{^r _i}\delta{^r _j}\nb\\
&&+\frac{1}{2}r\Big(2rA''f+rA'f' -Af'\Big)\delta{^\theta _i}\delta{^\theta _j},\nb\\
\lb{2.1g}
F_{ij} &=& -\frac{1}{2r^3f}\Bigg[\frac{g_2}{\zeta^{2}}\Big(4r f f''-rf'^2 -4ff'\Big)+2\Lambda r^3\Bigg]\delta{^r _i}\delta{^r _j}\nb\\
&&-\frac{1}{2r}\Bigg[\frac{g_2}{\zeta^{2}}\Big(4r^2ff''' + 2r^2f''f'-8f''f\nb\\
&& ~~~~~~~~~ -3r f'^2 +8ff'\Big)+2\Lambda r^3\Bigg]\delta{^\theta _i}\delta{^\theta _j},\nb\\
\lb{2.1h}
{\cal{L}}_{K} &=& -\frac{1}{r^2}\Bigg[\lambda r hf(2h'f- hf') + (\lambda-1) \Bigg(r^2h'^2f^2\nb\\
&& ~~~~~ ~~~~ +h^2f^2-r^2hh'ff'+\frac{1}{4}r^2 h^2f'^2\Bigg)\Bigg],\nb\\
\lb{2.1i}
{\cal{L}}_{\varphi} &=& 0,\nb\\
\lb{2.1j}
{\cal{L}}_{A} &=& A\left(2\Lambda_{g} +\frac{f'}{r}\right),\nb\\
\lb{2.1k}
{\cal{L}}_{V} &=&2\Lambda + \frac{g_2 f'^2}{\zeta^{2} r^2}.
\eqn

For the static spacetimes described by Eq.(\ref{8}),   Eqs.(\ref{eq1}), (\ref{eq2}), (\ref{eq4a}), (\ref{eq4b}), (\ref{eq3}), (\ref{eq5a}), and (\ref{eq5b}) reduce, respectively, to
 \bqn
\lb{B-2}
&& \int f^{3/2}  {dr}\Bigg\{\lambda h^2\left(2 \frac{h'}{h}-\frac{f'}{f}\right)\nb\\
&& + (\lambda-1) r h^2\left(\frac{h'^2}{h^2} - \frac{f'h'}{fh} + \frac{f'^2}{4f^2} + \frac{1}{r^2}\right)\nb\\
&&- \frac{g_2 f'^2 +2\zeta^2\Lambda r^2}{\zeta^2 r f^2}\Bigg\} = - 8\pi G \int{\frac{r J^{t} dr}{\sqrt{f}}}, \\
 \lb{2.2b}
&&(\lambda-1)h\left(\frac{h''}{h}-\frac{f''}{2f}-\frac{f'h'}{2fh}+\frac{f'^2}{2f^2}+\frac{h'}{rh}-\frac{1}{r^2}\right)\nb\\
&&~~~~~~~~~~~~~~ +\frac{ h f'}{2rf}=8\pi G J^r,\\
\lb{2.2c}
&&\Bigg[h''' + 2\frac{h''}{r}-  h'\left(\frac{f''}{f}- \frac{3f'^2}{4f^2} + \frac{1}{r^2}\right) \nb\\
&& - h\left(\frac{f''}{2rf}
 + \frac{f'''}{2f} - \frac{5f''f'}{4f^2} + \frac{f'}{2r^2f}+ \frac{3f'^3}{4f^3} \right.\nb\\
 && ~~~~~~~~ \left. - \frac{f'^2}{2rf^2}  +  \frac{1}{r^3}\right)\Bigg] (\lambda-1)\nb\\
&&- \Lambda_g\left(\frac{h'}{f}+\frac{h}{rf}-\frac{f'h}{2f^2}\right)  =- \frac{8 \pi G}{f} J_\varphi, \\
\lb{2.2d}
&& {f'}+2\Lambda_g r =  - 8\pi G r J_A,  \\
\lb{2.2e1}
&&(\lambda-1)h^2\left(2\frac{h''}{h}-\frac{f''}{f}+\frac{5f'^2}{4f^2}+\frac{h'^2}{h^2}-\frac{2f'h'}{fh}\right.\nb\\
&&~~~~~~~~~~~~~~~ \left. -\frac{f'}{rf}-\frac{1}{r^2}\right) +2(2\lambda-1) \frac{hh'}{r} \nb\\
&&~~~ - \frac{g_2f}{\zeta^2 r^2}\Bigg(4f''-\frac{f'^2}{f}-4\frac{f'}{r}\Bigg)\nb\\
&&~~~ +2\left(\frac{A'f}{r} +A\Lambda_g-\Lambda\right)=\frac{16\pi G}{f}{\cal{N}}_r,  \\
\lb{2.2e2}
&&(\lambda+1)h'^2 -(2\lambda+1)\frac{hh'f'}{f}+(5\lambda+1)h^2\frac{f'^2}{4f^2}\nb\\
&&~~ +(\lambda-1)h^2\left(\frac{4h'}{rh}-\frac{f'}{rf}-\frac{1}{r^2}\right)\nb\\
&&~~ -\frac{g_2f^2}{\zeta^{2}}\left(\frac{4f'''}{rf}+2\frac{f''f'}{rf^2}-\frac{8f''}{r^2f}-\frac{3f'^2}{r^2f^2}+\frac{8f'}{r^3f}\right)\nb\\
&&~~  +2\left(A''f+\frac{A'f'}{2}+\Lambda_g A-\Lambda\right)\nb\\
&&~~~ +\lambda h^2\left(\frac{2h''}{h}-\frac{f''}{f}\right) =16\pi G r^2{\cal{N}}_\theta,
\eqn
where
\bqn
\lb{2.2e3}
J^{i} &=& (J^r, 0, 0),\nb\\
{\cal{N}}_{ij} &=&\frac{1}{f} {\cal{N}}_{r} \delta^{r}_{i} \delta^{r}_{j}  + r^2 {\cal{N}}_{\theta}\delta^{\theta}_{i} \delta^{\theta}_{j}.
\eqn

The functions  $a, b, c, d, P, Q, U, V$ and $W$ appearing in Eqs.(\ref{2.2bA})-(\ref{2.2e2A}) are given by
\bqn
\lb{coefficients}
a(r) &\equiv&
\frac{f'}{2f}-\frac{1}{r}, \nb\\
b(r) &\equiv&  \frac{f'}{2rf} - (\lambda-1)\left(\frac{f''}{2f} - \frac{f'^2}{2f^2}+\frac{1}{r^2}\right),\nb\\
c(r) &\equiv& - (\lambda-1)\left(\frac{f''}{f}-\frac{3f'^2}{4f^2}+\frac{1}{r^2}\right)-\frac{\Lambda_g}{f}, \nb\\
d(r) &\equiv& - (\lambda-1)\Bigg(\frac{f'''}{2f}-\frac{5f'f''}{4f^2}+\frac{f''}{2rf}+\frac{3f'^3}{4f^3}\nb\\
&&-\frac{f'^2}{2rf^2} +\frac{f'}{2r^2f}-\frac{1}{r^3}\Bigg)+\Lambda_g\left(\frac{f'}{2f^2}-\frac{1}{rf}\right),\nb\\
P(r) &\equiv& \Lambda_g \frac{r}{f},\nb\\
Q(r) &\equiv& (\lambda-1)\Bigg(\frac{rhh''}{f}-\frac{rf'hh'}{f^2}+\frac{rh'^2}{2f}+\frac{5rf'^2h^2}{8f^3}\nb\\
&&-\frac{rf''h^2}{2f^2}-\frac{f'h^2}{2f^2}-\frac{h^2}{2f}\Bigg)+(2\lambda-1)\frac{hh'}{f}\nb\\
&&-\frac{g_2}{\zeta^{2}}\left(\frac{2f''}{r}-\frac{2f'}{r^2}+\frac{f'^2}{2rf}\right)-\Lambda\frac{r}{f},\nb\\
U(r) &\equiv& \frac{f'}{2f},\nb\\
V(r) &\equiv& \frac{\Lambda_g}{f},\nb\\
W(r) &\equiv& (\lambda-1)\left(\frac{2hh'}{rf}-\frac{f'h^2}{2rf^2}-\frac{h^2}{2r^2f}\right)\nb\\
&& +(5\lambda+1)\frac{f'^2h^2}{8f^3}
+(2\lambda+1)\frac{f'hh'}{2f^2}\nb\\
&& -\frac{g_2}{\zeta^{2}}\Bigg(\frac{2f'''}{r}
 -\frac{4f''}{r^2}+\frac{f''f'}{rf} -\frac{3f'^2}{2r^2f}+\frac{4f'}{r^3}\Bigg)\nb\\
&& +(\lambda+1)\frac{h'^2}{2f} +\lambda\left(\frac{hh''}{f}-\frac{f''h^2}{2f^2}\right) -\frac{\Lambda}{f}.
\eqn

\section*{Appendix C:   The main properties of the solution with $\Lambda_g\not=0$, $\lambda=1$ and $A_1\not=-1$}
 \renewcommand{\theequation}{C.\arabic{equation}} \setcounter{equation}{0}

In this case, the metric (\ref{eq8-2}) with $a_1 = 1,\; a_2 = 0$  reads,
 \bqn
 \lb{D1}
 ds^2&=&-\left(\sqrt{C_1-\Lambda_gr^2}-\bar{A}_1\right)^2dt^2+\frac{dr^2}{C_1-\Lambda_gr^2}\nb\\
 && +r^2d\theta^2,
 \eqn
where $\bar{A}_1=-(A_1+1)/{A_0}\not=0$. To have the metric coefficients real for $r \in [0, \infty)$, we must assume that $C_1 \ge 0,$ and $\Lambda_g < 0$.
Then, we find that the scalar curvature $R$ is given by
 \bqn
 \lb{D4}
 R&=&2\Lambda_g\frac{3C_1-3\Lambda_gr^2-\bar{A}_1\sqrt{C_1-\Lambda_gr^2}}{\sqrt{C_1-\Lambda_gr^2}\left(\sqrt{C_1-\Lambda_gr^2}-\bar{A}_1\right)},
 \eqn
 which shows that the corresponding spaceitme is singular at both $C_1-\Lambda_gr^2 = 0$ and $\sqrt{C_1-\Lambda_gr^2} - \bar{A}_1 =0$. Then, the physical
 meaning of the solution is unclear, if there is any. 

\end{document}